\def\V{{ \langle \widetilde{V} \rangle}}
\def\tV{{\widetilde{V}_{\alpha}}}
\def\cV{{\widetilde{V}_{\text{calc}}}}
\def\Pt{{P_{mean}(t)}}

\documentclass[a4paper,fleqn]{cas-dc}

\usepackage[numbers]{natbib}

\def\tsc#1{\csdef{#1}{\textsc{\lowercase{#1}}\xspace}}
\tsc{WGM}
\tsc{QE}
\tsc{EP}
\tsc{PMS}
\tsc{BEC}
\tsc{DE}
\begin{document}
\let\WriteBookmarks\relax
\def\floatpagepagefraction{1}
\def\textpagefraction{.001}
\shorttitle{}
\shortauthors{}

\title [mode = title]{L\'evy noise effects on Josephson junctions}

\author[1,2]{C. Guarcello}[orcid=0000-0002-3683-2509]
%\cormark[1]
\ead{cguarcello@unisa.it}
\address[1]{Dipartimento di Fisica ``E.R. Caianiello'', Universit\`a di Salerno, Via Giovanni Paolo II, 132, I-84084 Fisciano (SA), Italy}
\address[2]{INFN, Gruppo Collegato Salerno, I-84084 Fisciano (SA), Italy}

%\cortext[cor1]{Corresponding author}

\begin{abstract}
We review three different approaches to investigate the non-equilibrium stochastic dynamics of a Josephson junction affected by L\'evy-distributed current fluctuations. 
First, we study the lifetime in the metastable superconducting state of current-biased short and long junctions, in the presence of Gaussian and L\'evy noise sources. We highlight the noise-induced nonmonotonic behavior of the mean switching time as a function of noise intensity and driving frequency, that is the noise enhanced stability and the stochastic resonant activation, respectively. 
Then, we characterize the L\'evy noise source through the average voltage drop across a current-biased junction. The voltage measurement versus the noise intensity allows to infer the value of the stability index that characterizes L\'evy-distributed fluctuations. The numerical calculation of the average voltage drop across the junction well agrees with the analytical estimate of the average velocity for L\'evy-driven escape processes from a metastable state. 
Finally, we look at the distribution of switching currents out of the zero-voltage state, when a L\'evy noise signal is added to a linearly ramped bias current. The analysis of the cumulative distribution function of the switching currents gives information on both the  L\'evy stability index and the intensity of fluctuations. We present also a theoretical model to catch the features of the L\'evy signal from a measured distribution of switching currents.
The phenomena discussed in this work can pave the way for an effective and reliable Josephson-based scheme to characterize L\'evy components eventually embedded in an unknown noisy signal.
\end{abstract}

\begin{keywords}
Josephson junction \sep Nonequilibrium Stochastic dynamics \sep Multistable Systems \sep Non-Gaussian noise \sep L\'evy noise \sep Switching current distribution
\end{keywords}

\maketitle

\section{Introduction}
\label{Sec01}

A Josephson junction (JJ) is a natural threshold detector for current fluctuations, being essentially a current-controlled metastable system working on an activation mechanism at cryogenic temperatures~\cite{Bar82,Lik86}. In common setups, the junction is biased by a fixed or a linearly ramped current until the system switches to the finite voltage state. When a measurable voltage appears, one can record also the current, or the time, at which the passage to the voltage-state has occurred. Thus, Josephson devices are nowadays often employed for sensing and detection applications~\cite{Pea98,Ber13,Ull15,Wal17,Oel17,GuaBra19,GuaBraSol19,Bra19,Rev20,Pie21,Gol21,Yab21,Gua21}. 

In the past two decades, the seminal cue of Refs.~\cite{Pek04,Ank05,Tob04} has prompted several experimental approaches for Josephson-based noise detection~\cite{Pek05,Ank07,Suk07,Tim07,Pel07,Hua07,Gra08,Nov09,LeM09,Urb09,Fil10,Add12,Oel13,Sol15,Sai16}. 
In these examples, the interesting information content is commonly obtained from the highest moments of electrical noise, despite the discrepancies eventually observed from a typical Gaussian response can be small and an experimental measurement of higher moments, beyond the variance, can be demanding.

Thus, in this work we review alternative methods to characterize a specific kind of non-Gaussian fluctuations, namely, the $\alpha$-stable L\'evy noise, which correspond to stochastic processes that exhibit very long {\it distance} in a single displacement, namely, a {\it flight}, by looking at the switches from the superconducting to the dissipative regime of a JJ. The proposed setup assumes the presence of a L\'evy noise source together with an intrinsic, Gaussian thermal noise source. We show that the useful information content can be effectively retrieved from three different Josephson observables: the distributions of times the system takes to leave the superconducting regime, the average value of the voltage drop across the device, and the probability/cumulative distribution functions of the bias current values at which the junction switches. Accordingly, three different strategies for the characterization of a L\'evy noise source can be designed.

Nowadays JJs are used in different application fields~\cite{Taf19}.
For instance, superconducting circuits that store and manipulate information are particularly appealing also in view of their low-energy operation. Among these, Josephson-based \emph{memdevices} were recently suggested~\cite{Chu03,Peo14,Pfe16,She16,Sal17,GuaSol17,She18}. These are circuit elements, that is, resistors, capacitors, and inductors, with memory~\cite{Chu71,Chu76,DiV09,Per11}, \emph{i.e.}, with characteristics that depend on the past states through which the system has evolved. Memdevices can be also combined in complex circuits to perform logic and unconventional computing operations in massive parallel schemes, and in the same physical location where storing occurs. 
Thermal noise plays an important role in the response of these Josephson-based elements~\cite{Peo14,Pfe16,GuaSol17,She18}, as much as in the behavior of every solid-state memdevice. Indeed, the investigation of the complex stochastic multistable response, the nonlinear relaxation phenomena, and the role of noise inherent in condensed matter systems can improve the understanding of the fundamental conduction and switching mechanism of memristive devices~\cite{Fil19,Mik16,Mik20}. 
Moreover, the insight of the diffusion processes, the instability, and the degradation of memristive system is fundamental for the comprehension of the switching mechanism~\cite{Pan05,Yak19}.

In this work we demonstrate how a noise source of a different nature from the usual Gaussian thermal statistic can have a significant effect on the response of a JJ. In particular, our findings are important to understand the role of a non-Gaussian noise source on the metastable dynamics of out-of-equilibrium processes in Josephson devices. This is relevant not only in the general framework of the nonequilibrium statistical mechanics and interdisciplinary physics models, where the positive role of noise is well-established~\cite{Dub05,Giu09,Piz10,Den13,DenVal13,Gua16,Car18,Car21}, but also to improve the performance of this device. 

The paper is organized as follows. 
In Sec.~\ref{Sec02}, we discuss the operating principles of a JJ. 
Section~\ref{Sec03} describes the statistical properties of the L\'evy noise and the method employed for the stochastic simulations. 
In Sec.~\ref{Sec04}, the results in the case of escape processes driven by L\'evy flights from a washboard potential well are shown and analyzed. 
In detail, in Sec.~\ref{Sec04a} the theoretical results for the behaviors of the mean switching time as a function of the frequency of the external driving current and the intensity of the noise fluctuations are discussed, both in the cases of short and long JJs. 
In Sec.~\ref{Sec04b} we show the evolution of the average voltage drop across a short JJ; we illustrate also the comparison between the numerical results and the analytical estimate of the average velocity of the phase particle, in the case of a L\'evy--driven escape process from a washboard-potential well.
In Sec.~\ref{Sec04c}, we present the theoretical distributions of switching currents, both calculated numerically and estimated analytically, in the presence of L\'evy fluctuations.
In Sec.~\ref{Sec05}, conclusions are drawn.

\section{The Josephson phase dynamics}
\label{Sec02}

When talking about a JJ one refers to a device formed by two superconducting electrodes connected by a weak link, which can be a tunnel insulating barrier, a point contact, a Dayem bridge, or a proximity connection. The behavior of a tunnel JJ is described in terms of the two well-known Josephson equations~\cite{Jos62,Jos74}:
\begin{equation}\label{JJvoltage}
I_J = I_c\sin\varphi \qquad\text{and}\qquad V = \frac{\hbar}{2e}\frac{d\varphi}{dt},
\end{equation}
where $e$ is the charge of the electron and $\hbar$ is the reduced Planck constant. The first equation gives the supercurrent through a JJ, with $I_c$ being the critical current of the device and $\varphi$ is the difference between the phases of the superconducting wave function in two weakly connected superconductors. At currents above $I_c$, a voltage drop, $V$, appears across the system and the phase difference evolves at a rate determined by Eq.~\eqref{JJvoltage}.
This means that, in the voltage state the JJ emits radiation with a frequency $\nu_J=2eV/h\simeq483V(\text{MHz}/\mu\text{V})$~\cite{Bar82,Lik86}. When a voltage-biased JJ is exposed to an external microwave source with frequency $\nu$, its I-V characteristic is characterized by sudden current steps, the so-called Shapiro steps, placed at voltages $V=nh\nu/2e$, where $n=0,1,2,3\ldots$.
Commonly, Josephson-based devices are thin-film planar, single--layered or multilayered structures, and the most used are superconductor -- insulator -- superconductor (SIS) Nb-- or Al--based junction.

%Fig. 1
\begin{figure}[t!!]
\includegraphics[width=\columnwidth]{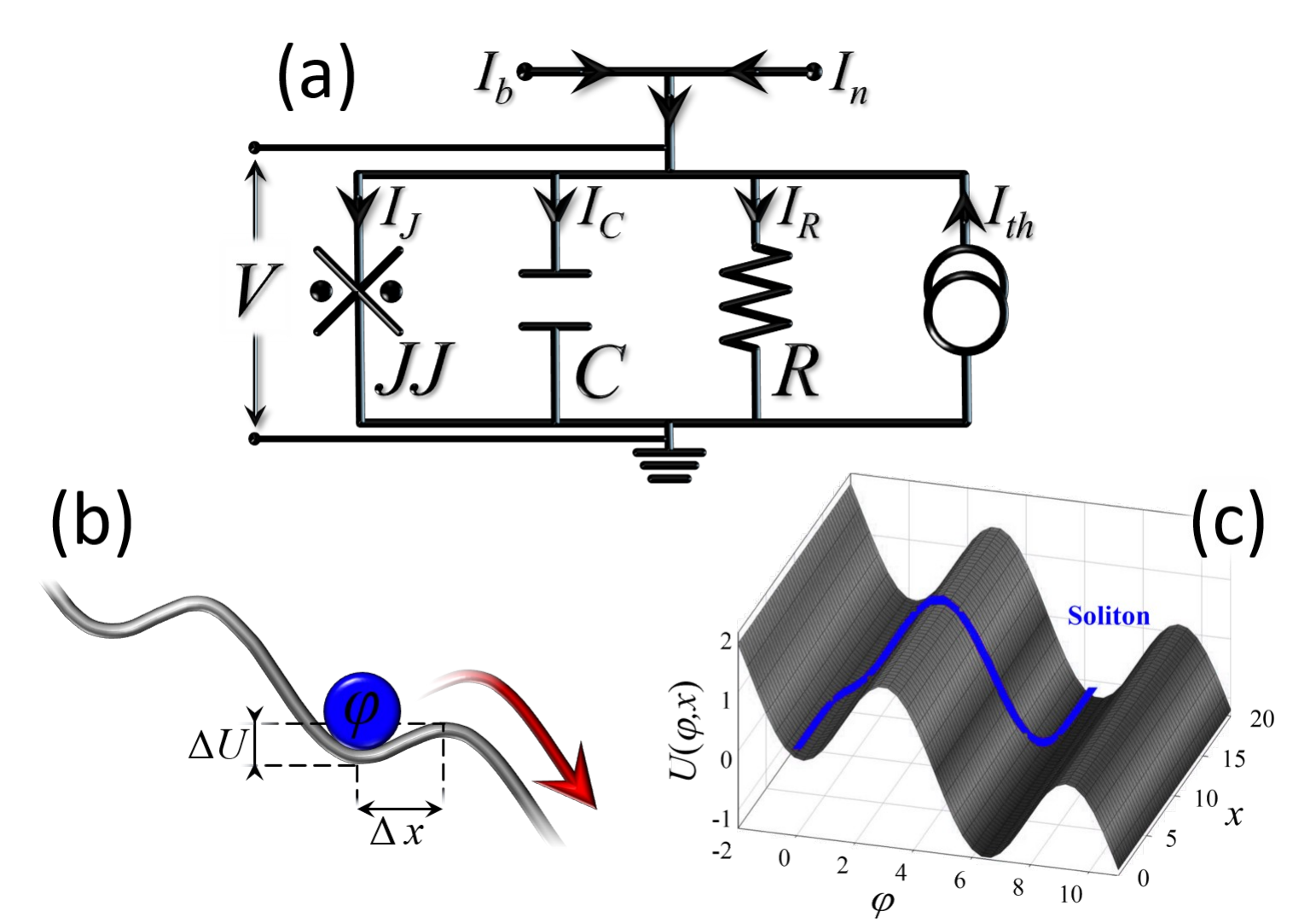}
\caption{(a) Electrical model of a short JJ. The bias current, $I_b$, the non-Gaussian current source, $I_n$, and the thermal noise current, $I_{th}$ are included in the diagram. $R$ represents the quasiparticle tunneling resistance and $C$ the junction capacitance. (b) Sketch of the phase particle within a potential minimum of the tilted washboard potential $U$. The barrier height, $\Delta U$, and the distance between the minimum and maximum of the potential, $\Delta x$, are also shown. (c) Tilted washboard potential in the case of a long JJ. A 2$\pi$-twist of the phase, that is, a soliton, lying on two adjacent valleys of the potential profile is also depicted.}
\label{Fig01}
\end{figure}

Equation~\eqref{JJvoltage} refers only to the Cooper pair supercurrent, but if $V >0$ and $T>0$ also a quasiparticle current component flows through the system. In fact, the full time--dependent response of a JJ is appropriately described in terms of the equivalent circuit depicted in Fig.~\ref{Fig01}, which comprises the capacitance $C = \epsilon_r \epsilon_0 A / t_{ox}$ (where $A$ is the junction area, $t_{ox}$ and $\epsilon_r$ are, respectively, the thickness and the relative permittivity of the oxide layer, and $\epsilon_0$ is the vacuum permittivity) between the superconducting plates and the normal resistance $R$ of the weak link itself. Due to the capacitive effects, for a typical quite high $R$ value of tunnel junction the static I-V characteristics are hysteretic. If the current exceeds the critical value, the JJ switches to the voltage state and can remain there even when the bias voltage drops below the characteristics gap voltage. In this case, for removing the emerging hysteresis an additional shunt resistance $R_{sh}$ can be eventually used, so that the effective resistance becomes $RR_{sh}/(R+R_{sh})$.

The resistor $R$ depends nonlinearly both on voltage and temperature, although, as it is related to the dissipation in the system, in the case of a moderate or weak damping its nonlinearity is often not taken into account.
Associated to the Ohmic resistor $R$ there is a thermal noise current source, indicated with $I_{th}$ in Fig.~\ref{Fig01}, whose spectral power density is frequency-independent but temperature-dependent (Johnson noise)~\cite{Kogan96}.
In Fig.~\ref{Fig01} we included also the electric current, $I_b$, biasing the junction and an additional current term $I_n$, that in this work will have the role of the additional noise source affecting the junction. We observe also that the Josephson element can be alternatively seen as a nonlinear inductance $L_J=L_c/\cos\varphi$, where we have defined $L_c=\frac{\hbar}{2e}\frac{1}{I_c}$.

Thus, taking into account all the current contributions and Josephson equations, from Kirchhoff's laws it is straightforward to derive the following Langevin equation~\cite{Ben84}
\begin{equation}\label{eqJJ}
C\frac{\hbar}{2e}\frac{d^2\varphi}{dt^2}+\frac{1}{R}\frac{\hbar}{2e}\frac{d\varphi}{dt}+I_c\sin\varphi = I_b + I_s + I_{th}\left(t \right).
\end{equation}
This equation is usually handled in two different ways, according to the chosen time normalization. In fact, we can define two different timescales, and therefore two different frequencies, 
\begin{equation}\label{frequencies}
\omega_p=\frac{1}{\sqrt{L_cC}}=\sqrt{\frac{2e}{\hbar}\frac{I_c}{C}} \;\quad\text{and}\;\quad\omega_c=\frac{R}{L_c}=\frac{2e}{\hbar}\frac{I_c}{R},
\end{equation}
that is, the plasma and the characteristic Josephson frequencies, respectively, through which we can conveniently recast Eq.~\eqref{eqJJ} in a quite compact form
\begin{equation}\label{eqJJ_2}
\frac{1}{\omega_p^2}\frac{d^2\varphi}{dt^2}+\frac{1}{\omega_c}\frac{d\varphi}{dt}+\sin\varphi = i_b +i_n + i_{th} (t).
\end{equation}
The current contributions on the rhs of this equation are normalized to the critical current $I_c$. Therefore, introducing the normalized times $\tau_p=\omega_p t$ and $\tau_c=\omega_c t$ one can directly obtain the following two equations
\begin{eqnarray}\label{eqJJ_norm_p}
\frac{d^2\varphi}{d\tau_p^2}+\beta_J\frac{d\varphi}{d\tau_p}+\sin\varphi = i_b +i_n + i_{th} (\tau_p)\\
\beta_c\frac{d^2\varphi}{d\tau_c^2}+\frac{d\varphi}{d\tau_c}+\sin\varphi = i_b +i_n + i_{th} (\tau_c),\label{eqJJ_norm_c}
\end{eqnarray}
including the Johnson and Stewart-McCumber damping parameters~\cite{Bar82} 
\begin{equation}\label{dampangparameters}
\beta_J=\frac{\omega_p}{\omega_c}\qquad\text{and}\qquad\beta_c=\frac{\omega_c^2}{\omega_p^2},
\end{equation}
so that $\beta_c=\beta_J^{-2}$. A highly damped (or overdamped) JJ has $\beta_c\ll 1$ (or $\beta_J\gg1$), that is a small capacitance and/or resistance, while a weakly damped (underdamped) JJ has $\beta_c\gg 1$ (or $\beta_J\ll1$), that is large capacitance and/or resistance.

One might wonder which of the two normalizations is best suited to tackle a specific physical problem. Certainly, the final outcome cannot depend on this choice, which should only provide an advantage from a purely computational point of view. In fact, choosing the normalization wisely can allow to achieve the desired result in a shorter simulation time: the key parameter to address this point is the damping.
Indeed, being $t=\tau_p/\omega_p =\tau_c/\omega_c$, the normalized times are linked by the relation $\tau_p/\tau_c=\beta_J $. This means that, at a given $t$, $\tau_p$ will be larger (smaller) than $\tau_c$ in the case of an overdamped (underdamped) JJ. In other words, it is more convenient to study an overdamped system ($\beta_J\gg1$) in terms of the Stewart-McCumber time-normalization Eq.~\eqref{eqJJ_norm_c}, and an underdamped system ($\beta_J\ll1$) in terms of the Johnson normalization Eq.~\eqref{eqJJ_norm_p}.

Thermal effects are taken into account through the stochastic noise term $I_{th}$, which has the following statistical properties
\begin{equation}\label{correl}
\langle I_{th}(t) \rangle = 0 \quad\text{and} \quad \langle I_{th}(t),I_{th}(t') \rangle = 2 \frac{k_BT}{R} \delta(t-t'),
\end{equation}
where $k_B$ is the Boltzmann constant, $\delta()$ is the Dirac delta function, and $T$ is the temperature of the system. According to the time normalization adopted, the previous equations can be rewritten as
\begin{equation}\label{correl_norm_1}
\langle i_{th}(\tau_p) \rangle = 0 \;\;\text{and} \;\; \langle i_{th}(\tau_p),i_{th}(\tau_p') \rangle = 2 \beta_J D \delta(\tau_p-\tau_p')
\end{equation}
or
\begin{equation}\label{correl_norm_2}
\langle i_{th}(\tau_c) \rangle = 0 \;\;\text{and} \;\; \langle i_{th}(\tau_c),i_{th}(\tau_c') \rangle = 2 D \delta(\tau_c-\tau_c').
\end{equation}
Here, we defined the noise intensity,
\begin{equation}\label{noiseampl}
D=\frac{k_B T}{E_{J_0}},
\end{equation}
as the ratio between the thermal energy and the Josephson coupling energy $E_{J_0}=(\hbar/2e)I_c$. The noise intensity can be also rewritten as $D=I_{\scriptscriptstyle{th}}/I_c$, where $I_{\scriptscriptstyle{th}}=\frac{2e}{\hbar}k_BT$ is the \emph{equivalent thermal noise current}, and takes the value $ I_{\scriptscriptstyle{th}}\simeq 42\;\text{nA}$ at $T=1\;K$.
The role of Gaussian thermal noise on a short junction was throughly studied~\cite{Ort04,Gor06,Sun07,Gor08}, even on the magnetization switching phenomenon occurring in a current-biased ferromagnetic anomalous junction~\cite{GuaBer20,GuaBer21}.

Interestingly, Eq.~\eqref{eqJJ_2} serves also to describe the motion of a damped pendulum with torque, or even the dynamics of a particle in a tilted washboard-like potential with friction~\cite{Bar82}, see Fig.~\ref{Fig01}(b). The washboard potential responsible for the phase-particle dynamics can be written as~\cite{Bar82, Ben84}
\begin {equation}\label{eq:potential}
U(\varphi,i_b) = E_{J_0}[1 - \cos (\varphi) - i_b\; \varphi].
\end {equation}
In this picture, the plasma frequency $\omega_p$ can be seen as a small-amplitude oscillation frequency of the phase particle on the bottom of a potential well.
The washboard potential is composed by a periodical sequence of minima and maxima, located at
\begin{equation}\label{min_max}\nonumber
\varphi^n_{min}=\arcsin( i_b)+2n\pi\quad\varphi^n_{\text{max}}=(2n-1)\pi-\arcsin( i_b),
\end{equation}
with $n=0,\pm 1,\pm 2,\ldots$. In the following, for the initial condition to solve the differential equation we assume the phase particle at rest in the bottom of a potential well, that is we impose $\varphi(0)=\varphi^0_{min}$ and $d\!\varphi(t=0)/dt=0$, while the threshold values to recognize a switching event will be the $\varphi^{n}_{\text{max}}$ maxima with $n=0,1$.

The bias current $i_b$ flowing through the junction acts as a tilting of $U(\varphi,i_b)$, so that for $i_b < 1$ the potential shows metastable wells with a barrier height given by~\cite{Ben84}
\begin {equation}\label{eq:barrier}
\Delta\mathcal{U}(i_b)=\frac{\Delta U(i_b)}{E_{J_0}}=2 \left[ \sqrt{1-i_b^2} - i_b \cos^{-1}(i_b) \right].
\end {equation}
Instead, for $i_b \geq 1$ the potential profile has no maxima and minima. Besides skewing the potential, the bias current changes the shape of the metastable wells, so that also the value of the plasma frequency changes. In the harmonic approximation, it can be written as $\widetilde{\omega}_J(i_b)=\omega_J(i_b)/\omega_p=\left(1-i_b^2\right)^{1/4}$.

Thermal fluctuations can drive the system out of a washboard potential minimum. In fact, in the presence of noise the phase solution in the potential minima becomes metastable, and the resulting thermally activated escape rate $\Gamma_{\textsc{ta}}$ is given by the Kramers approximation~\cite{Kra40, Risken89}, for moderate damping, as
\begin{equation}\label{r0_full}
 \Gamma_{\textsc{ta}}(\alpha,i_b,D) = \left( \sqrt{\frac{\alpha^2}{4}+\widetilde{\omega}_J^2(i_b)}-\frac{\alpha}{2} \right) \frac{e^{-\frac{\Delta\, \mathcal{U}(i_b)}{ D}}}{2\pi}.
\end {equation}

Another escape mechanism can drive the particle out of the metastable state, that is the macroscopic quantum tunneling (MQT), with a rate given by~\cite{Dev85,Bla16}
\begin{equation}\label{MQTescape}
\Gamma_{\textsc{mqt}}(\alpha,i_b) =a_q\frac{\widetilde{\omega}_J(i_b)}{2\pi}e^{ \left [ -7.2\frac{\Delta U(i_b)}{\hbar \omega_J(i_b)}\left ( 1+0.87\alpha \right ) \right ]},\end{equation}
where $a_q=\sqrt{120\pi\left [ \frac{7.2 \Delta U(i_b)}{\hbar \omega_J(i_b)} \right ]}$.
The tunneling rate is independent of the temperature, being mainly a function of the height $\Delta U$ of the potential barrier to overcome. Thus, decreasing the temperature, the thermal activation rate can reduce so much that MQT processes dominate the escape dynamics.
The temperature that separates the thermal regimes in which the rates of the two processes, thermal activation and quantum tunneling, equal each other is called \emph{crossover temperature}~\cite{Gra84,Bla16}, $T_{\text{cr}}$. In the case of $\alpha\ll1$ and $a_q\approx1$, this threshold temperature can be simply estimated as $T_{\text{cr}}\approx\frac{\hbar\omega_p}{7.2k_B}$, and acquires a value of $T_{\text{cr}}\simeq106\;\text{mK}$ if $\omega_p = 0.1\;\textup{THz}$.

What has been said so far holds for a short tunnel JJ, that is a junction in which the magnetic field due to the Josephson current is negligible compared to the externally applied magnetic field. In this case, the dimensions of the device are smaller than the characteristics length-scale for this kind of systems, that is the Josephson penetration depth $\lambda_J=\sqrt{\frac{\Phi_0}{2\pi \mu_0}\frac{1}{t_d J_c}}$~\cite{Bar82}. Here, $t_d=\lambda_{L,1}+\lambda_{L,2}+d$ is the effective magnetic thickness (with $\lambda_{L,i}$ and $d$ being the London penetration depth of the $i$-th electrodes and the insulating layer thickness, respectively), $\mu_0$ is the vacuum permeability, and $J_c$ is the critical current area density. 
To give a realistic number, for a Nb junction we estimate a low-temperature critical current area density~\cite{Bar82} $J_c=\frac{\pi}{2}\frac{1.764kT_c}{eR_a}\simeq44\;\mu\text{A}/{\mu\text{m}}^2$ (with a normal-state resistance per area $R_a=50\;\Omega{\mu\text{m}}^2$) and an effective magnetic thickness $t_d\simeq161\;\text{nm}$ (assuming $T_c=9.2\;\text{K}$ and $\lambda_{L}^0=80\;\text{nm}$ for Nb), so that $\lambda_J\sim 6\;\mu\text{m}$.

In the case of a long junction one of the lateral dimensions $L$ or $W$ is greater than the Josephson penetration depth, specifically $L>\lambda_J$ and $W\ll\lambda_J$, and the magnetic field generated by the Josephson current itself is no longer negligible. The behavior of a long JJ can be described in terms of a partial differential equation, namely, the perturbed sine-Gordon (SG) equation, for the phase $\varphi(x,t)$~\cite{Bar82,Lik86}:
\begin{equation}\label{eqJJ_e}
\frac{1}{\omega_p^2}\frac{d^2\varphi}{dt^2}+\frac{1}{\omega_c}\frac{d\varphi}{dt}-\lambda_J^2\frac{d^2\varphi}{dx^2}+\sin\varphi = i_b +i_n + i_{th} (x,t).
\end{equation}
The boundary conditions take into account an external magnetic field $H_{\text{ext}}$
\begin{equation}
\frac{d \varphi (0,t)}{dx} = \frac{d \varphi (L,t)}{dx} = 2\frac{H_{\text{ext}}}{H_{c_1}}\frac{1}{\lambda_J},
\label{boundary}
\end{equation}
where $H_{c_1}=\frac{\Phi_0}{\pi \mu_0 t_d\lambda_J}$ is the so-called first critical field of the junction~\cite{Gold01}.

The SG equation admits a traveling wave solution, called \emph{soliton}~\cite{Ust98}, \emph{i.e.}, a phase twist from 0 to 2$\pi$, which corresponds to a flux quantum $\Phi_0=h/(2e)$ along the junction~\cite{McL82}, see Fig.~\ref{Fig01}(c). Thus, speaking about Josephson devices a soliton is usually referred to as a \emph{fluxon}, or Josephson vortex, with a width of the order of the Josephson penetration depth $\lambda_J$, which is surrounded by circulating supercurrent. The interplay between thermal effects and solitons excited in a long JJ was already extensively investigated, both as the noise influence on the soliton dynamics~\cite{Fed07,Fed08,Fed09,Aug09,Pan12,sol13,Val14,GuaVal15,Pan17} and thermal transport sustained by solitons across a long junction~\cite{GuaGia16,GuaSol18,GuaSolBra2018,Gua18,GuaSol19}. Instead, in this work we make a step forward showing how soliton-sustained switching processes in long JJs modify in the presence of a non-Gaussian noise source.
In the following, we use also the words \emph{string} when we refer to the whole long junction phase, see Fig.~\ref{Fig01}(c), and \emph{cell} to indicate each element, with a size $\Delta \chi$, forming the string.

\section{The L\'evy noise}
\label{Sec03}

In this section we summarize the features of $\alpha$-stable L\'evy distributions~\cite{Khi36,Khi38,Gne54,Fel71,Ber96,Sat99}. A random non-degenerate variable $X$ is stable if
\begin{equation}
\forall n\in\mathbb{N}, \exists (a_n,b_n) \in \mathbb{R}^+\times\mathbb{R}: \qquad X+b_n=a_n\sum_{j=1}^{n} X_j, \label{AlfaStable}
\end{equation}
with $X_j$ being independent copies of $X$. Furthermore, $X$ is ``strictly stable'' if and only if $b_n=0 \,\,\, \forall n$. The well
known Gaussian distribution belongs to this class. This definition does not give a parametric form of stable
distributions, but the characteristic function, instead, permits to manage them. The generic definition of the characteristic function for a
random variable $X$ with an given distribution function $F(x)$ is
\begin{equation}\label{GenerealCharFunc}
\phi (u) = \left < e^{iuX} \right > = \int_{-\infty}^{+\infty}e^{iuX}dF(x).
\end{equation}
Accordingly, a random variable $X$ is stable if, and only if,
\begin{equation}
\exists (\alpha ,\sigma, \beta, \mu )\in\,\,(0, 2]\times \mathbb{R}^+\times [-1, 1]\times \mathbb{R}: \qquad X\overset{d}{=}\sigma Z+\mu,
\label{XStableFunc}
\end{equation}
with $Z$ being a random variable with a characteristic function equal to
\begin{eqnarray}
\phi(u)=\left\{\begin{matrix}
\exp \left \{-\left | u \right |^\alpha\left [ 1-i\beta \tan\frac{\pi\alpha}{2}(\textup{sign}u) \right ]\right \} \,\,\,\, \alpha\neq 1\\
\exp \left \{-\left | u \right |\left [ 1+i\beta\frac{2}{\pi}(\textup{sign}u)\log \left | u \right |\right ]\right \} \,\,\,\, \alpha=1
\end{matrix}\right. .\label{XCharFunc}
\end{eqnarray}
If $\alpha=1$, $0\cdot \log0$ is taken as $\lim_{x\to0} x\log x=0$, from which $\phi(0)=1$.

The definition in Eq.~\eqref{XStableFunc} requires four parameters: a \emph{stability index} (or characteristic exponent) $\alpha\in(0,2]$,
an \emph{asymmetry parameter} $\beta\in[-1,1]$, a scale parameter $\sigma>0$, and a location parameter $\mu$. 
The names of these parameters reflects their physical meaning. 
Stable distributions are often indicated with $S_{\alpha}(\sigma, \beta, \mu)$. In particular, if $\sigma=1$ and $\mu=0$ we deal with \emph{standard} distributions. All stable distributions are unimodal, but there is no known formula for the location of the mode. If we set $\beta=0$, the resulting distribution is symmetric. Instead, the $\alpha$ value influences how the tails of the distribution go to zero: in fact, the stability index entails the asymptotic long-tail power law for the $x$-distribution, which for $\alpha<2$ is of the $\left | x \right |^{-\left ( 1+\alpha \right )}$ type, while $\alpha=2$ results in the Gaussian distribution. In Eq.~(\ref{XStableFunc}), $\tan \left ( \scriptstyle{\frac{\pi \alpha }{2}} \right )=0$ for $\alpha=2$, that is the characteristic function is real and the distribution is symmetric, independently of $\beta$. As $\alpha$ diminishes, the influence of $\beta$ gets more pronounced, and the left (right) tail gets lighter and lighter if $\beta\to+1 (\beta\to-1)$.

%Fig.2
\begin{figure}[t!!]
\includegraphics[width=\columnwidth]{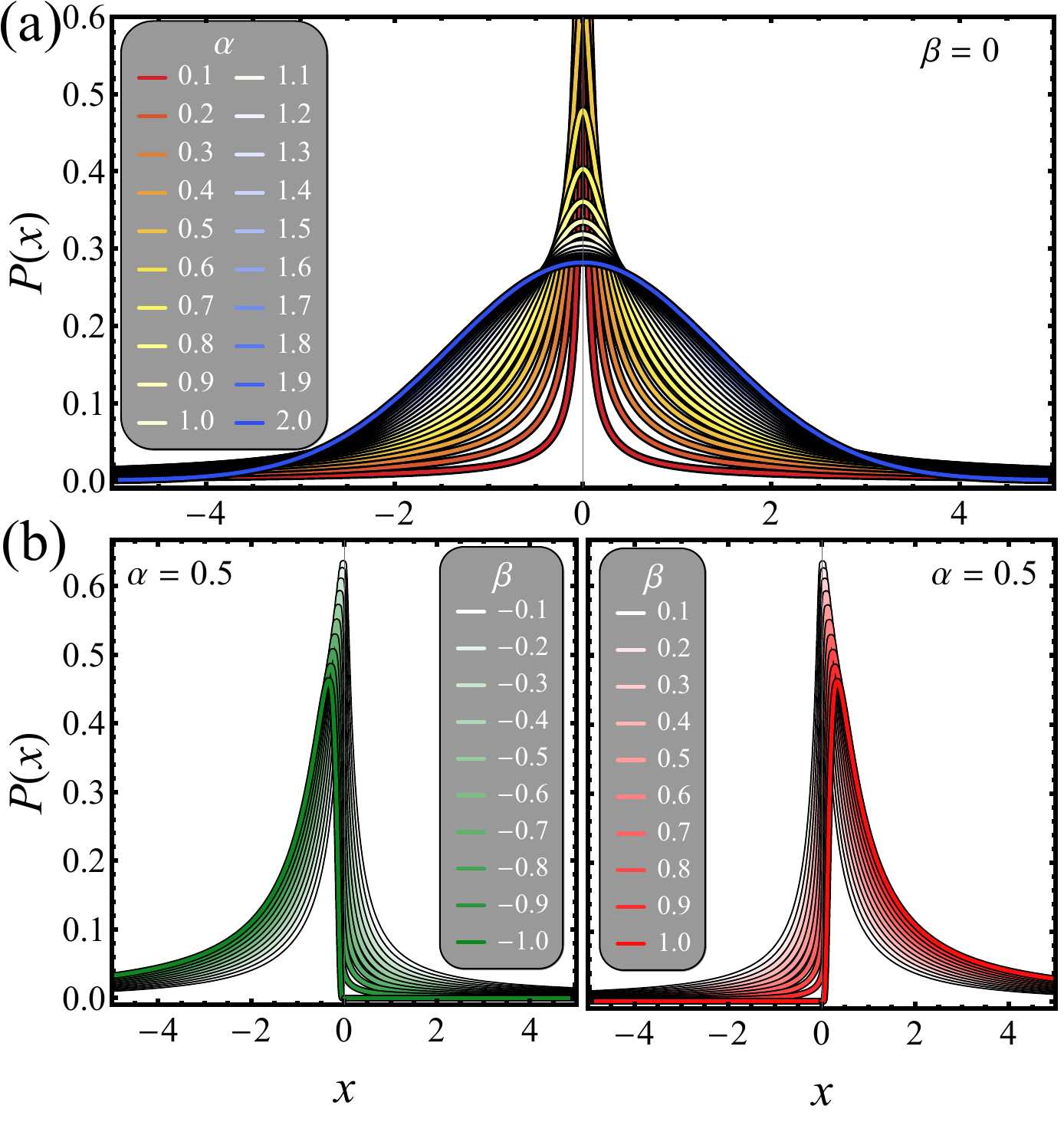}
\caption{a) Symmetric stable densities of $S_{\alpha}(1, 0, 0)$ with $\alpha\in(0,2]$. b) Skewed stable densities of $S_{0.5}(1, \beta, 0)$ with $\beta\in[-1,0)$ and $\beta\in(0,1]$, left and right panels, respectively.}
\label{Fig2}
\end{figure}

Only a few number of L\'evy distributions has a probability density function known in explicit form:
the Gaussian and Cauchy-Lorentz distributions (both with $\beta=0$ and $\alpha=2$ and $\alpha=1$, respectively) are symmetric with respect to $x=0$, while L\'evy-Smirnov distributions (normal and reflected) are skewed to the right ($\alpha=0.5$ and $\beta=+1$) or to the left ($\alpha=0.5$ and $\beta=-1$). The corresponding probability density functions are:
\begin{eqnarray}\nonumber
P(x)=\frac1{\sqrt{2\pi}\sigma}{e^{-\frac{(x-\mu)^2}{2\sigma^2}}}\quad&\text{if}&\quad S_2(\sigma, 0, \mu)\\
P(x)=\frac{\sigma/\pi }{\sigma ^2+(x-\mu)^2}\quad&\text{if}&\quad S_1(\sigma, 0, \mu)\\\nonumber
P(x)=\sqrt{\frac{\sigma}{2\pi}}\frac{e^{-\frac{\sigma}{2(x-\mu)}}}{(x-\mu)^{3/2}}\quad&\text{if}&\quad S_{0.5}(\sigma, 1, \mu).
\end{eqnarray}

Figure~\ref{Fig2}(a) illustrates the bell-shaped probability density functions for the symmetric stable distributions $S_{\alpha}(1, 0, 0)$ with $\alpha\in(0,2]$. When $\alpha$ reduces, the peak of the distribution get higher and narrower (the so called \emph{limited space displacement}~\cite{Aug10,Val14}), and correspondingly the tails become heavier. 

Figure~\ref{Fig2}(b) displays the probability density functions for the stable distributions $S_{0.5}(1, \beta, 0)$ with $\beta<0$ (left panel) and $\beta>0$ (right panel). These skewed distributions are characterized by the right (or left) tails heavier than the left (or right) ones. If $\beta=+1$ ($-1$), the distribution is totally skewed to the right (or left).

The presence of heavy tails of the probability distribution indicates the occurrence of events with large values of $x$, whose probability densities are not negligible. Heavy-tailed statistics permits to invoke rare events, corresponding to large values of $x$, that is the L\'evy flights previously discussed. 

The algorithm used in this work for simulating L\'evy noise sources is that proposed by Weron~\cite{Wer96} for the implementation of the Chambers method~\cite{Cha76}. In this work we consider only additive L\'evy noise sources~\cite{Che02,Dyb07}, despite efforts have been also done for studying stochastic processes driven by multiplicative L\'evy noise~\cite{LaC10,Sro10}, and recently even for investigating the problem of the on-off intermittency phenomenon~\cite{Kan21}.
When we consider both Gaussian and L\'evy-distributed fluctuations with intensities $D$ and $D_L$, respectively, the stochastic independent increment reads
\begin{equation}
\Delta i_N \simeq 
 \sqrt{ 2D \Delta t\; }\; N\left(0,1 \right) + 
 \left( D_L \Delta t \right) ^{1/\alpha}S_{\alpha}\left(1,\beta,0\right).
\label{GLFincr}
\end{equation}
Here, the symbol $N\left(0, 1 \right)$ indicates a random function Gaussianly distributed with zero mean and unit standard deviation, while $S_{\alpha}(1,\beta, 0)$ denotes a standard $\alpha$-stable random L\'evy variable.

For the L\'evy statistics, the mean escape time $\tau_L$, in the case of $\Delta x = \Delta U = 1$ with $\Delta x$ being the distance between
neighbor minimum and maximum and $\Delta U$ given by Eq.~\eqref{eq:barrier} [see Fig.~\ref{Fig01}(b)], can be expressed as a function of the noise parameters $D_L$ as~\cite{Che05}
\begin{equation}
\tau_L\left ( \alpha,D_L \right )=\frac{{\cal C}_{\alpha}} {D_L^{\mu_{\alpha}} },
\label{tau_Sw}
\end{equation}
where both the coefficient ${\cal C}_{\alpha}$ and the power-law exponent $\mu_{\alpha}$ are a function of $\alpha$ and are assumed to behave universally for overdamped systems~\cite{Che05,Che07,Dyb06,Dyb07,Dub09}.
For arbitrary spatial and energy scale, by rescaling time, energy, and space, in the overdamped case~\cite{Che05} and considering $\mu_\alpha\simeq1$, one obtains
\begin{equation}
\tau_L\left ( \alpha,D_L \right ) = 
\left ( \frac{ \Delta x } {2} \right )^{\alpha} 
\frac{{\cal C}_{ \alpha}} {D_L^{\mu_{\alpha} }},
\label{tau_Levy}
\end{equation}
that is, the mean escape time only depends on the minimum-to-maximum distance $\Delta x$ of the potential. We stress that this is quite diverse from the Gaussian noise case, where the probability to cross the potential barrier depends exponentially on its height, see Eq.~\eqref{r0_full}. Considering a tilted washboard potential $U(\varphi,i_b)$, see Eq.~\eqref{eq:potential}, the distance $\Delta x$ between a minimum and the closest maximum depends on $i_b$ as $\Delta x(i_b) = \pi -2\arcsin{i_b}$.

Results on the dynamics of systems driven by L\'evy flights have been recently reviewed in Refs.~\cite{Dub08, Zab15}. 
L\'evy flights are often invoked to describe transport phenomena in many condensed matter frameworks and also practical applications. 
For instance, the occurrence of L\'evy--distributed fluctuations has been recently considered in graphene~\cite{Bri14,Gat16,Kis19}, and it has been even speculated that in graphene-based Josephson devices anomalous premature switches, that are likely to be unrelated to thermal fluctuations~\cite{Cos12}, could be attributed to L\'evy-fluctuating phenomena~\cite{Gua17}. 
%In such a case, the behavior of any graphene-based device could be intrinsically influenced by L\'evy-type noisy fluctuations.

Also photoluminescence experiments in moderately doped $n$-InP samples unveil anomalous L\'evy-type distributions~\cite{Lur10,Sem12,Sub14}, and L\'evy statistics was also brought up to describe interstellar scintillations~\cite{Bol03,Bol05,Bol06,Gwi07}.
Moreover, L\'evy processes come also out in the electron transport~\cite{Nov05} and optical properties~\cite{Kun00,Kun01,Shi01,Mes01,Bro03} of semiconducting nanocrystals quantum dots.
About thermal properties of materials, the quasiballistic heat conduction in semiconductor alloys was demonstrated to be ruled by L\'evy superdiffusion~\cite{Ver15I,Ver15II,Moh15,Upa16}.

L\'evy noise often occurs also in telecommunications and networks~\cite{Yan03,Bha06,Cor10}, for noise exhibits impulsive, L\'evy--type, as well as Gaussian, features in some communication channels. The source of this noise may be either natural or man-made, can include atmospheric noise or ambient noise, and it might results from relay contacts, electromagnetic devices, electronic apparatus, or transportation systems, switching transients, and accidental hits in telephone lines~\cite{Bha06}. The need to identify these disturbances gave rise to several suggestions for models and detection schemes~\cite{Tsi95,Sub15,Sho15,Kar20}.
Finally, L\'evy fluctuations have been also proposed for vibration data in industrial bearings~\cite{LiYu10,Cho14,Saa15} and in wind turbines rotation parts~\cite{Ely16}. In fact, as a rolling element bearing runs in fault conditions, the measured vibration signal is impulsive and non-Gaussian~\cite{Whi84,McF84}, and a detection mean for these cases can give a route for rotating--machine fault diagnosis.

The situations illustrated heretofore confirm the usefulness of a reliable tool for investigating fluctuating signals distributed according to L\'evy statistics. 
In all cases in which a signal that can be transduced in an electric current feeding to the JJ, we can even conceive a Josephson-based noise detector suitable for studying and detect this type of fluctuations. In the following, we present different strategy for this purpose.

\section{L\'evy noise effects on the Josephson response}
\label{Sec04}
\vskip 0.2cm

Here we seek the effects produced by a L\'evy noise source on different observables of a JJ. In particular, we show that a non-Gaussian noise source can influence the junction response in a quite different way with respect to a thermal Gaussian source. The results discussed in the following prove, among others, that a Josephson-based noise detector can be suitable for this type of fluctuations.

In the following we review three different approaches based on different experimentally measurable quantities. First, we show the effect of a L\'evy-distributed current signal on the mean switching time (MST) in both the cases of a short and a long JJ (SJJ and LJJ, respectively), in the presence of both dc and ac bias currents flowing through the junction~\cite{Gua13,Val14,Gua16}. In particular, we discuss peculiar noise-induced effects, such as the \emph{resonant activation} (RA) and the \emph{noise enhanced stability} (NES), in the presence of a L\'evy noise source. 

Then, we discuss the impact of L\'evy-distributed fluctuations on the mean voltage drop (MVD) across a current-biased SJJ~\cite{Gua20}. In fact, as the junction switches from the superconducting to the resistive state, a measurable voltage appears across the junction, according to Eq.~\eqref{JJvoltage}. We propose to characterize the features of a L\'evy noise conveyed to the junction by looking at this voltage signal, which is quite sensitive to the presence of such a non-Gaussian noise source.

Finally, we investigate the switching currents distributions (SCDs) in an SJJ biased by a linearly ramping current, in the presence of a L\'evy noise source~\cite{Gua19}. The SCDs exhibit a characteristic behavior markedly different from the Gaussian noise case, thus allowing to eventually infer the characteristics of the L\'evy noise source.

%FigureMFPT_SJJ
\begin{figure*}[t!!]
\centering
\includegraphics[width=2\columnwidth]{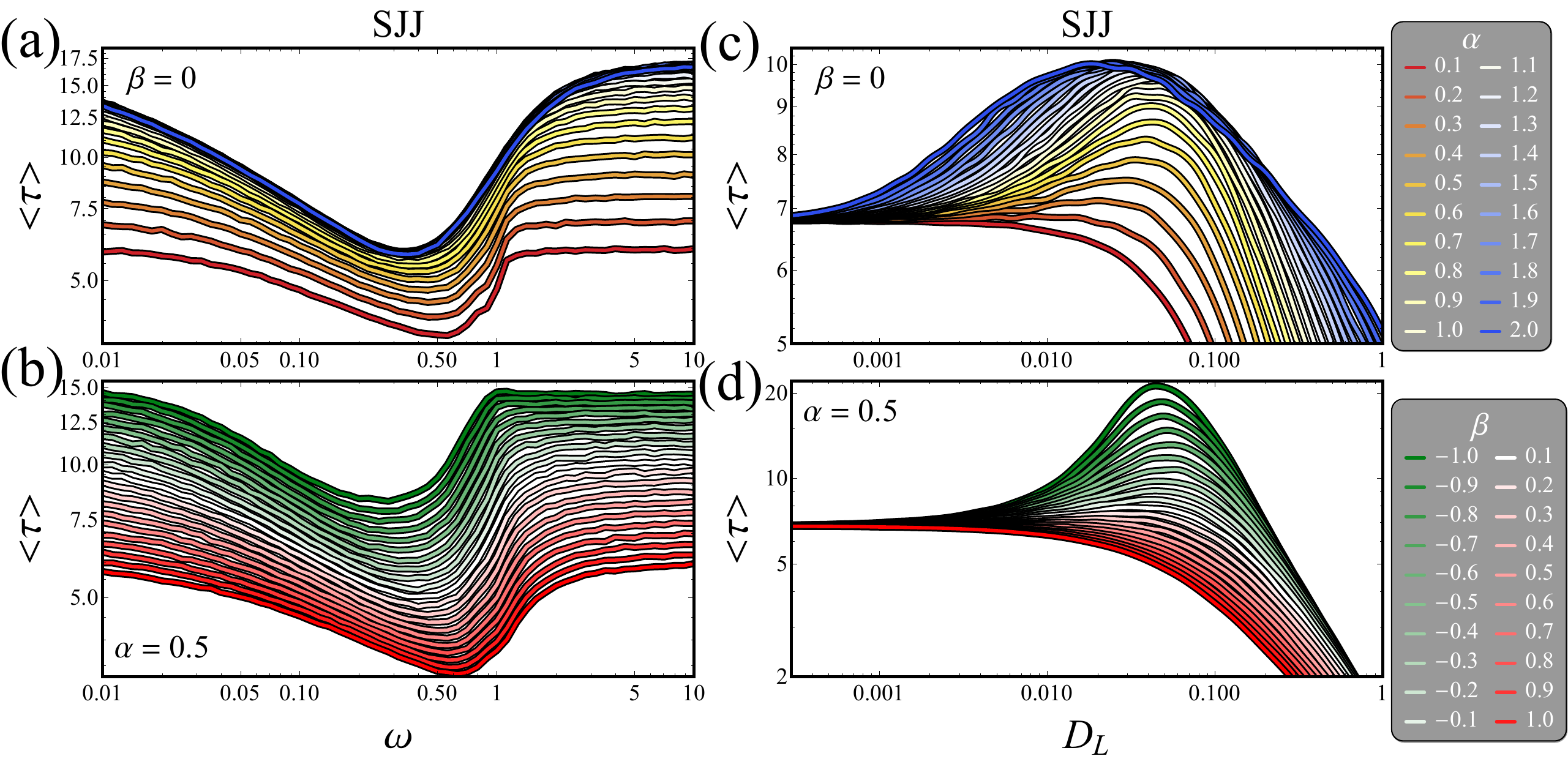}
\caption{Short JJ: (a) and (b), MST, $\left \langle \tau \right \rangle$, as a function of the noise intensity, $D_L$, setting $\omega=0.9$ for $S_{\alpha}(1, 0, 0)$ and $\alpha\in(0,2]$ and $S_{0.5}(1, \beta, 0)$ and $\beta\in[-1,1]$. (c) and (d), MST, $\left \langle \tau \right \rangle$, as a function of the driving frequency, $\omega$, setting $D_L=0.1$ for $S_{\alpha}(1, 0, 0)$ and $\alpha\in(0,2]$ and $S_{0.5}(1, \beta, 0)$ and $\beta\in[-1,1]$. The values of the other parameters are $\beta_c=0.01$, $i_{dc}=0.9$, and $i_{ac}=0.7$.}
\label{FigureMFPT_SJJ}
\end{figure*}

\subsection{L\'evy noise effects on the MSTs}
\label{Sec04a}
\vskip 0.2cm

Here, we study the effects induced by a L\'evy noise source on the MST towards the dissipative regime from the superconducting metastable state, that is the bottom of a potential minimum chosen as initial condition. Here, we investigate the responses of both an SJJ and an LJJ: in particular, for the switching time measurements we assume overdamped junctions, \emph{i.e.}, we set $\beta_c=0.01$. 

The MSTs are estimated in two different ways, according to the kind of junction we are looking for.
For an SJJ the MST is obtained as a \emph{mean first passage time}: as the phase particle overcomes the closest potential barrier the time is recorded and the numerical realization is interrupted. Then, the MST is obtained by averaging the switching times stored in all the independent numerical realizations, $N_{\text{exp}}=5\times10^4$, as 
\begin{equation}\label{average_tau}
\langle \tau \rangle = \frac{1}{N_{\text{exp}}} \sum_{i=1}^{N_{\text{exp}}} \tau_i .
\end{equation}

In the case of an LJJ, we calculate the MST as a \emph{nonlinear relaxation time}~\cite{Dub04}: we compute the mean value of the permanence times of the phase $\varphi$ within the first valley of the washboard potential, \emph{i.e.}, $\varphi\in\left[\varphi^0_{\text{max}}-\varphi^1_{\text{max}}\right]$. There are no absorbing barriers; in this way, during the whole measurement time, $t_{\text{max}}$, all the ``temporary'' trapping events contribute for calculating $\left \langle \tau \right \rangle$. We recall that the phase of an LJJ can be thought of as a string made up of cells with a width $\Delta \chi$.
In the \textit{i}-th numerical realization and for the \textit{j}-th cell, the probability $P_{ij}$ that $\varphi_j\in\left [ \varphi^0_{\text{max}}-\varphi^1_{\text{max}} \right ]$ is 
\begin{eqnarray} \label{P_t}
P_{ij}(t) = \left\{\begin{matrix}
1 \iff \varphi_j\in\left [ \varphi^0_{\text{max}}-\varphi^1_{\text{max}} \right ]\,\,\,\,\\
0 \iff \varphi_j\notin\left [ \varphi^0_{\text{max}}-\varphi^1_{\text{max}} \right ].
\end{matrix}\right.
\end{eqnarray}
If we sum $P_{ij}(t)$ over the total number of cells, $N_{c}=L/\Delta \chi$ (with $L$ being the junction length), and over the number of independent numerical realizations, $N_{\text{exp}}=5\times10^3$, we obtain the average probability that the whole string is in the superconducting state at time $t$ as
\begin{equation}\label{P_averaged}
\Pt = \frac{1}{N_{\text{exp}} N_{c}}\sum_{i=1}^{N_{\text{exp}}}\sum_{j=1}^{N_{c}}P_{ij}(t).
\end{equation}
The MST is finally obtained as
\begin{equation}\label{tauNLRT}
\left \langle \tau \right \rangle = \int_{0}^{t_{\text{max}}} \Pt dt.
\end{equation}
As there are no absorbent barriers, each numerical simulation is interrupted only once the preset measurement time $t_{\text{max}}$ is reached. This is quite different from the SJJ case, in which the numerical simulation is stopped just when the phase particle exceeds the potential maximum. 

The junction is driven by a bias current given by the sum of two contributions
\begin{equation}
i_b(t) = i_{dc} + i_{ac} \thinspace \sin(\omega t),
 \label{DrivingCurrent}
\end{equation}
where $i_{ac}$ and $\omega$ are amplitude and frequency (normalized to $I_c$ and $\omega_p$, respectively) of the dimensionless oscillating driving current. The presence of the ac term makes the potential oscillating with a frequency $\omega$. In this condition, we can reasonably expect the emergence of resonances in the phase dynamics as the driving frequency matches natural frequencies of the system. 

In the following we show the MST, $\left \langle \tau \right \rangle$, in the presence of different sources of L\'{e}vy noise, modeled by changing the values of $\alpha$ and $\beta$. In particular, we discuss two quite interesting noise-induced effects on the MST behavior, which develop as the system is affected by an oscillating drive (with a proper frequency) and a stochastic force (with a specific fluctuation intensity).

First, we demonstrate the occurrence of RA~\cite{Doe92,Pec94,Mar96,Man00,Dub04,Dyb09,Miy10,Has11,Fia11,Val14}, a noise-induced phenomenon whose signature is a minimum in the MST \emph{vs} $\omega$ profile when the driving frequency is close to a characteristic frequency of the system. 
In fact, the escape of the particle from a potential well can be preferentially induced when the potential barrier oscillates on a time-scale characteristic of the particle escape itself. 
In the Josephson framework, we are referring to the plasma frequency, that in the presence of a bias current modifies as $\widetilde{\omega}_J(i_b)=\left(1-i_b^2\right)^{1/4}$. 
For the sake of precision, the one just described is the \emph{dynamic RA}~\cite{Dev84,Dev85,Mar87,Gua15}, which is distinguished from another type of resonant phenomenon, \emph{i.e.}, the \emph{stochastic RA}~\cite{Pan09,Add12,Val14,Gua15}. This is instead triggered when the driving frequency is close to the average escape rate at the minimum, see Eq.~\eqref{r0_full} in the lowest configuration.

Still in the presence of a resonant drive, at a proper intensity of the noise fluctuations the MST curves can exhibit what is called NES~\cite{Man96,Agu01,Spa04,Dub04,DOd05,Fia05,Hur06,Spa07,Man08,Yos08,Fia09,Tra09,Fia10,Li10,Smi10,Val14}. This is a noise--induced phenomenon consisting in a nonmonotonic behavior of the MST as a function of the noise intensity, with the appearance of a pronounced maximum; this peculiar response clearly departs from the expected monotonic Kramers-like behavior~\cite{Kra40,Mel91,Han90}. In other words, the noise intensity can be such as to enhance the stability of the system and, therefore, to increase the average lifetime of the metastable state. This enhancement of stability has been observed in many physical and biological systems, and belongs to a highly topical interdisciplinary field of research, extending from condensed matter physics to molecular biology and to cancer growth dynamics~\cite{Spa07,Spa12}.

%FigureMFPT_LJJ
\begin{figure*}[t!!]
\centering
\includegraphics[width=2\columnwidth]{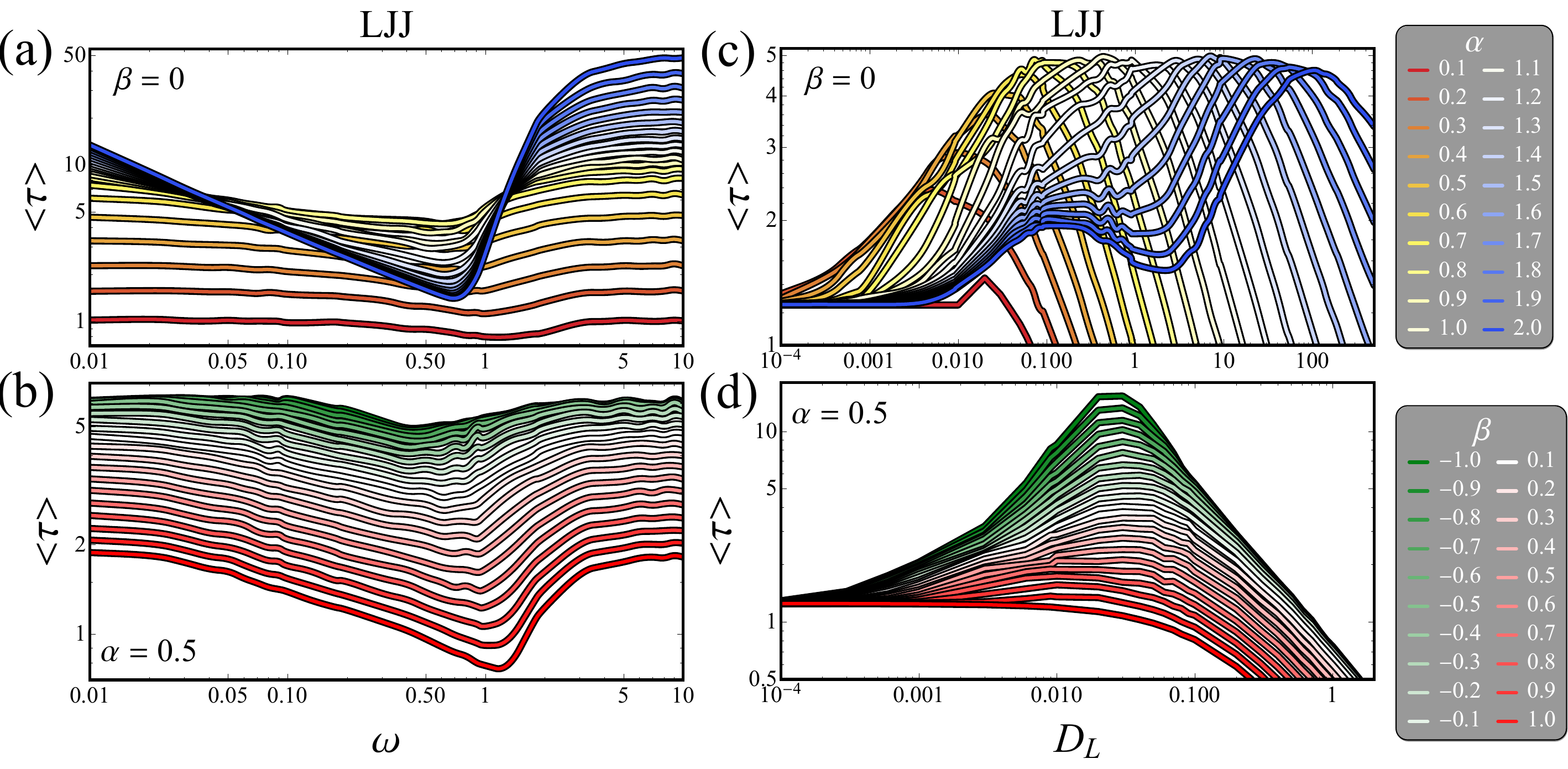}
\caption{Long JJ: (a) and (b), MST, $\left \langle \tau \right \rangle$, as a function of the noise intensity, $D_L$, setting $\omega=0.9$ for $S_{\alpha}(1, 0, 0)$ and $\alpha\in(0,2]$ and $S_{0.5}(1, \beta, 0)$ and $\beta\in[-1,1]$. (c) and (d), MST, $\left \langle \tau \right \rangle$, as a function of the driving frequency, $\omega$, setting $D_L=0.1$ for $S_{\alpha}(1, 0, 0)$ and $\alpha\in(0,2]$ and $S_{0.5}(1, \beta, 0)$ and $\beta\in[-1,1]$. The values of the other parameters are $L=10$, $\beta_c=0.01$, $i_{dc}=0.9$, and $i_{ac}=0.7$.}
\label{FigureMFPT_LJJ}
\end{figure*}

In Fig.~\ref{FigureMFPT_SJJ} we explore the behavior of the MSTs as a function of the driving frequency $\omega$ and noise intensity $D_L$ in the case of an SJJ. To ensure the noise-induced effects to emerge clearly, in the following we choose $i_{dc}=0.9$ and $i_{ac}=0.7$. 

First we discuss the $\left \langle \tau \right \rangle$ \emph{vs} $\omega$ behavior shown in Figs.~\ref{FigureMFPT_SJJ}(a) and (b) obtained, respectively, at different $\alpha\in(0,2]$, setting $\beta=0$, and at different $\beta\in[-1,1]$, setting $\alpha=0.5$, by keeping fixed the L\'evy noise intensity at $D_L=0.1$. 
All $\left \langle \tau \right \rangle$ \emph{vs} $\omega$ curves in Fig.~\ref{FigureMFPT_SJJ} clearly demonstrate the presence of RA. In particular, the RA minimum indicates a resonant escape of the phase particle from the potential well as the driving frequency is close to $\widetilde{\omega}_J(i_b=0.9)\simeq0.66$. At frequencies $\omega\gtrsim1$ we instead observe a plateau due to the very fast oscillations of the potential. 

The RA phenomenon is robust enough to be observed in the presence of L\'{e}vy noise sources, that is by changing $\alpha$ we still notice the formation of a minimum in the $\left \langle \tau \right \rangle$ \emph{vs} $\omega$ profiles, see Fig.~\ref{FigureMFPT_SJJ}(a). Interestingly, by reducing $\alpha$ we observe that the whole MST curve tends to reduce due to the influence of L\'evy flights that speed up the escape process and that the position of the minimum slightly shifts towards higher frequencies. The noise dependence suggests the stochastic nature of the RA phenomenon.
The same insight stems from the $\left \langle \tau \right \rangle$ \emph{vs} $\omega$ curves at different $\beta$, see Fig.~\ref{FigureMFPT_SJJ}(b). We note that $\left \langle \tau \right \rangle$ tends to reduce by increasing $\beta$. This is due to the asymmetric fluctuations that these noise sources induce for $\beta\neq 0$, see Fig.~\ref{Fig2}(b). In particular, for $\beta<0$, the L\'evy flights push $\varphi$ towards negative values, that is in the opposite direction with respect to the tilting set by the positive bias current. This is why the more the fluctuation distribution is skewed to the left, \emph{i.e.}, $\beta=-1$, the higher the MSTs are. Conversely, a positive value of $\beta$ gives fluctuations supporting the rolling of the phase particle along the potential, and this makes the $\left \langle \tau \right \rangle$ values lower. Also in this case, the position of the $\left \langle \tau \right \rangle$ minimum slightly moves towards higher frequencies by increasing $\beta$.

Interestingly, if we consider an LJJ, the $\left \langle \tau \right \rangle$ \emph{vs} $\omega$ profiles differ substantially from those of an SJJ, see Figs.~\ref{FigureMFPT_LJJ}(a)-(b) for an LJJ with $L=10$, $D_L=0.1$, $i_{dc}=0.9$, and $i_{ac}=0.7$. In fact, if the junction is long enough, solitons can be generated and this influences the MST, which is intimately connected to the formation of solitons along the junction. In particular, we observe that all curves in Fig.~\ref{FigureMFPT_LJJ}(a), for $\alpha\in(0,2]$ and $\beta=0$, show the RA minimum, but also a non-monotonic behavior of $\left \langle \tau \right \rangle$ as a function of $\alpha$ for $\omega\in(0.04, 1.3)$. The behavior of $\left \langle \tau \right \rangle$ \emph{vs} $\alpha$ at a given $\omega$ within this range reflects the evolution of the mean soliton density as a function of $\alpha$, as demonstrated in Ref.~\cite{Gua16}.

The MSTs \emph{vs} $\omega$ curves of an LJJ, at different values of $\beta\in[-1,1]$ and $\alpha=0.5$, are shown in Fig.~\ref{FigureMFPT_LJJ}(b), setting $L=10$, $D_L=0.1$, $i_{dc}=0.9$, and $i_{ac}=0.7$. By increasing $\beta$, we observe that the RA phenomenon can be still observed, despite the frequency in correspondence of the minimum tends to grow, while $\left \langle \tau \right \rangle$ tends to reduce.

We can now take a look at the behavior of the MST as a function of the L\'evy noise intensity, by keeping fixed the frequency of the driving signal. In particular, in Figs.~\ref{FigureMFPT_SJJ}(c) and (d) we illustrate the $\left \langle \tau \right \rangle$ \emph{vs} $D_L$ behavior for an SJJ, obtained, respectively, at different $\alpha\in(0,2]$, setting $\beta=0$, and at different $\beta\in[-1,1]$, setting $\alpha=0.5$, by keeping fixed the driving frequency at $\omega=0.9$. 

First, we discuss the behavior by changing $\alpha\in(0,2]$ shown in Fig.~\ref{FigureMFPT_SJJ}(c). We initially observe that for $D_L\to 0$ all curves converge to the same value, \emph{i.e.}, the deterministic lifetime in the superconducting state.
Increasing the intensity of noise, the MST curves exhibit the NES effect, which is demonstrated by the presence of a maximum, see Fig.~\ref{FigureMFPT_SJJ}(c). For $\alpha=2$ this maximum is located roughly at $\left. D_L(\left \langle \tau \right \rangle_{\text{max}}) \right|_{\textsc{sjj}}\sim0.018$, whereas by reducing $\alpha$ we observe that its position slightly shifts towards higher $D_L$ and contemporaneously the whole $\left \langle \tau \right \rangle$ \emph{vs} $D_L$ curve tends to reduce. Similarly, the NES phenomenon appears also in the $\left \langle \tau \right \rangle$ \emph{vs} $D_L$ by changing $\beta$, see Fig.~\ref{FigureMFPT_SJJ}(d); in this case, all curves show a maximum at $\left. D_L(\left \langle \tau \right \rangle_{\text{max}}) \right|_{\textsc{sjj}}\sim0.04$. The highest MST values occur for $\beta = -1 $ while the lowest values occur for $\beta=1$, and this is again due to the symmetry of the distribution of the fluctuations, as previously discussed.

In the case of an LJJ, we observe that the MST curves show two maxima instead of one, as in the case of SJJ, see Fig.~\ref{FigureMFPT_LJJ}(c). In particular, the $\left \langle \tau \right \rangle$ \emph{vs} $D_L$ curve for $\alpha=2$ these two maxima are in $\left. D_L(\left \langle \tau \right \rangle_{\text{max}}) \right|_{\textsc{ljj}}\sim0.14$ and $\left. D_L(\left \langle \tau \right \rangle_{\text{max}}) \right|_{\textsc{ljj}}\sim100$. Reducing the value of $\alpha$ the higher maximum shifts towards smaller $D$, still maintaining its height, up to merge with the first maximum. Specifically, the position $D_L(\left \langle \tau \right \rangle_{\text{max}})$ of the second NES peak decreases exponentially towards lower noise intensities as the value of $\alpha$ reduces.

The appearance of a double-NES maxima is, in such a way, connected to the procedure for calculating the MST, that is as an NLRT for an LJJ. In fact, the double-maxima NES for LJJs with different lengths in the presence of a L\'evy noise source was also observed in Refs.~\cite{Val14,Gua16}, where it was demonstrated that in correspondence of the high-$D_L$ maximum the NES effect is due to the possibility that the phase string comes back into the first valley after a first escape event. In particular, the second NES maximum tends to appear at noise intensities high enough to allow the phase string to move leftwards, that is even in the opposite direction with respect to the tilting due to the positive bias current. On the contrary, the low-$D_L$ maximum of $\left \langle \tau \right \rangle$ is of the same nature as that appearing in the SJJ case. In particular, in the case of $\alpha=2$ these maxima are settled at noise intensities $\left. D_L(\left \langle \tau \right \rangle_{\text{max}}) \right|_{\textsc{sjj}}\simeq0.018$ and $\left. D_L(\left \langle \tau \right \rangle_{\text{max}}) \right|_{\textsc{ljj}}\simeq0.14$. We observe that the noise intensity giving the NES maximum for an LJJ is roughly eight times larger than that for an SJJ. This reflects the facts that the activation energy of a soliton along a phase string is $\sim8\Delta U$~\cite{Castellano96}; in other words, when NES effect emerges the thermal energy to trigger a soliton is eight times greater than that required to cause a phase particle to overcome the potential barrier $\Delta U$. 

The second NES maximum is evident also for $\alpha<2$, see Fig.~\ref{FigureMFPT_LJJ}(c). In fact, if one reduces $\alpha$, the probability to intense noise fluctuations grows, so that, after a first escape, temporary confinements within the initial metastable well can still happen, but at lower noise intensities. 

Figure~\ref{FigureMFPT_LJJ}(d) shows the MSTs as a function of $D_L$, at $\omega=0.9$, still in the case of an LJJ, for different values of $\beta\in[-1,1]$ and $\alpha=0.5$. The NES effect is still present, and it is evidently influenced by the $\beta$ value. In fact, the time that the string ``spends'' in the initial metastable well is longer for a L\'evy noise strongly skewed in the opposite direction with respect to the direction imposed by a positive $i_b$. We finally observe that by changing $\beta$ only the height, but not the position, of the NES peak changes. 

In summary, a L\'evy source has clear effects on the switching time distributions, and the observed noise-induced phenomena, such as stochastic resonant activation and noise-enhanced stability, emerge even in the case of L\'evy fluctuations, but showing different characteristics.

%FigureVoltage_01
\begin{figure}[t!!]
\includegraphics[width=\columnwidth]{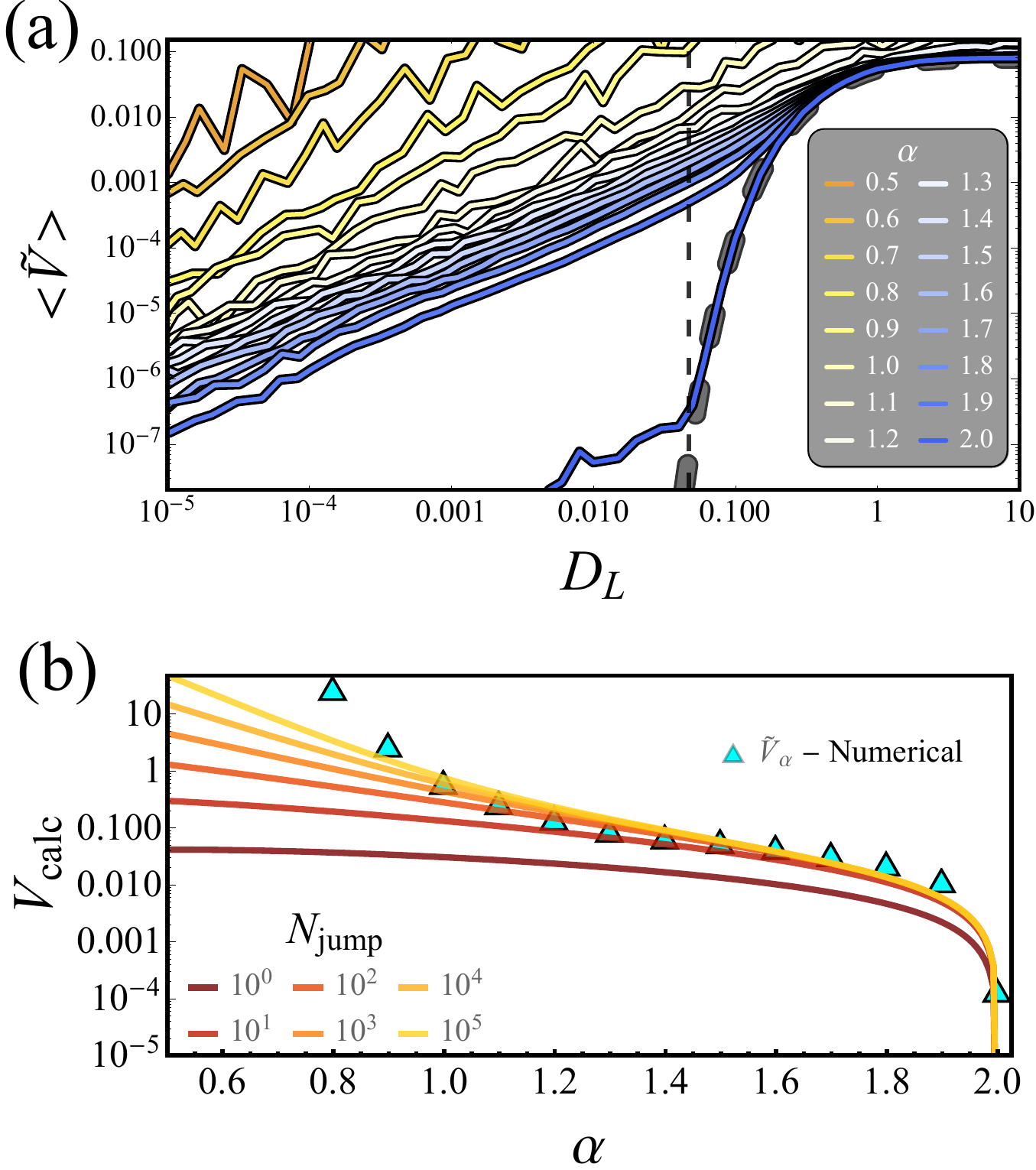}
\caption{(a) Normalized average voltage drop, $\V$, as a function of the L\'evy noise intensity, $D_L$, for different values of the parameter $\alpha$ and $i_b=0.5$. The black dashed vertical line indicates the noise intensity $D^{th}$ at which the 
inverse Kramers rate, see Eq.~\eqref{r0_full}, equals the measurement time $t_{\text{max}}=10^4$. The gray dashed curve indicates the average voltage drop \emph{vs} the noise intensity, for thermal fluctuations, analytically calculated from Eq.\eqref{a1}. (b) Behavior of $\tV$ (cyan triangles) and $\cV(\alpha,i_b,{\cal N}_{\text{jump}})$ (solid lines), see Eq.~\eqref{Vcalc}, as a function of $\alpha$ at different values of ${\cal N}_{\text{jump}}$ and $i_b=0.5$.}
\label{FigureVoltage_01}
\end{figure}

\subsection{L\'evy noise effects on the MVDs}
\label{Sec04b}
\vskip 0.2cm

In this section we discuss a different approach based on the measurement of the MVD across the junction. In this scheme, the injected bias current is kept fixed at a value lower than the critical current, to steady maintain the system in the superconducting metastable regime until the noise eventually pushes it out, thus inducing the switching from the zero-voltage state to the finite voltage ``running'' state. 
In fact, the voltage in a JJ is proportional to the time derivative of the phase difference $\varphi$, according to the a.c. Josephson relation, see Eq.~\eqref{JJvoltage}. We pursue for an analysis of the MVD which allows to catch some features of the noise source affecting the phase dynamics.

Here the average is intended as a double, \emph{i.e.}, ensemble and time, averaging. 
In the \textit{i}-th numerical realization, the time average of the voltage difference across the junction can be obtained as follows
\begin{equation}\label{MeanV_i}
\left \langle V_i \right \rangle=\frac{1}{\tau_{\text{max}}}\int_{0}^{\tau_{\text{max}}}\frac{\Phi_0}{2\pi}\frac{\mathrm{d} \varphi_i \left ( \tau \right )}{\mathrm{d} \tau}d\tau,
\end{equation}
with $\varphi(0)=\arcsin (i_b)$ being the initial phase and $t_{\text{max}}=\omega_c\tau_{\text{max}}$ the normalized measurement time. The MVD across the 
junction is finally obtained by averaging over the total number of independent numerical repetitions, $N_{\text{exp}}$. In units of $\Phi_0\omega_c$, the MVD reads
\begin{equation}\label{MeanV}
\V=\frac{\left \langle V \right \rangle}{\Phi_0\omega_c}=\frac{1}{N_{\text{exp}}}\sum_{i=1}^{N_{\text{exp}}}\frac{\left \langle V_i \right \rangle}{\Phi_0\omega_c}.
\end{equation}
For a typical $\omega_c\simeq100\;\text{GHz}$ the normalizing factor reads $\Phi_0\omega_c\simeq0.2\;\text{mV}$.
In the following, the value of $\V$ is estimated averaging over a normalized time $t_{\text{max}}=10^4$ and $N_{\text{exp}}=10^4$ independent numerical repetitions. Here, we investigate an overdamped SJJ with $\beta_c=0.01$. In order to highlight the influence of L\'evy flights we initially assume vanishingly small thermal fluctuations ($D\to0$). 
In this work, we consider only symmetric (\emph{i.e.}, $\beta=0$), bell-shaped, standard (\emph{i.e.}, with $\sigma=1$ and $\lambda=0$), stable distributions $S_{\alpha}(1, 0, 0)$, with $\alpha\in[0.1,2]$.

In Fig.~\ref{FigureVoltage_01}(a) we show the normalized MVD, $\V$, versus the L\'evy noise intensity, $D_L$, changing the value of the stability index, $\alpha$, and imposing $i_b=0.5$.

We start discussing the curve for $\alpha = 2$, since in this case the L\'evy distribution amounts to the Gaussian situation. In this case, L\'evy flights, which push out the particle from the metastable well making its speed rapidly increasing, are indeed missing and the $\V$ curve is several orders of magnitude lower than the $\alpha\leq1.9$ profiles. The data for $\alpha = 2$ show two distinct behaviors, respectively above and below a specific threshold value $D^{th}$, which is marked in Fig.~\ref{FigureVoltage_01}(a) with a black vertical dashed line. This threshold can be estimated as the noise intensity at which the inverse Kramers rate~\cite{Kra40} matches the measurement time, that is 
$\Gamma_{\textsc{ta}}(i_b,D^{th})^{-1}=\tau_{\text{max}}$. 

Figure~\ref{FigureVoltage_01}(a) tells us that at low noise intensities, \emph{i.e.}, below the threshold, the Gaussianly-distributed fluctuations are not intense enough to trigger escapes within the time $t_{\text{max}}$, and this is why $\V$ remains vanishingly small. Conversely, for higher intensities, \emph{i.e.}, above the threshold, noise-induced switches can be induced; in this case, the phase particle can leave the initial metastable state rolling down along the potential, thus its speed increases and a non--negligible MVD arises. 

Interestingly, we observe that the curve obtained numerically for $\alpha = 2$ perfectly matches the MVD analytically calculated for a finite--junction capacitance and in the presence of thermal fluctuations~\cite{Lee71,Bar82}, see the gray-dashed curve in Fig.~\ref{FigureVoltage_01}(a), obtained as
\begin{equation}\label{a1}
\left \langle V \right \rangle=2DRI_{c}\frac{\exp(\pi i_b/D)-1}{\exp(\pi i_b/D)} T^{-1}_{1} \left ( 1+\beta_c \frac{T_{2}}{T_{1}} \right )
\end{equation}
where
\begin{eqnarray}\label{a2}\nonumber
&&T_{1}=\int_{0}^{2\pi}\!\!d\varphi \,\text{I}_{0}\bigg(D^{-1} \sin\frac{\varphi}{2}\bigg) \exp\left ( -\frac{i_b\varphi}{2D} \right )\\\label{a3}\nonumber
&&T_{2}=\int_{0}^{2\pi}\!\!d\varphi \sin\left ( \frac{\varphi}{2} \right ) \text{I}_{1}\left ( D^{-1} \sin\frac{\varphi}{2} \right )\exp\left ( -\frac{i_b\varphi}{2D} \right ),
\end{eqnarray}
and $\text{I}_0(x)$ and $\text{I}_1(x)$ are modified Bessel functions.

Looking at the curves for $\alpha<2$, we note that for $D_L$ values below the threshold, all the $\V$ \emph{vs} $D_L$ curves behave quite similarly. In fact, changing the index $\alpha$, the average voltage data organize in well-distinct parallel lines, in a log-log plot, with a positive slope. A further increase in the noise intensity has little consequences, as the fluctuations are intense enough to force the phase particle to overcome the potential barrier for any $\alpha$. This is why, for $D_L\gtrsim1$, the $\V$ all curves tend to a common plateau.

%FigureVoltage_02
\begin{figure}[t!!]
\includegraphics[width=\columnwidth]{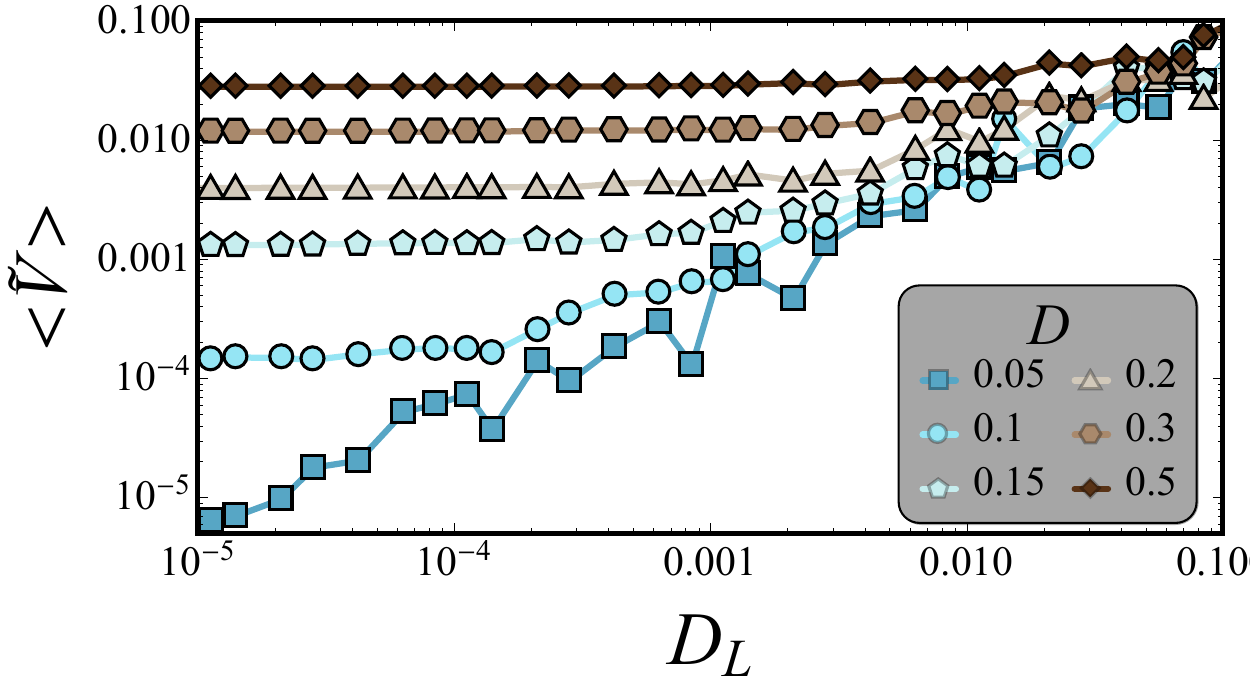}
\caption{Normalized average voltage drop as a function of the L\'evy noise intensity $D_L$, at $\alpha=1$ and $i_b=0.5$, in the presence 
of a Gaussian noise source with different intensities $D$. The lines in the figure are guides for the eye.}
\label{FigureVoltage_02}
\end{figure}

In the linear portion of the $\V$ \emph{vs} $D_L$ data, all curves of Fig.~\ref{FigureVoltage_01}(a) can be fitted with 
the function $\tV\times D_L^{\mu_{\alpha}}$, with $\tV$ being the fitting parameter and $\mu_{\alpha} \simeq 1$ for the L\'evy noise escapes \cite{Che07}. Thus, Fig.~\ref{FigureVoltage_01}(b) illustrates the monotonically reducing behavior by increasing $\alpha$ of the fitting parameter $\tV$ extracted from Fig.~\ref{FigureVoltage_01}(a) in the range of noise intensity $D_L\in[10^{-4},10^{-2}]$, see cyan triangles. 

In the following, we demonstrate that with few simple assumptions it is possible to give an estimation of the average velocity 
of a particle moving from a metastable state of a cosine potential with friction, in the presence of a driving force and L\'evy fluctuations.
This allows to roughly estimate the fitting parameter $\tV$. First, we observe that the phase particle can make $2\pi$ jumps between the minima of the washboard potential and that the MST for the L\'evy statistics follows the power-law asymptotic behaviour given by Eq.~\eqref{tau_Sw}.
We observe that the normalized MVD in Eq.~\eqref{MeanV} represents the average speed for escape processes from a metastable state
\begin{equation}\label{averagespeed}
\left \langle v_i \right \rangle=\frac{1}{t_{\text{max}}}\frac{\varphi_i(t_{\text{max}})-\arcsin (i_b)}{2\pi}=\frac{{\cal N}_{\text{jump}}}{t_{\text{max}}},
\end{equation}
where ${\cal N}_{\text{jump}}$ counts the $2\pi$-jumps that the phase particle undergoes to reach the position $\varphi_i(t_{\text{max}})$, starting from the initial state $\varphi(0)=\arcsin (i_b)$. 

If the phase particle takes a time $\tau_L$ to cover ${\cal N}$ potential minima with a single jump, the average speed can be evaluated as $\left \langle v_{\cal N}\right \rangle=\frac{{\cal N}}{\tau_L}=\left [\frac{{\cal N}}{{\cal C}_\alpha}\left ( \frac{2}{\Delta x} \right )^{\alpha} \right ]\times D_L$. 
Since the particle can leave a metastable well going left or rightward, the distances $\Delta x_l$ and $\Delta x_r$ respectively traversed with a jump over ${\cal N}$ minima can be estimated as $\Delta x_{r/l}(i_b,{\cal N})=(2{\cal N}+1)\pi\mp2\text{arcsin}(i_b)$. Finally, including all possible jumps up to ${\cal N}_{\text{jump}}$, the average speed can be calculated as $\left \langle v \right \rangle=\cV\times D_L$, where
\begin{eqnarray}\label{Vcalc}
&&\cV(\alpha,i_b,{\cal N}_{\text{jump}})=\\\nonumber
&&\sum_{{\cal N}=1}^{{\cal N}_{\text{jump}}}
\Bigg \{ \frac{{\cal N}}{{\cal C}_\alpha} \left \{ \left [ \frac{2}{\Delta x_{r}(i_b,{\cal N})} \right ]^{\alpha}- \left [ \frac{2}{\Delta x_{l}(i_b,{\cal N})} \right ]^{\alpha} \right \}\Bigg \}.
\end{eqnarray}

The behavior of $\cV(\alpha,i_b,{\cal N}_{\text{jump}})$ \emph{vs} $\alpha$ at different ${\cal N}_{\text{jump}}$ and $i_b=0.5$ is indicated by the solid curves in Fig.~\ref{FigureVoltage_01}(b). For simplicity, we used here the coefficient ${\cal C}_\alpha=\Gamma (1-\alpha) \cos(\pi \alpha/2 )$ given in Ref.~\cite{Che05} for an overdamped escape dynamics through a fixed-height barrier of a cubic potential.
The simple analytical approach given in Eq.~\eqref{Vcalc} well agrees with numerical outcomes, for $\alpha\gtrsim1$. 

Up to now, we assumed a negligible thermal influence. Here, we discuss what happens if the L\'evy component is embedded in a thermal noise background. The temperature of the system is, in such a way, a disturbance, since the concomitant presence of a L\'evy and a Gaussian noise source with a non-negligible intensity ($D\ne0$) results in a modification of the expected linear behavior of the average voltage as a function of the L\'evy noise intensity. The $\V$ \emph{vs} $D_L$ curves shown in Fig.~\ref{FigureVoltage_02} are drawn at a fixed L\'evy index $\alpha=1$ and a bias current $i_b=0.5$, and at different Gaussian noise intensities $D$. These curves demonstrate how the $\V$ behavior depends on the additional Gaussian source in the case of $i_b=0.5$. For $D\lesssim 0.05$, thermal noise has no effects on $\V$, which still shows the linear behavior already discussed in Fig.~\ref{FigureVoltage_01}. Conversely, for thermal noise intensities $D > 0.05$ we observe a plateau of $\V$, whose value grows with increasing $D$. The presence of a plateau at low $D_L$ values implies that, in this range of L\'evy noise intensities, the phase dynamics is mainly guided by the Gaussian fluctuations, being therefore independent of $D_L$.

However, the amount of Gaussian noise that make the system instead ruled by L\'evy noise depends on the time taken for the voltage measurement. In fact, within the time $t_{\text{max}}$ during which the voltage is measured, the junction is exposed to thermal noise. The longer this exposure, the lower the temperature at which a significant number of thermal escapes can occur.

\subsection{L\'evy noise effects on the SCDs}
\label{Sec04c}
\vskip 0.2cm

In this section the L\'evy noise current is added to a linearly ramped bias current flowing through a SJJ. In this way, we aim to induce clear changes in the distribution of switching currents out of the zero-voltage state of the junction.
In particular, we consider the contemporaneous presence of a thermal noise source, with the usual Gaussian white-noise statistics given in Eq.~\eqref{correl_norm_1}, and a L\'evy noise source $S_{\alpha}(1, 0, 0)$ with $\alpha\in[0.1-2]$. Moreover, we consider a linearly increasing bias current $i_b(t)=I_b(t)/I_c=v_bt$, where $v_b=t_{\text{max}}^{-1}$ is the ramp speed. 
Figure~\ref{FigureSCD_01} shows the probability distribution functions (PDFs) and the cumulative distribution functions (CDFs) of the switching currents, setting the Gaussian and L\'evy noise intensities at $D=10^{-2}$ and $D_L=5\times10^{-7}$, respectively, and the other parameters at $N_{\text{exp}} =10^{4}$ and $t_{\text{max}} =10^{7}$. In this case, we consider an underdamped JJ, thus we set $\beta_J = 0.1$.

%FigureSCD_03
\begin{figure}[t!!]
\centering
\includegraphics[width=\columnwidth]{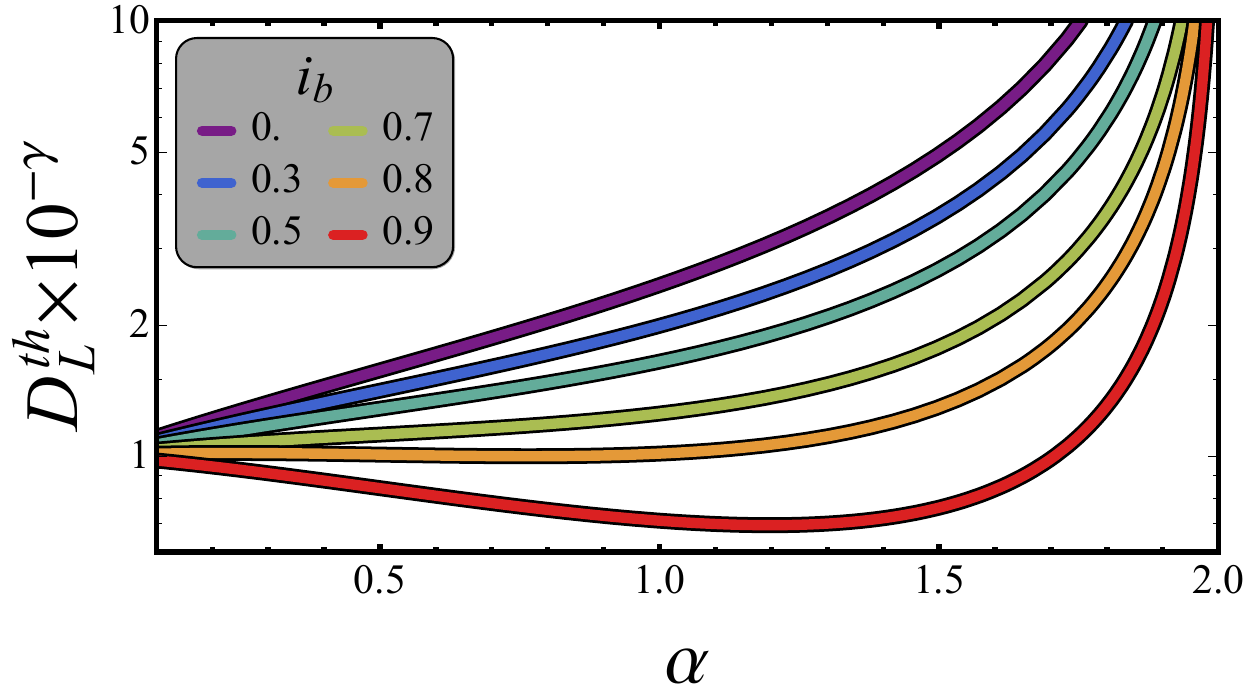}
\caption{Behavior of the threshold noise intensity, $D_L^{th}$, at which the inverse L\'evy rate, see Eq.~\eqref{tau_Levy}, equals the measurement time $t_{\text{max}}=10^{\gamma}$ (with a generic $\gamma$), as a function of $\alpha$ at different values of $i_b$.}
\label{FigureSCD_03}
\end{figure}

The choice of the intensity of the L\'evy fluctuations is a delicate point~\footnote{This statement might seem misleading, as in a concrete situation it is not said that one can ``set'' the intensity of the noise at will, as in numerical simulations. However, the reference parameter is the measurement time, which instead can be reasonably decided.}. In fact, if the noise intensity is too high, the escape process will be strongly characterized by premature switches due to L\'evy flights and we can reasonably expect SCDs formed by very intense peaks at low bias currents, while if it is too low we will obtain SCDs very close to the thermal ones. To decide a suitable noise level, we could estimate the L\'evy noise intensity $D_L^{th}$ such that the escape time given in Eq.~\eqref{tau_Levy} coincides with the measurement time, that is $\tau_L\left ( \alpha,i_b,D_L^{th} \right ) \equiv t_{\text {max}}$. This specific noise--intensity threshold gives an estimate of practical noise ranges to obtain reasonable SCDs accounting for the L\'evy flights. In particular, if we generically assume $t_{\text {max}}=10^\gamma$, the behavior of the threshold intensity $D_L^{th}$ as a function of $\alpha$ at different $i_b$ is drawn in Fig.~\ref{FigureSCD_03}: we observe that $D_L^{th}\gtrsim 10^{-\gamma}$. For this reason, having chose $t_{\text{max}} =10^{7}$, in this work we impose $D_L=5\times10^{-7}$ to be sure that L\'evy flights play a non-destructive role in the creation of the SCD. Interestingly, since $t_{\text{max}}$ is substantially the time to ramp the bias current, we could in principle conveniently tune this ramp time to the optimal value for making the noise--induced effects on the SCDs more evident. In other words, we can adjust the ramp time to highlight the effect on a SCD in the presence of a L\'evy source even without knowing the fluctuations intensity. In this way, one could even envisage to infer the intensity of the L\'evy source from the SCDs measured at different sweep rates.

The PDF for the pure Gaussian noise case, indicated by a blue thick curve in Fig.~\ref{FigureSCD_01}(a), is characterized by a peak at high values of the bias current. In this case L\'evy flights are missing and the Gaussian noise source induces switching events only at current values close to the critical current. The position and the width of the peak of a SCD due to thermally induced switching processes depend only on the value of the noise intensity $D$, that is, according to Eq.~\eqref{noiseampl}, on the value of the temperature $T$ at which the junction resides. 
In fact, if thermal noise is taken into account, the phase particle can ``hop out'' of the washboard potential well and slip down the potential profile, with an escape rate $\Gamma_{\textsc{ta}}$ at temperature $T$, according to the Kramers theory~\cite{Kra40}, see Eq.~\eqref{r0_full}.
Thus, for any finite bias current value $i_b$ a switching from the metastable superconducting state is unavoidable, which will happen sooner for a shallow well.

%FigureSCD_01
\begin{figure}[t!!]
\includegraphics[width=\columnwidth]{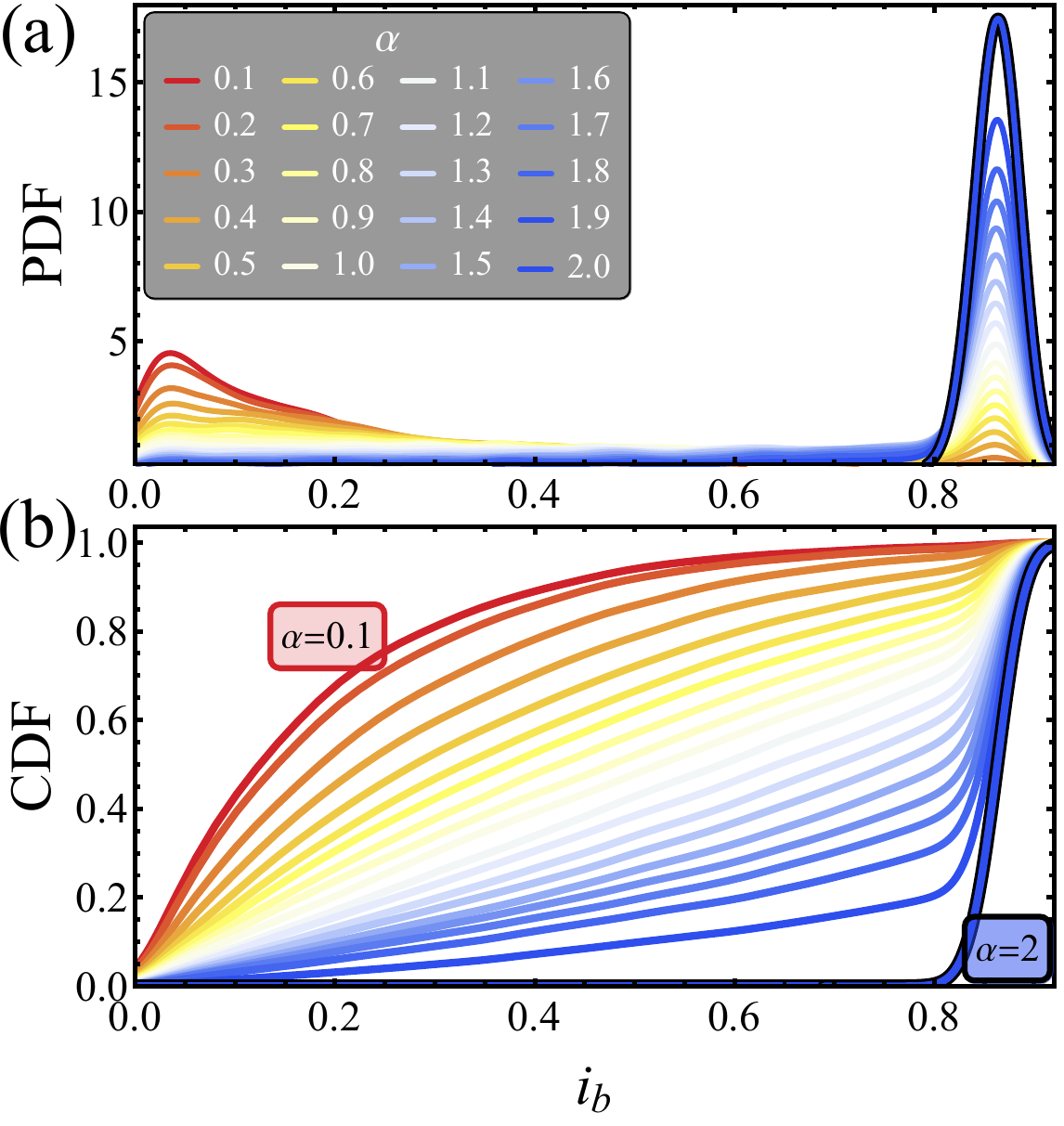}
\caption{(a) Probability distribution function (PDF) and (b) cumulative distribution function (CDF) for $\alpha\in[0.1,2]$. The values of the other parameters are: $D = 10^{-2}$, $D_L= 5\times 10^{-7}$, $N_{\text{exp}} =10^{4}$, $t_{\text{max}} =10^{7}$, and $\beta_J = 0.1$. The legend in panel (a) refers to both panels.}
\label{FigureSCD_01}
\end{figure}

The typical experimental scheme for measuring switching events is to sweep the bias current upward from zero. 
Early in the sweep, the washboard well is deep, but late in the sweep the potential well gets shallow and, as a consequence, the thermal activation is easier. 
Moreover, as the bias current is swept, the shape of the potential well changes so that the plasma frequency reduces, and so does the attempt rate $\omega_J$. 
The interplay of these two mechanisms determines the net escape rate as a function of the bias current.

The escape process from a potential well depends on the temperature, so that at the crossover temperature $T_{\text{cr}}$ thermal activation becomes tantamount to tunneling rate. As the MQT rate is temperature independent, both the position and the width of SCD peaks saturate at low temperatures, \emph{i.e.}, the SCD peaks at temperatures $T<T_{\text{cr}}$ tend to superimpose and are placed at a bias value close to the critical current. Instead, at higher temperatures, $T>T_{\text{cr}}$, thermally activated processes rule the switching dynamics, the SCD peaks move at smaller $i_b$ and become larger. 
Here, we consider a working temperature high enough so that thermal activation prevails and MQT processes can be ignored.

When we take into account a L\'evy noise source with $\alpha<2$, the SCDs depart considerably from the pure thermal noise distribution, see Fig.~\ref{FigureSCD_01}(a). 
In fact, in this case L\'evy flights drive ``premature'' switches from the superconducting metastable state. We observe that as $\alpha$ decreases the low-current tail of the SCDs grows, and correspondingly the switching probability becomes sizable, at the expense of the SCD peak at high $i_b$ that is depleted. Moreover, we observe that the position of the high-$i_b$ SCD peak is not influenced by the L\'evy parameter $\alpha$, demonstrating its thermal origin.

We can even look at the CDFs, indicating the probability that the switching current $i_{\text{sw}}$ takes a value less than or equal to $i_b$, which highlights clearly the differences with the pure thermal case, see Fig.~\ref{FigureSCD_01}(b). 
In fact, the CDFs obtained changing $\alpha$ are evidently organized in well separate curves. In particular, it is evident that at a given bias $i_b$ the CDFs enlarge while decreasing $\alpha$. 
Moreover, as $i_b$ increases the CDF curves tend to saturate at the unitary value as the thermally activated processes are triggered.

We also give the analytical estimation of both PDF and CDF, in the case of a L\'evy noise~\cite{Gua19}. 
The CDF of $i_{\text{sw}}$ as a function of $i_b$ for a given initial bias ramp value, $i_0$, can be written as~\cite{Gua19}
\begin{equation}
\text{CDF}(i_b |i_0)=1- \mathrm{Prob} \left[ i_{\text{sw}} > i_b |i_0 \right ].
\label{eq:CDF_ib}
\end{equation}
Since the escape time distribution is exponential with rate $\tau^{-1} (i_b)$ even for L\'evy-flight noise~\cite{Che07}, the PDF related to Eq.~\eqref{eq:CDF_ib} \emph{vs} the average escape time $\tau(i_b)$ can be calculated as~\cite{Gua19}
\begin{equation}
\mathcal{P}(i_{b}|i_0) = \frac{\mathcal{N}}{v_b\tau(i_b)}\exp\left [ -\frac{1}{v_b} \int_{i_0}^{i_b} \frac{d i}{\tau(i) } \right ].
\label{Pib}
\end{equation}
Inserting in this equation the specific $\tau_L$ for a L\'evy noise, see Eq.~\eqref{tau_Levy}, at the first order in $i_b$ we attain the relation
\begin{equation}
\mathcal{P}(i_b|i_0) \propto \exp \left [ 
- \left( \frac {2} { \pi } \right)^{\alpha} 
\frac{i_b D^{\mu_{\alpha}} } {{\cal C}_{\alpha}v_b} \right ].
\label{P_t}
\end{equation}
Thus, the PDF of a current-biased junction can be written as
\begin{eqnarray}
\mathcal{P}(i_b|i_0) = \frac{1}{{\cal N}} 
\frac{\mathrm{d} {\cal F}_{\alpha}}{\mathrm{d} i_b}
\exp \left\{ -\frac{D^{\mu_{\alpha}}}{{\cal C}_{\alpha} v_{b}} \Big [ {\cal F}_{\alpha}(i_b)-{\cal F}_{\alpha}(i_0) \Big ] \right\},
\label{PDFib}
\end{eqnarray}
where
\begin{eqnarray}
&{\cal F}_{\alpha} (i_b) = 2^{\alpha } \bigg\{ i \frac{\pi ^{1-\alpha } }{4} \bigg [E_{\alpha }\left(-\frac{i \pi }{2}\right)-E_{\alpha }\left(\frac{i \pi }{2}\right)\bigg ]+\\
&\frac{\cosh^{-1}\left(i_b\right)/2}{\left[\pi -2 \arcsin \left(i_b\right)\right]^{\alpha }} \bigg [E_{\alpha }\Big(\!\cosh^{-1}i_b\!\Big)\!-\!E_{\alpha }\Big(\!-\cosh^{-1}i_b\!\Big)\bigg ]\!\! \bigg\},\nonumber
\label{Faux}
\end{eqnarray}
with $E_{\alpha} (\;)$ being the exponential integral with $\alpha$ argument and ${\cal N}$ the normalizing factor reading
\begin{equation}
{\cal N} = 1 - \exp{ \left[ -\frac{D^{\mu_{\alpha}}}
{{\cal C}_{\alpha} v_{b}} \Big ( {\cal F}_{\alpha}(1)-{\cal F}_{\alpha}(i_0) \Big ) \right] } .
\end{equation}
Finally, the resulting CDF can be written in a compact form as
\begin{equation}
\text{CDF}(i_b|i_0)= \frac{1}{{\cal N}} \left\{ \!\!1 - \exp \left[ -\frac{D^{\mu_{\alpha}}}{{\cal C}_{\alpha} v_{b}} \Big ( {\cal F}_{\alpha}(i_b)-{\cal F}_{\alpha}(i_0) \Big ) \right]\!\! \right\}.
\label{CDFib}
\end{equation}
This equation links the features of L\'evy flights, \emph{i.e.}, the exponent $\alpha$, with an experimentally accessible quantity, i.e, the SCD.

\begin{figure}[t!!]
\centering
\includegraphics[width=0.7\columnwidth]{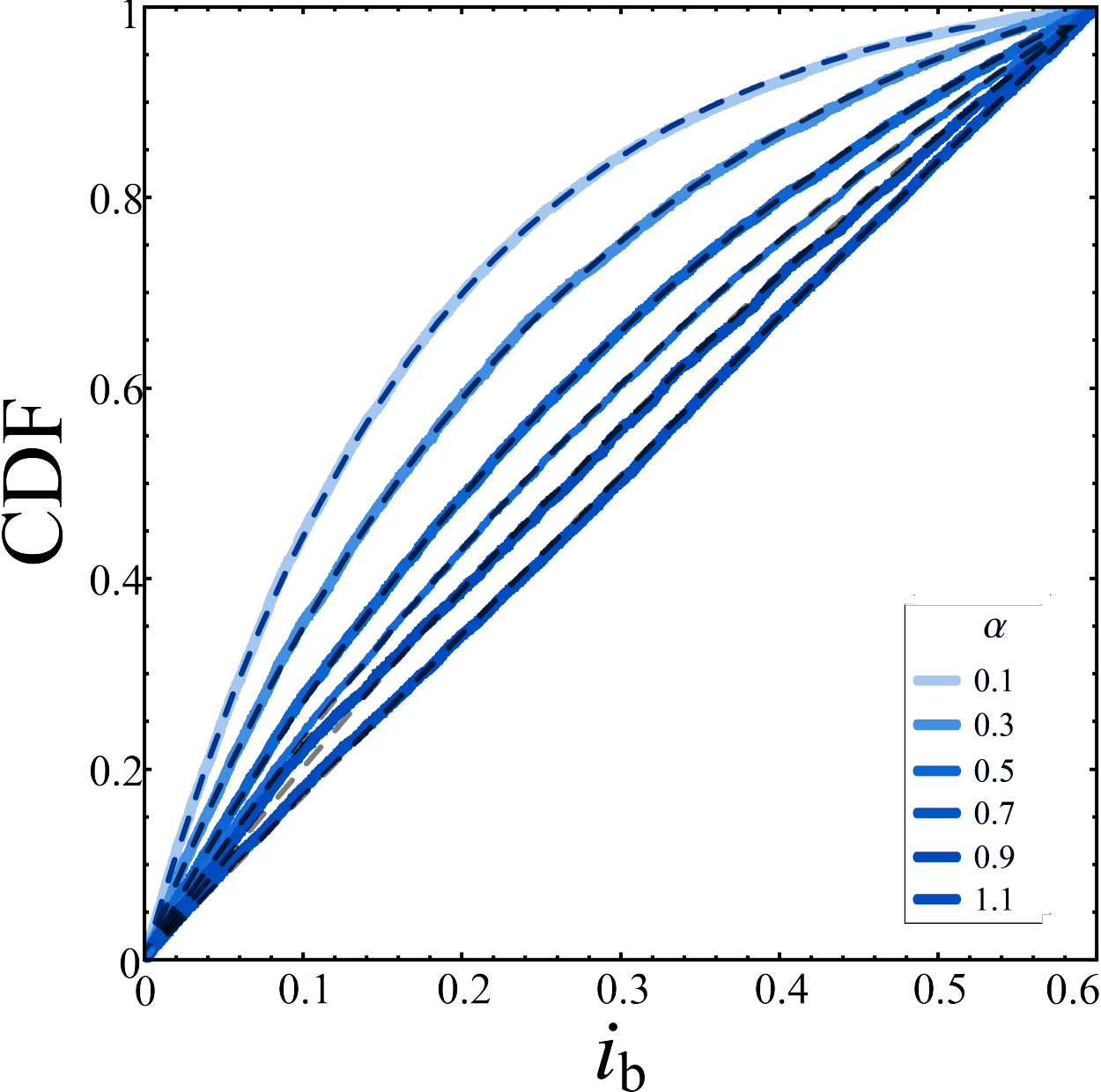}
\caption{CDF \emph{vs} $i_b$: theoretical curves obtained through Eq.~\eqref{CDFib} (dashed lines) and the theoretical marginal, \emph{i.e.}, for $i_b \leq 0.6$, obtained by solving Eq.~\eqref{eqJJ_norm_p} numerically (solid lines). The values of the other parameters are: $\alpha\in[0.1 , 1.1]$, $D_L=5\times10^{-7}$, $D=0$, $N_{\text{exp}} =10^{4}$, $t_{\text{max}} =10^{7}$, and $\beta_J = 0.1$.}
\label{FigureSCD_02}
\end{figure}

The effectiveness of the analytical procedure discussed so far can be confirmed by comparing the numerical results calculated considered the L\'evy noise only, that is imposing $D=0$, and Eq.~\eqref{CDFib}. Thus, Fig.~\ref{FigureSCD_02} shows the L\'evy-induced marginal CDFs (the term ``marginal'' refers to having imposed a maximum bias current $i_b=0.6$), calculated by solving Eq.~\eqref{eqJJ_norm_p} numerically (solid lines) and that obtained analytically from Eq.~\eqref{CDFib} (dashed line), for $\alpha\in[0.1-1.1]$ and $D_L=5\times10^{-7}$. In the limited range of $i_b$ values taken into account, a Gaussian noise eventually affecting the system could be in any case safely ignored and this justifies having set a zero-intensity Gaussian noise source. Finally, we observe that the computational outcomes well agree with the theoretical analysis for $\alpha\leq1$.
 %
%FigureSCD_02

\section{Conclusions}
\label{Sec05}

We explore the response of a Josephson junction (JJ) affected by a L\'evy noise source, \emph{i.e.}, a non-Gaussian noise characterized by a fat-tails fluctuations distribution. In particular, we discuss the effects on three different quantities that can be effectively measured experimentally: the mean switching time from the superconducting to the resistive state, the mean voltage drop across the junction, and the switching current distribution. We demonstrate that all these quantities carries information on a non-Gaussian background noise eventually interesting the system.

First, we study how the contemporaneous presence of stochastic non-Gaussian fluctuations and an oscillating driving contribute to the switching dynamics both in the case of a short and a long JJ. In this scheme, we consider both a constant and an alternating current signals feeding the junction and we analyze the behavior of the mean switching time (MST) as a function of the values of the characteristic L\'evy parameters $\alpha$ and $\beta$. The MST from the initial metastable state, \emph{i.e.}, a minimum of the tilted washboard potential, is calculated as a mean first passage time, in the case of short JJ, and as a nonlinear relaxation time, in the case of a long JJ. 
Analyzing the behavior of the MST as a function of the driving frequency $\omega$, we observe the emergence of resonance activation, that is a noise--induced phenomenon whose signature is the appearance of a minimum in the MST \emph{vs} $\omega$ curves, as the driving frequency matches a natural frequency of the system. 
We also find evidence of a nonmonotonic response of the MSTs as a function of the noise intensity, and the phenomenon of noise enhanced stability, whose characteristics strongly depend on $\alpha$ and $\beta$.

We also demonstrate that the mean voltage drop (MVD) across a short JJ exhibits a peculiar behavior as a function of the L\'evy noise intensity, which is markedly different from the Gaussian noise case for the occurrence of L\'evy flights, that is scale-free jumps. In this case, an accurate and stable dc input current is required, in order to bias the junction with a constant current below the critical value. We observe that the voltage grows linearly as a function of the L\'evy noise intensity. 
Moreover, we show that the slope of the linear behavior of the MVDs depends on the L\'evy index $\alpha$, and it is therefore possible to 
discriminate the feature of the noise source from the analysis of the voltage drop across the junction. 
Moreover, thermal effects can be kept at bay if the temperature of the device is maintained below a certain threshold.
Finally, we also present the analytical estimate of the average speed, $\left \langle v \right \rangle$, of a particle in a metastable 
washboard potential in the presence of L\'evy-distributed fluctuations. This estimation agrees with our numerical results, being $\left \langle v \right \rangle$ proportional to the voltage in the Josephson framework. 

Last but not least, we demonstrate that the switching currents distributions (SCDs) of a short JJ, embedded in a thermal noise background and in the presence of a L\'evy noise source, disclose a peculiar behavior that diverges significantly from the Gaussian noise case. In this setup, the bias current is linearly swept upward from zero, and the current at which the system switches to the resistive state is recorded, so that reiterating the experiment many times, a distribution can be constructed. 
We show that the investigation of both the PDFs and the CDFs of the switching currents permits to reveal the presence of a L\'evy noise signal by the occurrence of anomalous premature switches in the low-current tail of the distributions. 
Moreover, we demonstrate that the specific features of the L\'evy noise signal can be inferred from the SCD profile, even considering a thermal noise disturbance. In fact, if we assume both L\'evy and Gaussian (thermal) noises, they do not interfere. In fact, they produce switching at different bias levels, \emph{i.e.}, the L\'evy noise in the low-bias part of the distribution, while the Gaussian noise when the energy barrier becomes comparable to the noise energy, that is at high bias currents close to the critical value. Sweeping the bias is a very effective method, for the bias increment reduces the confining potential barrier, so that in a predetermined ramp time the energy barrier disappears and a switching event is definitively recorded, even for a vanishing noise intensity. Moreover, the ramp time of the bias current can be even adjusted to make the L\'evy source effects on the SCDs more evident, according to the intensity of the noise fluctuations.

By way of conclusion, we demonstrated that the study of the Josephson response can pave the way to concrete applications of Josephson devices for characterizing L\'evy noise sources. The proposed methods are quite robust in recognizing the L\'evy component in a thermal noisy background, for the probability of a particle to pass a barrier when exposed to L\'evy noise is independent of the barrier energy. This differs significantly from the Gaussian noise case, where the probability to exceed the barrier depends exponentially on the barrier height.

\section*{Acknowledgments }
The idea of considering this problem was suggested by Bernardo Spagnolo and Davide Valenti, to whom the author is grateful for seminal discussions of the work findings. The author would also like to acknowledge interesting and fruitful discussions with Giovanni Filatrella.

\printcredits

%% Loading bibliography style file
%\bibliographystyle{model1-num-names}
\bibliographystyle{cas-model2-names}

% Loading bibliography database
%\bibliography{biblio.bib}

\begin{thebibliography}{180}
\expandafter\ifx\csname natexlab\endcsname\relax\def\natexlab#1{#1}\fi
\providecommand{\url}[1]{\texttt{#1}}
\providecommand{\href}[2]{#2}
\providecommand{\path}[1]{#1}
\providecommand{\DOIprefix}{doi:}
\providecommand{\ArXivprefix}{arXiv:}
\providecommand{\URLprefix}{URL: }
\providecommand{\Pubmedprefix}{pmid:}
\providecommand{\doi}[1]{\href{http://dx.doi.org/#1}{\path{#1}}}
\providecommand{\Pubmed}[1]{\href{pmid:#1}{\path{#1}}}
\providecommand{\bibinfo}[2]{#2}
\ifx\xfnm\relax \def\xfnm[#1]{\unskip,\space#1}\fi
%Type = Article
\bibitem[{Addesso et~al.(2012)Addesso, Filatrella and Pierro}]{Add12}
\bibinfo{author}{Addesso, P.}, \bibinfo{author}{Filatrella, G.},
  \bibinfo{author}{Pierro, V.}, \bibinfo{year}{2012}.
\newblock \bibinfo{title}{Characterization of escape times of Josephson
  junctions for signal detection}.
\newblock \bibinfo{journal}{Phys. Rev. E} \bibinfo{volume}{85},
  \bibinfo{pages}{016708}.
\newblock \URLprefix \url{https://link.aps.org/doi/10.1103/PhysRevE.85.016708},
  \DOIprefix\doi{10.1103/PhysRevE.85.016708}.
%Type = Article
\bibitem[{Agudov and Spagnolo(2001)}]{Agu01}
\bibinfo{author}{Agudov, N.V.}, \bibinfo{author}{Spagnolo, B.},
  \bibinfo{year}{2001}.
\newblock \bibinfo{title}{Noise-enhanced stability of periodically driven
  metastable states}.
\newblock \bibinfo{journal}{Phys. Rev. E} \bibinfo{volume}{64},
  \bibinfo{pages}{351021--351024}.
%Type = Article
\bibitem[{Ankerhold(2007)}]{Ank07}
\bibinfo{author}{Ankerhold, J.}, \bibinfo{year}{2007}.
\newblock \bibinfo{title}{Detecting charge noise with a Josephson junction: A
  problem of thermal escape in presence of non-Gaussian fluctuations}.
\newblock \bibinfo{journal}{Phys. Rev. Lett.} \bibinfo{volume}{98},
  \bibinfo{pages}{036601}.
\newblock \URLprefix
  \url{http://link.aps.org/doi/10.1103/PhysRevLett.98.036601},
  \DOIprefix\doi{10.1103/PhysRevLett.98.036601}.
%Type = Article
\bibitem[{Ankerhold and Grabert(2005)}]{Ank05}
\bibinfo{author}{Ankerhold, J.}, \bibinfo{author}{Grabert, H.},
  \bibinfo{year}{2005}.
\newblock \bibinfo{title}{How to detect the fourth-order cumulant of electrical
  noise}.
\newblock \bibinfo{journal}{Phys. Rev. Lett.} \bibinfo{volume}{95},
  \bibinfo{pages}{186601}.
\newblock \URLprefix
  \url{http://link.aps.org/doi/10.1103/PhysRevLett.95.186601},
  \DOIprefix\doi{10.1103/PhysRevLett.95.186601}.
%Type = Article
\bibitem[{Augello et~al.(2009)Augello, Valenti, Pankratov and Spagnolo}]{Aug09}
\bibinfo{author}{Augello, G.}, \bibinfo{author}{Valenti, D.},
  \bibinfo{author}{Pankratov, A.L.}, \bibinfo{author}{Spagnolo, B.},
  \bibinfo{year}{2009}.
\newblock \bibinfo{title}{Lifetime of the superconductive state in short and
  long Josephson junctions}.
\newblock \bibinfo{journal}{The European Physical Journal B}
  \bibinfo{volume}{70}, \bibinfo{pages}{145--151}.
\newblock \URLprefix \url{https://doi.org/10.1140/epjb/e2009-00155-x},
  \DOIprefix\doi{10.1140/epjb/e2009-00155-x}.
%Type = Article
\bibitem[{Augello et~al.(2010)Augello, Valenti and Spagnolo}]{Aug10}
\bibinfo{author}{Augello, G.}, \bibinfo{author}{Valenti, D.},
  \bibinfo{author}{Spagnolo, B.}, \bibinfo{year}{2010}.
\newblock \bibinfo{title}{Non-Gaussian noise effects in the dynamics of a short
  overdamped Josephson junction}.
\newblock \bibinfo{journal}{Eur. Phys. J. B} \bibinfo{volume}{78},
  \bibinfo{pages}{225--234}.
\newblock \URLprefix \url{https://doi.org/10.1140/epjb/e2010-10106-1},
  \DOIprefix\doi{10.1140/epjb/e2010-10106-1}.
%Type = Book
\bibitem[{Barone and Patern\`{o}(1982)}]{Bar82}
\bibinfo{author}{Barone, A.}, \bibinfo{author}{Patern\`{o}, G.},
  \bibinfo{year}{1982}.
\newblock \bibinfo{title}{Physics and Applications of the Josephson Effect}.
\newblock \bibinfo{publisher}{Wiley, New York}.
%Type = Article
\bibitem[{Ben-Jacob and Bergman(1984)}]{Ben84}
\bibinfo{author}{Ben-Jacob, E.}, \bibinfo{author}{Bergman, D.J.},
  \bibinfo{year}{1984}.
\newblock \bibinfo{title}{Thermal noise effects on the microwave-induced steps
  of a current-driven Josephson junction}.
\newblock \bibinfo{journal}{Phys. Rev. A} \bibinfo{volume}{29},
  \bibinfo{pages}{2021--2028}.
\newblock \URLprefix \url{https://link.aps.org/doi/10.1103/PhysRevA.29.2021},
  \DOIprefix\doi{10.1103/PhysRevA.29.2021}.
%Type = Incollection
\bibitem[{Berggren et~al.(2013)Berggren, Dauler, Kerman, Nam and
  Rosenberg}]{Ber13}
\bibinfo{author}{Berggren, K.K.}, \bibinfo{author}{Dauler, E.A.},
  \bibinfo{author}{Kerman, A.J.}, \bibinfo{author}{Nam, S.W.},
  \bibinfo{author}{Rosenberg, D.}, \bibinfo{year}{2013}.
\newblock \bibinfo{title}{Detectors based on superconductors}, in:
  \bibinfo{booktitle}{Experimental Methods in the Physical Sciences}.
  \bibinfo{publisher}{Elsevier, New York}. volume~\bibinfo{volume}{45}, pp.
  \bibinfo{pages}{185--216}.
%Type = Book
\bibitem[{Bertoin(1996)}]{Ber96}
\bibinfo{author}{Bertoin, J.}, \bibinfo{year}{1996}.
\newblock \bibinfo{title}{L\'evy Processes}.
\newblock \bibinfo{publisher}{Cambridge University Press, Cambridge}.
%Type = Article
\bibitem[{Bhatia et~al.(2006)Bhatia, Mulgrew and Georgiadis}]{Bha06}
\bibinfo{author}{Bhatia, V.}, \bibinfo{author}{Mulgrew, B.},
  \bibinfo{author}{Georgiadis, A.}, \bibinfo{year}{2006}.
\newblock \bibinfo{title}{Stochastic gradient algorithms for equalisation in
  $\alpha$--stable noise}.
\newblock \bibinfo{journal}{Signal Processing} \bibinfo{volume}{86},
  \bibinfo{pages}{835 -- 845}.
\newblock \URLprefix
  \url{http://www.sciencedirect.com/science/article/pii/S0165168405002380},
  \DOIprefix\doi{https://doi.org/10.1016/j.sigpro.2005.06.013}.
%Type = Article
\bibitem[{Blackburn et~al.(2016)Blackburn, Cirillo and
  Gr{\o}nbech-Jensen}]{Bla16}
\bibinfo{author}{Blackburn, J.A.}, \bibinfo{author}{Cirillo, M.},
  \bibinfo{author}{Gr{\o}nbech-Jensen, N.}, \bibinfo{year}{2016}.
\newblock \bibinfo{title}{A survey of classical and quantum interpretations of
  experiments on Josephson junctions at very low temperatures}.
\newblock \bibinfo{journal}{Physics Reports} \bibinfo{volume}{611},
  \bibinfo{pages}{1 -- 33}.
\newblock \URLprefix
  \url{http://www.sciencedirect.com/science/article/pii/S0370157315004433},
  \DOIprefix\doi{https://doi.org/10.1016/j.physrep.2015.10.010}.
%Type = Article
\bibitem[{Boldyrev and Gwinn(2003)}]{Bol03}
\bibinfo{author}{Boldyrev, S.}, \bibinfo{author}{Gwinn, C.R.},
  \bibinfo{year}{2003}.
\newblock \bibinfo{title}{L\'evy model for interstellar scintillations}.
\newblock \bibinfo{journal}{Phys. Rev. Lett.} \bibinfo{volume}{91},
  \bibinfo{pages}{131101}.
\newblock \URLprefix
  \url{https://link.aps.org/doi/10.1103/PhysRevLett.91.131101},
  \DOIprefix\doi{10.1103/PhysRevLett.91.131101}.
%Type = Article
\bibitem[{Boldyrev and Gwinn(2005)}]{Bol05}
\bibinfo{author}{Boldyrev, S.}, \bibinfo{author}{Gwinn, C.R.},
  \bibinfo{year}{2005}.
\newblock \bibinfo{title}{Radio-wave propagation in the non-Gaussian
  interstellar medium}.
\newblock \bibinfo{journal}{Astrophys. J} \bibinfo{volume}{624},
  \bibinfo{pages}{213--222}.
\newblock \URLprefix \url{https://doi.org/10.1086%2F428919},
  \DOIprefix\doi{10.1086/428919}.
%Type = Article
\bibitem[{Boldyrev and Konigl(2006)}]{Bol06}
\bibinfo{author}{Boldyrev, S.}, \bibinfo{author}{Konigl, A.},
  \bibinfo{year}{2006}.
\newblock \bibinfo{title}{Non-Gaussian radio-wave scattering in the
  interstellar medium}.
\newblock \bibinfo{journal}{Astrophys. J} \bibinfo{volume}{640},
  \bibinfo{pages}{344--352}.
\newblock \URLprefix \url{https://doi.org/10.1086%2F499219},
  \DOIprefix\doi{10.1086/499219}.
%Type = Article
\bibitem[{Braginski(2019)}]{Bra19}
\bibinfo{author}{Braginski, A.I.}, \bibinfo{year}{2019}.
\newblock \bibinfo{title}{Superconductor electronics: Status and outlook}.
\newblock \bibinfo{journal}{Journal of Superconductivity and Novel Magnetism}
  \bibinfo{volume}{32}, \bibinfo{pages}{23--44}.
\newblock \URLprefix \url{https://doi.org/10.1007/s10948-018-4884-4},
  \DOIprefix\doi{10.1007/s10948-018-4884-4}.
%Type = Article
\bibitem[{Briskot et~al.(2014)Briskot, Dmitriev and Mirlin}]{Bri14}
\bibinfo{author}{Briskot, U.}, \bibinfo{author}{Dmitriev, I.A.},
  \bibinfo{author}{Mirlin, A.D.}, \bibinfo{year}{2014}.
\newblock \bibinfo{title}{Relaxation of optically excited carriers in graphene:
  Anomalous diffusion and L\'evy flights}.
\newblock \bibinfo{journal}{Phys. Rev. B} \bibinfo{volume}{89},
  \bibinfo{pages}{075414}.
\newblock \URLprefix \url{http://link.aps.org/doi/10.1103/PhysRevB.89.075414},
  \DOIprefix\doi{10.1103/PhysRevB.89.075414}.
%Type = Article
\bibitem[{Brokmann et~al.(2003)Brokmann, Hermier, Messin, Desbiolles, Bouchaud
  and Dahan}]{Bro03}
\bibinfo{author}{Brokmann, X.}, \bibinfo{author}{Hermier, J.P.},
  \bibinfo{author}{Messin, G.}, \bibinfo{author}{Desbiolles, P.},
  \bibinfo{author}{Bouchaud, J.P.}, \bibinfo{author}{Dahan, M.},
  \bibinfo{year}{2003}.
\newblock \bibinfo{title}{Statistical aging and nonergodicity in the
  fluorescence of single nanocrystals}.
\newblock \bibinfo{journal}{Phys. Rev. Lett.} \bibinfo{volume}{90},
  \bibinfo{pages}{120601}.
\newblock \URLprefix
  \url{https://link.aps.org/doi/10.1103/PhysRevLett.90.120601},
  \DOIprefix\doi{10.1103/PhysRevLett.90.120601}.
%Type = Article
\bibitem[{Carollo et~al.(2018)Carollo, Spagnolo and Valenti}]{Car18}
\bibinfo{author}{Carollo, A.}, \bibinfo{author}{Spagnolo, B.},
  \bibinfo{author}{Valenti, D.}, \bibinfo{year}{2018}.
\newblock \bibinfo{title}{Uhlmann curvature in dissipative phase transitions}.
\newblock \bibinfo{journal}{Scientific Reports} \bibinfo{volume}{8},
  \bibinfo{pages}{9852}.
\newblock \URLprefix \url{https://doi.org/10.1038/s41598-018-27362-9},
  \DOIprefix\doi{10.1038/s41598-018-27362-9}.
%Type = Article
\bibitem[{Carollo et~al.(2020)Carollo, Valenti and Spagnolo}]{Car21}
\bibinfo{author}{Carollo, A.}, \bibinfo{author}{Valenti, D.},
  \bibinfo{author}{Spagnolo, B.}, \bibinfo{year}{2020}.
\newblock \bibinfo{title}{Geometry of quantum phase transitions}.
\newblock \bibinfo{journal}{Physics Reports} \bibinfo{volume}{838},
  \bibinfo{pages}{1--72}.
\newblock \URLprefix
  \url{https://www.sciencedirect.com/science/article/pii/S0370157319303655},
  \DOIprefix\doi{https://doi.org/10.1016/j.physrep.2019.11.002}.
  \bibinfo{note}{geometry of quantum phase transitions}.
%Type = Article
\bibitem[{Castellano et~al.(1996)Castellano, Leoni, Torrioli, Chiarello,
  Cosmelli, Costantini, Diambrini‐Palazzi, Carelli, Cristiano and
  Frunzio}]{Castellano96}
\bibinfo{author}{Castellano, M.G.}, \bibinfo{author}{Leoni, R.},
  \bibinfo{author}{Torrioli, G.}, \bibinfo{author}{Chiarello, F.},
  \bibinfo{author}{Cosmelli, C.}, \bibinfo{author}{Costantini, A.},
  \bibinfo{author}{Diambrini‐Palazzi, G.}, \bibinfo{author}{Carelli, P.},
  \bibinfo{author}{Cristiano, R.}, \bibinfo{author}{Frunzio, L.},
  \bibinfo{year}{1996}.
\newblock \bibinfo{title}{Switching dynamics of Nb/AlOx/Nb Josephson junctions:
  Measurements for an experiment of macroscopic quantum coherence}.
\newblock \bibinfo{journal}{Journal of Applied Physics} \bibinfo{volume}{80},
  \bibinfo{pages}{2922--2928}.
\newblock \URLprefix \url{https://doi.org/10.1063/1.363147},
  \DOIprefix\doi{10.1063/1.363147}.
%Type = Article
\bibitem[{Chambers et~al.(1976)Chambers, Mallows and Stuck}]{Cha76}
\bibinfo{author}{Chambers, J.M.}, \bibinfo{author}{Mallows, C.L.},
  \bibinfo{author}{Stuck, B.W.}, \bibinfo{year}{1976}.
\newblock \bibinfo{title}{A method for simulating stable random variables}.
\newblock \bibinfo{journal}{J. Amer. Statist. Assoc.} \bibinfo{volume}{71},
  \bibinfo{pages}{340--344}.
  \newblock \DOIprefix\doi{10.2307/2285309}.
%Type = Article
\bibitem[{Chechkin et~al.(2002)Chechkin, Gonchar, Klafter, Metzler and
  Tanatarov}]{Che02}
\bibinfo{author}{Chechkin, A.}, \bibinfo{author}{Gonchar, V.},
  \bibinfo{author}{Klafter, J.}, \bibinfo{author}{Metzler, R.},
  \bibinfo{author}{Tanatarov, L.}, \bibinfo{year}{2002}.
\newblock \bibinfo{title}{Stationary states of non-linear oscillators driven by
  L\'evy noise}.
\newblock \bibinfo{journal}{Chemical Physics} \bibinfo{volume}{284},
  \bibinfo{pages}{233--251}.
\newblock \URLprefix
  \url{https://www.sciencedirect.com/science/article/pii/S0301010402005517},
  \DOIprefix\doi{https://doi.org/10.1016/S0301-0104(02)00551-7}.
  \bibinfo{note}{strange Kinetics}.
%Type = Article
\bibitem[{Chechkin et~al.(2005)Chechkin, Gonchar, Klafter and Metzler}]{Che05}
\bibinfo{author}{Chechkin, A.V.}, \bibinfo{author}{Gonchar, V.Y.},
  \bibinfo{author}{Klafter, J.}, \bibinfo{author}{Metzler, R.},
  \bibinfo{year}{2005}.
\newblock \bibinfo{title}{Barrier crossing of a L\'evy flight}.
\newblock \bibinfo{journal}{Europhys. Lett.} \bibinfo{volume}{72},
  \bibinfo{pages}{348}.
\newblock \URLprefix \url{http://stacks.iop.org/0295-5075/72/i=3/a=348}.
%Type = Article
\bibitem[{Chechkin et~al.(2007)Chechkin, Sliusarenko, Metzler and
  Klafter}]{Che07}
\bibinfo{author}{Chechkin, A.V.}, \bibinfo{author}{Sliusarenko, O.Y.},
  \bibinfo{author}{Metzler, R.}, \bibinfo{author}{Klafter, J.},
  \bibinfo{year}{2007}.
\newblock \bibinfo{title}{Barrier crossing driven by L\'evy noise: Universality
  and the role of noise intensity}.
\newblock \bibinfo{journal}{Phys. Rev. E} \bibinfo{volume}{75},
  \bibinfo{pages}{041101}.
\newblock \URLprefix \url{http://link.aps.org/doi/10.1103/PhysRevE.75.041101},
  \DOIprefix\doi{10.1103/PhysRevE.75.041101}.
%Type = Inproceedings
\bibitem[{Chouri et~al.(2014)Chouri, Fabrice, Dandache, Aroussi and
  Saadane}]{Cho14}
\bibinfo{author}{Chouri, B.}, \bibinfo{author}{Fabrice, M.},
  \bibinfo{author}{Dandache, A.}, \bibinfo{author}{Aroussi, M.E.},
  \bibinfo{author}{Saadane, R.}, \bibinfo{year}{2014}.
\newblock \bibinfo{title}{Bearing fault diagnosis based on alpha-stable
  distribution feature extraction and svm classifier}, in:
  \bibinfo{booktitle}{2014 International Conference on Multimedia Computing and
  Systems (ICMCS)}, pp. \bibinfo{pages}{1545--1550}.
%Type = Article
\bibitem[{Chua(1971)}]{Chu71}
\bibinfo{author}{Chua, L.}, \bibinfo{year}{1971}.
\newblock \bibinfo{title}{Memristor-the missing circuit element}.
\newblock \bibinfo{journal}{IEEE Trans. Circuit Theory} \bibinfo{volume}{18},
  \bibinfo{pages}{507--519}.
\newblock \DOIprefix\doi{10.1109/TCT.1971.1083337}.
%Type = Article
\bibitem[{Chua(2003)}]{Chu03}
\bibinfo{author}{Chua, L.}, \bibinfo{year}{2003}.
\newblock \bibinfo{title}{Nonlinear circuit foundations for nanodevices. i. the
  four-element torus}.
\newblock \bibinfo{journal}{Proceedings of the IEEE} \bibinfo{volume}{91},
  \bibinfo{pages}{1830--1859}.
\newblock \DOIprefix\doi{10.1109/JPROC.2003.818319}.
%Type = Article
\bibitem[{Chua and Kang(1976)}]{Chu76}
\bibinfo{author}{Chua, L.O.}, \bibinfo{author}{Kang, S.M.},
  \bibinfo{year}{1976}.
\newblock \bibinfo{title}{Memristive devices and systems}.
\newblock \bibinfo{journal}{Proceedings of the IEEE} \bibinfo{volume}{64},
  \bibinfo{pages}{209--223}.
\newblock \DOIprefix\doi{10.1109/PROC.1976.10092}.
%Type = Article
\bibitem[{Cortes et~al.(2010)Cortes, Diez, Canete and Sanchez-Martinez}]{Cor10}
\bibinfo{author}{Cortes, J.A.}, \bibinfo{author}{Diez, L.},
  \bibinfo{author}{Canete, F.J.}, \bibinfo{author}{Sanchez-Martinez, J.J.},
  \bibinfo{year}{2010}.
\newblock \bibinfo{title}{Analysis of the indoor broadband power-line noise
  scenario}.
\newblock \bibinfo{journal}{IEEE Transactions on Electromagnetic Compatibility}
  \bibinfo{volume}{52}, \bibinfo{pages}{849--858}.
  \newblock \URLprefix
  \url{https://ieeexplore.ieee.org/document/5570980},
  \DOIprefix\doi{10.1109/TEMC.2010.2052463}.
%Type = Article
\bibitem[{Coskun et~al.(2012)Coskun, Brenner, Hymel, Vakaryuk, Levchenko and
  Bezryadin}]{Cos12}
\bibinfo{author}{Coskun, U.C.}, \bibinfo{author}{Brenner, M.},
  \bibinfo{author}{Hymel, T.}, \bibinfo{author}{Vakaryuk, V.},
  \bibinfo{author}{Levchenko, A.}, \bibinfo{author}{Bezryadin, A.},
  \bibinfo{year}{2012}.
\newblock \bibinfo{title}{Distribution of supercurrent switching in graphene
  under the proximity effect}.
\newblock \bibinfo{journal}{Phys. Rev. Lett.} \bibinfo{volume}{108},
  \bibinfo{pages}{097003}.
\newblock \URLprefix
  \url{http://link.aps.org/doi/10.1103/PhysRevLett.108.097003},
  \DOIprefix\doi{10.1103/PhysRevLett.108.097003}.
%Type = Article
\bibitem[{Denaro et~al.(2013a)Denaro, Valenti, {La Cognata}, Spagnolo, Bonanno,
  Basilone, Mazzola, Zgozi, Aronica and Brunet}]{Den13}
\bibinfo{author}{Denaro, G.}, \bibinfo{author}{Valenti, D.},
  \bibinfo{author}{{La Cognata}, A.}, \bibinfo{author}{Spagnolo, B.},
  \bibinfo{author}{Bonanno, A.}, \bibinfo{author}{Basilone, G.},
  \bibinfo{author}{Mazzola, S.}, \bibinfo{author}{Zgozi, S.},
  \bibinfo{author}{Aronica, S.}, \bibinfo{author}{Brunet, C.},
  \bibinfo{year}{2013}a.
\newblock \bibinfo{title}{Spatio-temporal behaviour of the deep chlorophyll
  maximum in Mediterranean Sea: Development of a stochastic model for
  picophytoplankton dynamics}.
\newblock \bibinfo{journal}{Ecological Complexity} \bibinfo{volume}{13},
  \bibinfo{pages}{21--34}.
\newblock \URLprefix
  \url{https://www.sciencedirect.com/science/article/pii/S1476945X12000748},
  \DOIprefix\doi{https://doi.org/10.1016/j.ecocom.2012.10.002}.
%Type = Article
\bibitem[{Denaro et~al.(2013b)Denaro, Valenti, Spagnolo, Basilone, Mazzola,
  Zgozi, Aronica and Bonanno}]{DenVal13}
\bibinfo{author}{Denaro, G.}, \bibinfo{author}{Valenti, D.},
  \bibinfo{author}{Spagnolo, B.}, \bibinfo{author}{Basilone, G.},
  \bibinfo{author}{Mazzola, S.}, \bibinfo{author}{Zgozi, S.W.},
  \bibinfo{author}{Aronica, S.}, \bibinfo{author}{Bonanno, A.},
  \bibinfo{year}{2013}b.
\newblock \bibinfo{title}{Dynamics of two picophytoplankton groups in
  Mediterranean Sea: Analysis of the deep chlorophyll maximum by a stochastic
  advection-reaction-diffusion model}.
\newblock \bibinfo{journal}{PLoS One} \bibinfo{volume}{8},
  \bibinfo{pages}{e66765}.
  \newblock \URLprefix
  \url{https://journals.plos.org/plosone/article?id=10.1371/journal.pone.0066765},
  \DOIprefix\doi{10.1371/journal.pone.0066765}.
%Type = Article
\bibitem[{Devoret et~al.(1985)Devoret, Martinis and Clarke}]{Dev85}
\bibinfo{author}{Devoret, M.H.}, \bibinfo{author}{Martinis, J.M.},
  \bibinfo{author}{Clarke, J.}, \bibinfo{year}{1985}.
\newblock \bibinfo{title}{Measurements of macroscopic quantum tunneling out of
  the zero-voltage state of a current-biased Josephson junction}.
\newblock \bibinfo{journal}{Phys. Rev. Lett.} \bibinfo{volume}{55},
  \bibinfo{pages}{1908--1911}.
\newblock \URLprefix
  \url{https://link.aps.org/doi/10.1103/PhysRevLett.55.1908},
  \DOIprefix\doi{10.1103/PhysRevLett.55.1908}.
%Type = Article
\bibitem[{Devoret et~al.(1984)Devoret, Martinis, Esteve and Clarke}]{Dev84}
\bibinfo{author}{Devoret, M.H.}, \bibinfo{author}{Martinis, J.M.},
  \bibinfo{author}{Esteve, D.}, \bibinfo{author}{Clarke, J.},
  \bibinfo{year}{1984}.
\newblock \bibinfo{title}{Resonant activation from the zero-voltage state of a
  current-biased Josephson junction}.
\newblock \bibinfo{journal}{Phys. Rev. Lett.} \bibinfo{volume}{53},
  \bibinfo{pages}{1260--1263}.
  \newblock \URLprefix
  \url{https://link.aps.org/doi/10.1103/PhysRevLett.53.1260},
  \DOIprefix\doi{10.1103/PhysRevLett.53.1260}.
%Type = Article
\bibitem[{Di~Ventra et~al.(2009)Di~Ventra, Pershin and Chua}]{DiV09}
\bibinfo{author}{Di~Ventra, M.}, \bibinfo{author}{Pershin, Y.V.},
  \bibinfo{author}{Chua, L.O.}, \bibinfo{year}{2009}.
\newblock \bibinfo{title}{Circuit elements with memory: Memristors,
  memcapacitors, and meminductors}.
\newblock \bibinfo{journal}{Proc. IEEE} \bibinfo{volume}{97},
  \bibinfo{pages}{1717--1724}.
\newblock \DOIprefix\doi{10.1109/JPROC.2009.2021077}.
%Type = Article
\bibitem[{D'Odorico et~al.(2005)D'Odorico, Laio and Ridolfi}]{DOd05}
\bibinfo{author}{D'Odorico, P.}, \bibinfo{author}{Laio, F.},
  \bibinfo{author}{Ridolfi, L.}, \bibinfo{year}{2005}.
\newblock \bibinfo{title}{Noise-induced stability in dryland plant ecosystems}.
\newblock \bibinfo{journal}{Proc. Natl. Acad. Sci. U.S.A.}
  \bibinfo{volume}{102}, \bibinfo{pages}{10819--10822}.
  \newblock \URLprefix \url{https://www.pnas.org/content/102/31/10819},
  \DOIprefix\doi{10.1073/pnas.0502884102}.
%Type = Article
\bibitem[{Doering and Gadoua(1992)}]{Doe92}
\bibinfo{author}{Doering, C.R.}, \bibinfo{author}{Gadoua, J.C.},
  \bibinfo{year}{1992}.
\newblock \bibinfo{title}{Resonant activation over a fluctuating barrier}.
\newblock \bibinfo{journal}{Phys. Rev. Lett.} \bibinfo{volume}{69},
  \bibinfo{pages}{2318--2321}.
\newblock \URLprefix \url{http://link.aps.org/doi/10.1103/PhysRevLett.69.2318},
  \DOIprefix\doi{10.1103/PhysRevLett.69.2318}.
%Type = Article
\bibitem[{Dubkov et~al.(2004)Dubkov, Agudov and Spagnolo}]{Dub04}
\bibinfo{author}{Dubkov, A.A.}, \bibinfo{author}{Agudov, N.V.},
  \bibinfo{author}{Spagnolo, B.}, \bibinfo{year}{2004}.
\newblock \bibinfo{title}{Noise-enhanced stability in fluctuating metastable
  states}.
\newblock \bibinfo{journal}{Phys. Rev. E} \bibinfo{volume}{69},
  \bibinfo{pages}{061103}.
\newblock \URLprefix \url{http://link.aps.org/doi/10.1103/PhysRevE.69.061103},
  \DOIprefix\doi{10.1103/PhysRevE.69.061103}.
%Type = Article
\bibitem[{Dubkov et~al.(2009)Dubkov, Cognata and Spagnolo}]{Dub09}
\bibinfo{author}{Dubkov, A.A.}, \bibinfo{author}{Cognata, A.L.},
  \bibinfo{author}{Spagnolo, B.}, \bibinfo{year}{2009}.
\newblock \bibinfo{title}{The problem of analytical calculation of barrier
  crossing characteristics for L\'evy flights}.
\newblock \bibinfo{journal}{J. Stat. Mech.: Theory Exp.}
  \bibinfo{volume}{2009}, \bibinfo{pages}{P01002}.
\newblock \URLprefix \url{http://stacks.iop.org/1742-5468/2009/i=01/a=P01002}.
%Type = Article
\bibitem[{Dubkov and Spagnolo(2005)}]{Dub05}
\bibinfo{author}{Dubkov, A.A.}, \bibinfo{author}{Spagnolo, B.},
  \bibinfo{year}{2005}.
\newblock \bibinfo{title}{Acceleration of diffusion in randomly switching
  potential with supersymmetry}.
\newblock \bibinfo{journal}{Phys. Rev. E} \bibinfo{volume}{72},
  \bibinfo{pages}{041104}.
\newblock \URLprefix \url{https://link.aps.org/doi/10.1103/PhysRevE.72.041104},
  \DOIprefix\doi{10.1103/PhysRevE.72.041104}.
%Type = Article
\bibitem[{Dubkov et~al.(2008)Dubkov, Spagnolo and Uchaikin}]{Dub08}
\bibinfo{author}{Dubkov, A.A.}, \bibinfo{author}{Spagnolo, B.},
  \bibinfo{author}{Uchaikin, V.V.}, \bibinfo{year}{2008}.
\newblock \bibinfo{title}{L{\'e}vy flight superdiffusion: an introduction}.
\newblock \bibinfo{journal}{Int. J. Bifurcation Chaos Appl. Sci. Eng.}
  \bibinfo{volume}{18}, \bibinfo{pages}{2649--2672}.
%Type = Article
\bibitem[{Dybiec and Gudowska-Nowak(2009)}]{Dyb09}
\bibinfo{author}{Dybiec, B.}, \bibinfo{author}{Gudowska-Nowak, E.},
  \bibinfo{year}{2009}.
\newblock \bibinfo{title}{L{\'{e}}vy stable noise-induced transitions:
  stochastic resonance, resonant activation and dynamic hysteresis}.
\newblock \bibinfo{journal}{Journal of Statistical Mechanics: Theory and
  Experiment} \bibinfo{volume}{2009}, \bibinfo{pages}{P05004}.
\newblock \URLprefix \url{https://doi.org/10.1088/1742-5468/2009/05/p05004},
  \DOIprefix\doi{10.1088/1742-5468/2009/05/p05004}.
%Type = Article
\bibitem[{Dybiec et~al.(2006)Dybiec, Gudowska-Nowak and H\"anggi}]{Dyb06}
\bibinfo{author}{Dybiec, B.}, \bibinfo{author}{Gudowska-Nowak, E.},
  \bibinfo{author}{H\"anggi, P.}, \bibinfo{year}{2006}.
\newblock \bibinfo{title}{L\'evy-brownian motion on finite intervals: Mean
  first passage time analysis}.
\newblock \bibinfo{journal}{Phys. Rev. E} \bibinfo{volume}{73},
  \bibinfo{pages}{046104}.
\newblock \URLprefix \url{http://link.aps.org/doi/10.1103/PhysRevE.73.046104},
  \DOIprefix\doi{10.1103/PhysRevE.73.046104}.
%Type = Article
\bibitem[{Dybiec et~al.(2007)Dybiec, Gudowska-Nowak and H\"anggi}]{Dyb07}
\bibinfo{author}{Dybiec, B.}, \bibinfo{author}{Gudowska-Nowak, E.},
  \bibinfo{author}{H\"anggi, P.}, \bibinfo{year}{2007}.
\newblock \bibinfo{title}{Escape driven by $\ensuremath{\alpha}$-stable white
  noises}.
\newblock \bibinfo{journal}{Phys. Rev. E} \bibinfo{volume}{75},
  \bibinfo{pages}{021109}.
\newblock \URLprefix \url{http://link.aps.org/doi/10.1103/PhysRevE.75.021109},
  \DOIprefix\doi{10.1103/PhysRevE.75.021109}.
%Type = Article
\bibitem[{Elyassami et~al.(2016)Elyassami, Benjelloun and El~Aroussi}]{Ely16}
\bibinfo{author}{Elyassami, Y.}, \bibinfo{author}{Benjelloun, K.},
  \bibinfo{author}{El~Aroussi, M.}, \bibinfo{year}{2016}.
\newblock \bibinfo{title}{Bearing fault diagnosis and classification based on
  KDA and alpha-stable fusion}.
\newblock \bibinfo{journal}{Contemp. Eng. Sci.} \bibinfo{volume}{9},
  \bibinfo{pages}{453--465}.
  \newblock \URLprefix \url{http://www.m-hikari.com/ces/ces2016/ces9-12-2016/512328.html},
  \DOIprefix\doi{http://dx.doi.org/10.12988/ces.2016.512328}.
%Type = Article
\bibitem[{Fedorov and Pankratov(2007)}]{Fed07}
\bibinfo{author}{Fedorov, K.G.}, \bibinfo{author}{Pankratov, A.L.},
  \bibinfo{year}{2007}.
\newblock \bibinfo{title}{Mean time of the thermal escape in a current-biased
  long-overlap Josephson junction}.
\newblock \bibinfo{journal}{Phys. Rev. B} \bibinfo{volume}{76},
  \bibinfo{pages}{024504}.
\newblock \URLprefix \url{https://link.aps.org/doi/10.1103/PhysRevB.76.024504},
  \DOIprefix\doi{10.1103/PhysRevB.76.024504}.
%Type = Article
\bibitem[{Fedorov and Pankratov(2009)}]{Fed09}
\bibinfo{author}{Fedorov, K.G.}, \bibinfo{author}{Pankratov, A.L.},
  \bibinfo{year}{2009}.
\newblock \bibinfo{title}{Crossover of the thermal escape problem in annular
  spatially distributed systems}.
\newblock \bibinfo{journal}{Phys. Rev. Lett.} \bibinfo{volume}{103},
  \bibinfo{pages}{260601}.
\newblock \URLprefix
  \url{https://link.aps.org/doi/10.1103/PhysRevLett.103.260601},
  \DOIprefix\doi{10.1103/PhysRevLett.103.260601}.
%Type = Article
\bibitem[{Fedorov et~al.(2008)Fedorov, Pankratov and Spagnolo}]{Fed08}
\bibinfo{author}{Fedorov, K.G.}, \bibinfo{author}{Pankratov, A.L.},
  \bibinfo{author}{Spagnolo, B.}, \bibinfo{year}{2008}.
\newblock \bibinfo{title}{Influence of length on the noise delaied switching of
  long Josephson junctions}.
\newblock \bibinfo{journal}{International Journal of Bifurcation and Chaos}
  \bibinfo{volume}{18}, \bibinfo{pages}{2857--2862}.
\newblock \URLprefix \url{https://doi.org/10.1142/S0218127408022111},
  \DOIprefix\doi{10.1142/S0218127408022111}.
%Type = Book
\bibitem[{Feller(1971)}]{Fel71}
\bibinfo{author}{Feller, W.}, \bibinfo{year}{1971}.
\newblock \bibinfo{title}{An introduction to probability theory and its
  applications}.
\newblock \bibinfo{publisher}{John Wiley \& Sons, New York}.
\newblock \bibinfo{note}{Vol.2}.
%Type = Article
\bibitem[{Fiasconaro et~al.(2010)Fiasconaro, Mazo and Spagnolo}]{Fia10}
\bibinfo{author}{Fiasconaro, A.}, \bibinfo{author}{Mazo, J.J.},
  \bibinfo{author}{Spagnolo, B.}, \bibinfo{year}{2010}.
\newblock \bibinfo{title}{Noise-induced enhancement of stability in a
  metastable system with damping}.
\newblock \bibinfo{journal}{Phys. Rev. E} \bibinfo{volume}{82},
  \bibinfo{pages}{041120}.
\newblock \URLprefix \url{http://link.aps.org/doi/10.1103/PhysRevE.82.041120},
  \DOIprefix\doi{10.1103/PhysRevE.82.041120}.
%Type = Article
\bibitem[{Fiasconaro and Spagnolo(2009)}]{Fia09}
\bibinfo{author}{Fiasconaro, A.}, \bibinfo{author}{Spagnolo, B.},
  \bibinfo{year}{2009}.
\newblock \bibinfo{title}{Stability measures in metastable states with Gaussian
  colored noise}.
\newblock \bibinfo{journal}{Phys. Rev. E} \bibinfo{volume}{80},
  \bibinfo{pages}{041110}.
\newblock \URLprefix \url{http://link.aps.org/doi/10.1103/PhysRevE.80.041110},
  \DOIprefix\doi{10.1103/PhysRevE.80.041110}.
%Type = Article
\bibitem[{Fiasconaro and Spagnolo(2011)}]{Fia11}
\bibinfo{author}{Fiasconaro, A.}, \bibinfo{author}{Spagnolo, B.},
  \bibinfo{year}{2011}.
\newblock \bibinfo{title}{Resonant activation in piecewise linear asymmetric
  potentials}.
\newblock \bibinfo{journal}{Phys. Rev. E} \bibinfo{volume}{83},
  \bibinfo{pages}{041122}.
\newblock \URLprefix \url{http://link.aps.org/doi/10.1103/PhysRevE.83.041122},
  \DOIprefix\doi{10.1103/PhysRevE.83.041122}.
%Type = Article
\bibitem[{Fiasconaro et~al.(2005)Fiasconaro, Spagnolo and Boccaletti}]{Fia05}
\bibinfo{author}{Fiasconaro, A.}, \bibinfo{author}{Spagnolo, B.},
  \bibinfo{author}{Boccaletti, S.}, \bibinfo{year}{2005}.
\newblock \bibinfo{title}{Signatures of noise-enhanced stability in metastable
  states}.
\newblock \bibinfo{journal}{Phys. Rev. E} \bibinfo{volume}{72},
  \bibinfo{pages}{061110}.
\newblock \URLprefix \url{http://link.aps.org/doi/10.1103/PhysRevE.72.061110},
  \DOIprefix\doi{10.1103/PhysRevE.72.061110}.
%Type = Article
\bibitem[{Filatov et~al.(2019)Filatov, Vrzheshch, Tabakov, Novikov, Belov,
  Antonov, Sharkov, Koryazhkina, Mikhaylov, Gorshkov, Dubkov, Carollo and
  Spagnolo}]{Fil19}
\bibinfo{author}{Filatov, D.O.}, \bibinfo{author}{Vrzheshch, D.V.},
  \bibinfo{author}{Tabakov, O.V.}, \bibinfo{author}{Novikov, A.S.},
  \bibinfo{author}{Belov, A.I.}, \bibinfo{author}{Antonov, I.N.},
  \bibinfo{author}{Sharkov, V.V.}, \bibinfo{author}{Koryazhkina, M.N.},
  \bibinfo{author}{Mikhaylov, A.N.}, \bibinfo{author}{Gorshkov, O.N.},
  \bibinfo{author}{Dubkov, A.A.}, \bibinfo{author}{Carollo, A.},
  \bibinfo{author}{Spagnolo, B.}, \bibinfo{year}{2019}.
\newblock \bibinfo{title}{Noise-induced resistive switching in a memristor
  based on ZrO$_2$(Y)/Ta$_2$O$_5$ stack}.
\newblock \bibinfo{journal}{Journal of Statistical Mechanics: Theory and
  Experiment} \bibinfo{volume}{2019}, \bibinfo{pages}{124026}.
\newblock \URLprefix \url{https://doi.org/10.1088/1742-5468/ab5704},
  \DOIprefix\doi{10.1088/1742-5468/ab5704}.
%Type = Article
\bibitem[{Filatrella and Pierro(2010)}]{Fil10}
\bibinfo{author}{Filatrella, G.}, \bibinfo{author}{Pierro, V.},
  \bibinfo{year}{2010}.
\newblock \bibinfo{title}{Detection of noise-corrupted sinusoidal signals with
  Josephson junctions}.
\newblock \bibinfo{journal}{Phys. Rev. E} \bibinfo{volume}{82},
  \bibinfo{pages}{046712}.
\newblock \URLprefix \url{https://link.aps.org/doi/10.1103/PhysRevE.82.046712},
  \DOIprefix\doi{10.1103/PhysRevE.82.046712}.
%Type = Article
\bibitem[{Gattenl\"ohner et~al.(2016)Gattenl\"ohner, Gornyi, Ostrovsky,
  Trauzettel, Mirlin and Titov}]{Gat16}
\bibinfo{author}{Gattenl\"ohner, S.}, \bibinfo{author}{Gornyi, I.V.},
  \bibinfo{author}{Ostrovsky, P.M.}, \bibinfo{author}{Trauzettel, B.},
  \bibinfo{author}{Mirlin, A.D.}, \bibinfo{author}{Titov, M.},
  \bibinfo{year}{2016}.
\newblock \bibinfo{title}{L\'evy flights due to anisotropic disorder in
  graphene}.
\newblock \bibinfo{journal}{Phys. Rev. Lett.} \bibinfo{volume}{117},
  \bibinfo{pages}{046603}.
\newblock \URLprefix
  \url{http://link.aps.org/doi/10.1103/PhysRevLett.117.046603},
  \DOIprefix\doi{10.1103/PhysRevLett.117.046603}.
%Type = Article
\bibitem[{Giuffrida et~al.(2009)Giuffrida, Valenti, Ziino, Spagnolo and
  Panebianco}]{Giu09}
\bibinfo{author}{Giuffrida, A.}, \bibinfo{author}{Valenti, D.},
  \bibinfo{author}{Ziino, G.}, \bibinfo{author}{Spagnolo, B.},
  \bibinfo{author}{Panebianco, A.}, \bibinfo{year}{2009}.
\newblock \bibinfo{title}{A stochastic interspecific competition model to
  predict the behaviour of \emph{Listeria monocytogenes} in the fermentation process
  of a traditional Sicilian salami}.
\newblock \bibinfo{journal}{European Food Research and Technology}
  \bibinfo{volume}{228}, \bibinfo{pages}{767--775}.
\newblock \URLprefix \url{https://doi.org/10.1007/s00217-008-0988-6},
  \DOIprefix\doi{10.1007/s00217-008-0988-6}.
%Type = Book
\bibitem[{Gnedenko and Kolmogorov(1954)}]{Gne54}
\bibinfo{author}{Gnedenko, B.V.}, \bibinfo{author}{Kolmogorov, A.N.},
  \bibinfo{year}{1954}.
\newblock \bibinfo{title}{Limit Distributions for Sums of Independent Random
  Variables}.
\newblock \bibinfo{publisher}{Addison-Wesley, Cambridge, MA}.
%Type = Incollection
\bibitem[{Goldobin(2001)}]{Gold01}
\bibinfo{author}{Goldobin, E.}, \bibinfo{year}{2001}.
\newblock \bibinfo{title}{Flux-flow oscillators and phenomenon of Cherenkov
  radiation from fast moving fluxons}, in: \bibinfo{booktitle}{Microwave
  Superconductivity}. \bibinfo{publisher}{Springer}, pp.
  \bibinfo{pages}{581--614}.
%Type = Article
\bibitem[{Golubev et~al.(2021)Golubev, Il'ichev and Kuzmin}]{Gol21}
\bibinfo{author}{Golubev, D.S.}, \bibinfo{author}{Il'ichev, E.V.},
  \bibinfo{author}{Kuzmin, L.S.}, \bibinfo{year}{2021}.
\newblock \bibinfo{title}{Single-photon detection with a Josephson junction
  coupled to a resonator}.
\newblock \bibinfo{journal}{Phys. Rev. Applied} \bibinfo{volume}{16},
  \bibinfo{pages}{014025}.
\newblock \URLprefix
  \url{https://link.aps.org/doi/10.1103/PhysRevApplied.16.014025},
  \DOIprefix\doi{10.1103/PhysRevApplied.16.014025}.
%Type = Article
\bibitem[{Gordeeva and Pankratov(2006)}]{Gor06}
\bibinfo{author}{Gordeeva, A.V.}, \bibinfo{author}{Pankratov, A.L.},
  \bibinfo{year}{2006}.
\newblock \bibinfo{title}{Minimization of timing errors in reproduction of
  single flux quantum pulses}.
\newblock \bibinfo{journal}{Applied Physics Letters} \bibinfo{volume}{88},
  \bibinfo{pages}{022505}.
\newblock \URLprefix \url{https://doi.org/10.1063/1.2164389},
  \DOIprefix\doi{10.1063/1.2164389}.
%Type = Article
\bibitem[{Gordeeva et~al.(2008)Gordeeva, Pankratov and Spagnolo}]{Gor08}
\bibinfo{author}{Gordeeva, A.V.}, \bibinfo{author}{Pankratov, A.L.},
  \bibinfo{author}{Spagnolo, B.}, \bibinfo{year}{2008}.
\newblock \bibinfo{title}{Noise induced phenomena in point Josephson
  junctions}.
\newblock \bibinfo{journal}{International Journal of Bifurcation and Chaos}
  \bibinfo{volume}{18}, \bibinfo{pages}{2825--2831}.
\newblock \URLprefix \url{https://doi.org/10.1142/S0218127408022068},
  \DOIprefix\doi{10.1142/S0218127408022068}.
%Type = Article
\bibitem[{Grabert(2008)}]{Gra08}
\bibinfo{author}{Grabert, H.}, \bibinfo{year}{2008}.
\newblock \bibinfo{title}{Theory of a Josephson junction detector of
  non-Gaussian noise}.
\newblock \bibinfo{journal}{Phys. Rev. B} \bibinfo{volume}{77},
  \bibinfo{pages}{205315}.
\newblock \URLprefix \url{http://link.aps.org/doi/10.1103/PhysRevB.77.205315},
  \DOIprefix\doi{10.1103/PhysRevB.77.205315}.
%Type = Article
\bibitem[{Grabert and Weiss(1984)}]{Gra84}
\bibinfo{author}{Grabert, H.}, \bibinfo{author}{Weiss, U.},
  \bibinfo{year}{1984}.
\newblock \bibinfo{title}{Crossover from thermal hopping to quantum tunneling}.
\newblock \bibinfo{journal}{Phys. Rev. Lett.} \bibinfo{volume}{53},
  \bibinfo{pages}{1787--1790}.
\newblock \URLprefix \url{http://link.aps.org/doi/10.1103/PhysRevLett.53.1787},
  \DOIprefix\doi{10.1103/PhysRevLett.53.1787}.
%Type = Article
\bibitem[{Guarcello and Bergeret(2020)}]{GuaBer20}
\bibinfo{author}{Guarcello, C.}, \bibinfo{author}{Bergeret, F.},
  \bibinfo{year}{2020}.
\newblock \bibinfo{title}{Cryogenic memory element based on an anomalous
  Josephson junction}.
\newblock \bibinfo{journal}{Phys. Rev. Applied} \bibinfo{volume}{13},
  \bibinfo{pages}{034012}.
\newblock \URLprefix
  \url{https://link.aps.org/doi/10.1103/PhysRevApplied.13.034012},
  \DOIprefix\doi{10.1103/PhysRevApplied.13.034012}.
%Type = Article
\bibitem[{Guarcello and Bergeret(2021)}]{GuaBer21}
\bibinfo{author}{Guarcello, C.}, \bibinfo{author}{Bergeret, F.},
  \bibinfo{year}{2021}.
\newblock \bibinfo{title}{Thermal noise effects on the magnetization switching
  of a ferromagnetic anomalous Josephson junction}.
\newblock \bibinfo{journal}{Chaos Solitons Fract} \bibinfo{volume}{142},
  \bibinfo{pages}{110384}.
\newblock \URLprefix
  \url{https://www.sciencedirect.com/science/article/pii/S0960077920307785},
  \DOIprefix\doi{https://doi.org/10.1016/j.chaos.2020.110384}.
%Type = Article
\bibitem[{Guarcello et~al.(2019a)Guarcello, Braggio, Solinas and
  Giazotto}]{GuaBra19}
\bibinfo{author}{Guarcello, C.}, \bibinfo{author}{Braggio, A.},
  \bibinfo{author}{Solinas, P.}, \bibinfo{author}{Giazotto, F.},
  \bibinfo{year}{2019}a.
\newblock \bibinfo{title}{Nonlinear critical-current thermal response of an
  asymmetric Josephson tunnel junction}.
\newblock \bibinfo{journal}{Phys. Rev. Applied} \bibinfo{volume}{11},
  \bibinfo{pages}{024002}.
\newblock \URLprefix
  \url{https://link.aps.org/doi/10.1103/PhysRevApplied.11.024002},
  \DOIprefix\doi{10.1103/PhysRevApplied.11.024002}.
%Type = Article
\bibitem[{Guarcello et~al.(2019b)Guarcello, Braggio, Solinas, Pepe and
  Giazotto}]{GuaBraSol19}
\bibinfo{author}{Guarcello, C.}, \bibinfo{author}{Braggio, A.},
  \bibinfo{author}{Solinas, P.}, \bibinfo{author}{Pepe, G.P.},
  \bibinfo{author}{Giazotto, F.}, \bibinfo{year}{2019}b.
\newblock \bibinfo{title}{Josephson-threshold calorimeter}.
\newblock \bibinfo{journal}{Phys. Rev. Applied} \bibinfo{volume}{11},
  \bibinfo{pages}{054074}.
\newblock \URLprefix
  \url{https://link.aps.org/doi/10.1103/PhysRevApplied.11.054074},
  \DOIprefix\doi{10.1103/PhysRevApplied.11.054074}.
%Type = Article
\bibitem[{Guarcello et~al.(2020)Guarcello, Filatrella, Spagnolo, Pierro and
  Valenti}]{Gua20}
\bibinfo{author}{Guarcello, C.}, \bibinfo{author}{Filatrella, G.},
  \bibinfo{author}{Spagnolo, B.}, \bibinfo{author}{Pierro, V.},
  \bibinfo{author}{Valenti, D.}, \bibinfo{year}{2020}.
\newblock \bibinfo{title}{Voltage drop across Josephson junctions for L\'evy
  noise detection}.
\newblock \bibinfo{journal}{Phys. Rev. Research} \bibinfo{volume}{2},
  \bibinfo{pages}{043332}.
\newblock \URLprefix
  \url{https://link.aps.org/doi/10.1103/PhysRevResearch.2.043332},
  \DOIprefix\doi{10.1103/PhysRevResearch.2.043332}.
%Type = Article
\bibitem[{Guarcello et~al.(2016a)Guarcello, Giazotto and Solinas}]{GuaGia16}
\bibinfo{author}{Guarcello, C.}, \bibinfo{author}{Giazotto, F.},
  \bibinfo{author}{Solinas, P.}, \bibinfo{year}{2016}a.
\newblock \bibinfo{title}{Coherent diffraction of thermal currents in long
  Josephson tunnel junctions}.
\newblock \bibinfo{journal}{Phys. Rev. B} \bibinfo{volume}{94},
  \bibinfo{pages}{054522}.
\newblock \URLprefix \url{https://link.aps.org/doi/10.1103/PhysRevB.94.054522},
  \DOIprefix\doi{10.1103/PhysRevB.94.054522}.
%Type = Misc
\bibitem[{Guarcello et~al.(2021)Guarcello, Komnang, Barone, Rettaroli, Gatti,
  Pagano and Filatrella}]{Gua21}
\bibinfo{author}{Guarcello, C.}, \bibinfo{author}{Komnang, A.S.P.},
  \bibinfo{author}{Barone, C.}, \bibinfo{author}{Rettaroli, A.},
  \bibinfo{author}{Gatti, C.}, \bibinfo{author}{Pagano, S.},
  \bibinfo{author}{Filatrella, G.}, \bibinfo{year}{2021}.
\newblock \bibinfo{title}{Josephson-based scheme for the detection of microwave
  photons}.
\newblock \bibinfo{journal}{Phys. Rev. Applied} \bibinfo{volume}{16},
  \bibinfo{pages}{054015}.
\newblock \URLprefix \url{https://journals.aps.org/prapplied/abstract/10.1103/PhysRevApplied.16.054015},
  \DOIprefix\doi{10.1103/PhysRevApplied.16.054015}.
%Type = Article
\bibitem[{Guarcello et~al.(2018a)Guarcello, Solinas, Braggio and
  Giazotto}]{GuaSolBra2018}
\bibinfo{author}{Guarcello, C.}, \bibinfo{author}{Solinas, P.},
  \bibinfo{author}{Braggio, A.}, \bibinfo{author}{Giazotto, F.},
  \bibinfo{year}{2018}a.
\newblock \bibinfo{title}{Phase-coherent solitonic Josephson heat oscillator}.
\newblock \bibinfo{journal}{Scientific Reports} \bibinfo{volume}{8},
  \bibinfo{pages}{12287}.
\newblock \URLprefix \url{https://doi.org/10.1038/s41598-018-30268-1},
  \DOIprefix\doi{10.1038/s41598-018-30268-1}.
%Type = Article
\bibitem[{Guarcello et~al.(2018b)Guarcello, Solinas, Braggio and
  Giazotto}]{GuaSol18}
\bibinfo{author}{Guarcello, C.}, \bibinfo{author}{Solinas, P.},
  \bibinfo{author}{Braggio, A.}, \bibinfo{author}{Giazotto, F.},
  \bibinfo{year}{2018}b.
\newblock \bibinfo{title}{Solitonic Josephson thermal transport}.
\newblock \bibinfo{journal}{Phys. Rev. Applied} \bibinfo{volume}{9},
  \bibinfo{pages}{034014}.
\newblock \URLprefix
  \url{https://link.aps.org/doi/10.1103/PhysRevApplied.9.034014},
  \DOIprefix\doi{10.1103/PhysRevApplied.9.034014}.
%Type = Article
\bibitem[{Guarcello et~al.(2018c)Guarcello, Solinas, Braggio and
  Giazotto}]{Gua18}
\bibinfo{author}{Guarcello, C.}, \bibinfo{author}{Solinas, P.},
  \bibinfo{author}{Braggio, A.}, \bibinfo{author}{Giazotto, F.},
  \bibinfo{year}{2018}c.
\newblock \bibinfo{title}{Solitonic thermal transport in a current-biased long
  Josephson junction}.
\newblock \bibinfo{journal}{Phys. Rev. B} \bibinfo{volume}{98},
  \bibinfo{pages}{104501}.
\newblock \URLprefix \url{https://link.aps.org/doi/10.1103/PhysRevB.98.104501},
  \DOIprefix\doi{10.1103/PhysRevB.98.104501}.
%Type = Article
\bibitem[{Guarcello et~al.(2017a)Guarcello, Solinas, Di~Ventra and
  Giazotto}]{GuaSol17}
\bibinfo{author}{Guarcello, C.}, \bibinfo{author}{Solinas, P.},
  \bibinfo{author}{Di~Ventra, M.}, \bibinfo{author}{Giazotto, F.},
  \bibinfo{year}{2017}a.
\newblock \bibinfo{title}{Solitonic Josephson-based meminductive systems}.
\newblock \bibinfo{journal}{Sci. Rep.} \bibinfo{volume}{7},
  \bibinfo{pages}{46736}.
%Type = Article
\bibitem[{Guarcello et~al.(2019c)Guarcello, Solinas, Giazotto and
  Braggio}]{GuaSol19}
\bibinfo{author}{Guarcello, C.}, \bibinfo{author}{Solinas, P.},
  \bibinfo{author}{Giazotto, F.}, \bibinfo{author}{Braggio, A.},
  \bibinfo{year}{2019}c.
\newblock \bibinfo{title}{Thermal flux-flow regime in long Josephson tunnel
  junctions}.
\newblock \bibinfo{journal}{Journal of Statistical Mechanics: Theory and
  Experiment} \bibinfo{volume}{2019}, \bibinfo{pages}{084006}.
\newblock \URLprefix \url{https://doi.org/10.1088/1742-5468/ab3194},
  \DOIprefix\doi{10.1088/1742-5468/ab3194}.
%Type = Article
\bibitem[{Guarcello et~al.(2013)Guarcello, Valenti, Augello and
  Spagnolo}]{Gua13}
\bibinfo{author}{Guarcello, C.}, \bibinfo{author}{Valenti, D.},
  \bibinfo{author}{Augello, G.}, \bibinfo{author}{Spagnolo, B.},
  \bibinfo{year}{2013}.
\newblock \bibinfo{title}{The role of non-Gaussian sources in the transient
  dynamics of long Josephson junctions}.
\newblock \bibinfo{journal}{Acta Phys. Pol. B} \bibinfo{volume}{44},
  \bibinfo{pages}{997--1005}.
  \newblock \URLprefix \url{https://www.actaphys.uj.edu.pl/R/44/5/997}.
%Type = Article
\bibitem[{Guarcello et~al.(2015a)Guarcello, Valenti, Carollo and
  Spagnolo}]{GuaVal15}
\bibinfo{author}{Guarcello, C.}, \bibinfo{author}{Valenti, D.},
  \bibinfo{author}{Carollo, A.}, \bibinfo{author}{Spagnolo, B.},
  \bibinfo{year}{2015}a.
\newblock \bibinfo{title}{Stabilization effects of dichotomous noise on the
  lifetime of the superconducting state in a long Josephson junction}.
\newblock \bibinfo{journal}{Entropy} \bibinfo{volume}{17},
  \bibinfo{pages}{2862}.
\newblock \URLprefix \url{http://www.mdpi.com/1099-4300/17/5/2862},
  \DOIprefix\doi{10.3390/e17052862}.
%Type = Article
\bibitem[{Guarcello et~al.(2016b)Guarcello, Valenti, Carollo and
  Spagnolo}]{Gua16}
\bibinfo{author}{Guarcello, C.}, \bibinfo{author}{Valenti, D.},
  \bibinfo{author}{Carollo, A.}, \bibinfo{author}{Spagnolo, B.},
  \bibinfo{year}{2016}b.
\newblock \bibinfo{title}{Effects of L\'evy noise on the dynamics of
  sine-Gordon solitons in long Josephson junctions}.
\newblock \bibinfo{journal}{J. Stat. Mech.: Theory Exp.}
  \bibinfo{volume}{2016}, \bibinfo{pages}{054012}.
\newblock \URLprefix \url{http://stacks.iop.org/1742-5468/2016/i=5/a=054012},
  \DOIprefix\doi{10.1088/1742-5468/2016/05/054012}.
%Type = Article
\bibitem[{Guarcello et~al.(2015b)Guarcello, Valenti and Spagnolo}]{Gua15}
\bibinfo{author}{Guarcello, C.}, \bibinfo{author}{Valenti, D.},
  \bibinfo{author}{Spagnolo, B.}, \bibinfo{year}{2015}b.
\newblock \bibinfo{title}{Phase dynamics in graphene-based Josephson junctions
  in the presence of thermal and correlated fluctuations}.
\newblock \bibinfo{journal}{Phys. Rev. B} \bibinfo{volume}{92},
  \bibinfo{pages}{174519}.
\newblock \URLprefix \url{https://link.aps.org/doi/10.1103/PhysRevB.92.174519},
  \DOIprefix\doi{10.1103/PhysRevB.92.174519}.
%Type = Article
\bibitem[{Guarcello et~al.(2017b)Guarcello, Valenti, Spagnolo, Pierro and
  Filatrella}]{Gua17}
\bibinfo{author}{Guarcello, C.}, \bibinfo{author}{Valenti, D.},
  \bibinfo{author}{Spagnolo, B.}, \bibinfo{author}{Pierro, V.},
  \bibinfo{author}{Filatrella, G.}, \bibinfo{year}{2017}b.
\newblock \bibinfo{title}{Anomalous transport effects on switching currents of
  graphene-based Josephson junctions}.
\newblock \bibinfo{journal}{Nanotechnology} \bibinfo{volume}{28},
  \bibinfo{pages}{134001}.
%Type = Article
\bibitem[{Guarcello et~al.(2019d)Guarcello, Valenti, Spagnolo, Pierro and
  Filatrella}]{Gua19}
\bibinfo{author}{Guarcello, C.}, \bibinfo{author}{Valenti, D.},
  \bibinfo{author}{Spagnolo, B.}, \bibinfo{author}{Pierro, V.},
  \bibinfo{author}{Filatrella, G.}, \bibinfo{year}{2019}d.
\newblock \bibinfo{title}{Josephson-based threshold detector for
  L\'evy-distributed current fluctuations}.
\newblock \bibinfo{journal}{Phys. Rev. Applied} \bibinfo{volume}{11},
  \bibinfo{pages}{044078}.
\newblock \URLprefix
  \url{https://link.aps.org/doi/10.1103/PhysRevApplied.11.044078},
  \DOIprefix\doi{10.1103/PhysRevApplied.11.044078}.
%Type = Article
\bibitem[{Gwinn(2007)}]{Gwi07}
\bibinfo{author}{Gwinn, C.R.}, \bibinfo{year}{2007}.
\newblock \bibinfo{title}{Observations and L\'evy statistics in interstellar
  scattering}.
\newblock \bibinfo{journal}{Astron. Astrophys. Trans.} \bibinfo{volume}{26},
  \bibinfo{pages}{525--533}.
\newblock \URLprefix \url{https://doi.org/10.1080/10556790701610506},
  \DOIprefix\doi{10.1080/10556790701610506}.
%Type = Article
\bibitem[{H\"anggi et~al.(1990)H\"anggi, Talkner and Borkovec}]{Han90}
\bibinfo{author}{H\"anggi, P.}, \bibinfo{author}{Talkner, P.},
  \bibinfo{author}{Borkovec, M.}, \bibinfo{year}{1990}.
\newblock \bibinfo{title}{Reaction-rate theory: fifty years after Kramers}.
\newblock \bibinfo{journal}{Rev. Mod. Phys.} \bibinfo{volume}{62},
  \bibinfo{pages}{251--341}.
\newblock \URLprefix \url{http://link.aps.org/doi/10.1103/RevModPhys.62.251},
  \DOIprefix\doi{10.1103/RevModPhys.62.251}.
%Type = Article
\bibitem[{Hasegawa and Arita(2011)}]{Has11}
\bibinfo{author}{Hasegawa, Y.}, \bibinfo{author}{Arita, M.},
  \bibinfo{year}{2011}.
\newblock \bibinfo{title}{Escape process and stochastic resonance under noise
  intensity fluctuation}.
\newblock \bibinfo{journal}{Phys. Lett. A} \bibinfo{volume}{375},
  \bibinfo{pages}{3450 -- 3458}.
\newblock \URLprefix
  \url{http://www.sciencedirect.com/science/article/pii/S0375960111009406},
  \DOIprefix\doi{http://dx.doi.org/10.1016/j.physleta.2011.07.054}.
%Type = Article
\bibitem[{Huard et~al.(2007)Huard, Pothier, Birge, Esteve, Waintal and
  Ankerhold}]{Hua07}
\bibinfo{author}{Huard, B.}, \bibinfo{author}{Pothier, H.},
  \bibinfo{author}{Birge, N.}, \bibinfo{author}{Esteve, D.},
  \bibinfo{author}{Waintal, X.}, \bibinfo{author}{Ankerhold, J.},
  \bibinfo{year}{2007}.
\newblock \bibinfo{title}{Josephson junctions as detectors for non-Gaussian
  noise}.
\newblock \bibinfo{journal}{Ann. Phys.} \bibinfo{volume}{16},
  \bibinfo{pages}{736--750}.
\newblock \URLprefix \url{http://dx.doi.org/10.1002/andp.200710263},
  \DOIprefix\doi{10.1002/andp.200710263}.
%Type = Article
\bibitem[{Hurtado et~al.(2006)Hurtado, Marro and Garrido}]{Hur06}
\bibinfo{author}{Hurtado, P.I.}, \bibinfo{author}{Marro, J.},
  \bibinfo{author}{Garrido, P.L.}, \bibinfo{year}{2006}.
\newblock \bibinfo{title}{Metastability, nucleation, and noise-enhanced
  stabilization out of equilibrium}.
\newblock \bibinfo{journal}{Phys. Rev. E} \bibinfo{volume}{74},
  \bibinfo{pages}{050101}.
\newblock \URLprefix \url{http://link.aps.org/doi/10.1103/PhysRevE.74.050101},
  \DOIprefix\doi{10.1103/PhysRevE.74.050101}.
%Type = Article
\bibitem[{Josephson(1962)}]{Jos62}
\bibinfo{author}{Josephson, B.D.}, \bibinfo{year}{1962}.
\newblock \bibinfo{title}{Possible new effects in superconductive tunnelling}.
\newblock \bibinfo{journal}{Physics Letters} \bibinfo{volume}{1},
  \bibinfo{pages}{251 -- 253}.
\newblock \URLprefix
  \url{http://www.sciencedirect.com/science/article/pii/0031916362913690},
  \DOIprefix\doi{http://dx.doi.org/10.1016/0031-9163(62)91369-0}.
%Type = Article
\bibitem[{Josephson(1974)}]{Jos74}
\bibinfo{author}{Josephson, B.D.}, \bibinfo{year}{1974}.
\newblock \bibinfo{title}{The discovery of tunnelling supercurrents}.
\newblock \bibinfo{journal}{Rev. Mod. Phys.} \bibinfo{volume}{46},
  \bibinfo{pages}{251--254}.
\newblock \URLprefix \url{http://link.aps.org/doi/10.1103/RevModPhys.46.251},
  \DOIprefix\doi{10.1103/RevModPhys.46.251}.
%Type = Article
\bibitem[{van Kan et~al.(2021)van Kan, Alexakis and Brachet}]{Kan21}
\bibinfo{author}{van Kan, A.}, \bibinfo{author}{Alexakis, A.},
  \bibinfo{author}{Brachet, M.E.}, \bibinfo{year}{2021}.
\newblock \bibinfo{title}{L\'evy on-off intermittency}.
\newblock \bibinfo{journal}{Phys. Rev. E} \bibinfo{volume}{103},
  \bibinfo{pages}{052115}.
\newblock \URLprefix
  \url{https://link.aps.org/doi/10.1103/PhysRevE.103.052115},
  \DOIprefix\doi{10.1103/PhysRevE.103.052115}.
%Type = Article
\bibitem[{Karaku{\c{s}} et~al.(2020)Karaku{\c{s}}, Kuruo{\u{g}}lu and
  Alt{\i}nkaya}]{Kar20}
\bibinfo{author}{Karaku{\c{s}}, O.}, \bibinfo{author}{Kuruo{\u{g}}lu, E.E.},
  \bibinfo{author}{Alt{\i}nkaya, M.A.}, \bibinfo{year}{2020}.
\newblock \bibinfo{title}{Modelling impulsive noise in indoor powerline
  communication systems}.
\newblock \bibinfo{journal}{Signal, Image and Video Processing}
  \bibinfo{volume}{14}, \bibinfo{pages}{1655--1661}.
\newblock \URLprefix \url{https://doi.org/10.1007/s11760-020-01708-1},
  \DOIprefix\doi{10.1007/s11760-020-01708-1}.
%Type = Article
\bibitem[{Khintchine and L\'evy(1936)}]{Khi36}
\bibinfo{author}{Khintchine, A.}, \bibinfo{author}{L\'evy, P.},
  \bibinfo{year}{1936}.
\newblock \bibinfo{title}{Sur les lois stables}.
\newblock \bibinfo{journal}{C. R. Acad. Sci. Paris} \bibinfo{volume}{202},
  \bibinfo{pages}{374}.
%Type = Book
\bibitem[{Khintchine(1938)}]{Khi38}
\bibinfo{author}{Khintchine, A.Y.}, \bibinfo{year}{1938}.
\newblock \bibinfo{title}{Limit Distributions for the Sum of Independent Random
  Variables}.
\newblock \bibinfo{publisher}{ONTI, Moscow}.
%Type = Article
\bibitem[{Kiselev and Schmalian(2019)}]{Kis19}
\bibinfo{author}{Kiselev, E.I.}, \bibinfo{author}{Schmalian, J.},
  \bibinfo{year}{2019}.
\newblock \bibinfo{title}{L\'evy flights and hydrodynamic superdiffusion on the
  Dirac cone of graphene}.
\newblock \bibinfo{journal}{Phys. Rev. Lett.} \bibinfo{volume}{123},
  \bibinfo{pages}{195302}.
\newblock \URLprefix
  \url{https://link.aps.org/doi/10.1103/PhysRevLett.123.195302},
  \DOIprefix\doi{10.1103/PhysRevLett.123.195302}.
%Type = Book
\bibitem[{Kogan(1996)}]{Kogan96}
\bibinfo{author}{Kogan, S.}, \bibinfo{year}{1996}.
\newblock \bibinfo{title}{Electronic Noise and Fluctuations in Solids}.
\newblock \bibinfo{publisher}{Cambridge University Press, Cambridge, England}.
%Type = Article
\bibitem[{Kramers(1940)}]{Kra40}
\bibinfo{author}{Kramers, H.}, \bibinfo{year}{1940}.
\newblock \bibinfo{title}{Brownian motion in a field of force and the diffusion
  model of chemical reactions}.
\newblock \bibinfo{journal}{Physica} \bibinfo{volume}{7}, \bibinfo{pages}{284
  -- 304}.
\newblock \URLprefix
  \url{http://www.sciencedirect.com/science/article/pii/S0031891440900982},
  \DOIprefix\doi{http://dx.doi.org/10.1016/S0031-8914(40)90098-2}.
%Type = Article
\bibitem[{Kuno et~al.(2000)Kuno, Fromm, Hamann, Gallagher and Nesbitt}]{Kun00}
\bibinfo{author}{Kuno, M.}, \bibinfo{author}{Fromm, D.P.},
  \bibinfo{author}{Hamann, H.F.}, \bibinfo{author}{Gallagher, A.},
  \bibinfo{author}{Nesbitt, D.J.}, \bibinfo{year}{2000}.
\newblock \bibinfo{title}{Nonexponential ``blinking'' kinetics of single CdSe
  quantum dots: A universal power law behavior}.
\newblock \bibinfo{journal}{J. Chem. Phys.} \bibinfo{volume}{112},
  \bibinfo{pages}{3117--3120}.
\newblock \URLprefix \url{https://doi.org/10.1063/1.480896},
  \DOIprefix\doi{10.1063/1.480896}.
%Type = Article
\bibitem[{Kuno et~al.(2001)Kuno, Fromm, Hamann, Gallagher and Nesbitt}]{Kun01}
\bibinfo{author}{Kuno, M.}, \bibinfo{author}{Fromm, D.P.},
  \bibinfo{author}{Hamann, H.F.}, \bibinfo{author}{Gallagher, A.},
  \bibinfo{author}{Nesbitt, D.J.}, \bibinfo{year}{2001}.
\newblock \bibinfo{title}{"on"/"off" fluorescence intermittency of single
  semiconductor quantum dots}.
\newblock \bibinfo{journal}{J. Chem. Phys.} \bibinfo{volume}{115},
  \bibinfo{pages}{1028--1040}.
\newblock \URLprefix \url{https://doi.org/10.1063/1.1377883},
  \DOIprefix\doi{10.1063/1.1377883}.
%Type = Article
\bibitem[{La~Cognata et~al.(2010)La~Cognata, Valenti, Dubkov and
  Spagnolo}]{LaC10}
\bibinfo{author}{La~Cognata, A.}, \bibinfo{author}{Valenti, D.},
  \bibinfo{author}{Dubkov, A.A.}, \bibinfo{author}{Spagnolo, B.},
  \bibinfo{year}{2010}.
\newblock \bibinfo{title}{Dynamics of two competing species in the presence of
  L\'evy noise sources}.
\newblock \bibinfo{journal}{Phys. Rev. E} \bibinfo{volume}{82},
  \bibinfo{pages}{011121}.
\newblock \URLprefix \url{https://link.aps.org/doi/10.1103/PhysRevE.82.011121},
  \DOIprefix\doi{10.1103/PhysRevE.82.011121}.
%Type = Article
\bibitem[{Le~Masne et~al.(2009)Le~Masne, Pothier, Birge, Urbina and
  Esteve}]{LeM09}
\bibinfo{author}{Le~Masne, Q.}, \bibinfo{author}{Pothier, H.},
  \bibinfo{author}{Birge, N.O.}, \bibinfo{author}{Urbina, C.},
  \bibinfo{author}{Esteve, D.}, \bibinfo{year}{2009}.
\newblock \bibinfo{title}{Asymmetric noise probed with a Josephson junction}.
\newblock \bibinfo{journal}{Phys. Rev. Lett.} \bibinfo{volume}{102},
  \bibinfo{pages}{067002}.
\newblock \URLprefix
  \url{http://link.aps.org/doi/10.1103/PhysRevLett.102.067002},
  \DOIprefix\doi{10.1103/PhysRevLett.102.067002}.
%Type = Article
\bibitem[{Lee(1971)}]{Lee71}
\bibinfo{author}{Lee, P.A.}, \bibinfo{year}{1971}.
\newblock \bibinfo{title}{Effect of noise on the current‐voltage
  characteristics of a Josephson junction}.
\newblock \bibinfo{journal}{J. Appl. Phys.} \bibinfo{volume}{42},
  \bibinfo{pages}{325--334}.
\newblock \URLprefix \url{https://doi.org/10.1063/1.1659596},
  \DOIprefix\doi{10.1063/1.1659596}.
%Type = Inproceedings
\bibitem[{Li and Yu(2010)}]{LiYu10}
\bibinfo{author}{Li, C.}, \bibinfo{author}{Yu, G.}, \bibinfo{year}{2010}.
\newblock \bibinfo{title}{A new statistical model for rolling element bearing
  fault signals based on alpha-stable distribution}, in:
  \bibinfo{booktitle}{2010 Second International Conference on Computer Modeling
  and Simulation}, pp. \bibinfo{pages}{386--390}.
\newblock \DOIprefix\doi{10.1109/ICCMS.2010.309}.
%Type = Article
\bibitem[{Li and \L{}uczka(2010)}]{Li10}
\bibinfo{author}{Li, J.H.}, \bibinfo{author}{\L{}uczka, J.},
  \bibinfo{year}{2010}.
\newblock \bibinfo{title}{Thermal-inertial ratchet effects: Negative mobility,
  resonant activation, noise-enhanced stability, and noise-weakened stability}.
\newblock \bibinfo{journal}{Phys. Rev. E} \bibinfo{volume}{82},
  \bibinfo{pages}{041104}.
\newblock \URLprefix \url{http://link.aps.org/doi/10.1103/PhysRevE.82.041104},
  \DOIprefix\doi{10.1103/PhysRevE.82.041104}.
%Type = Book
\bibitem[{Likharev(1986)}]{Lik86}
\bibinfo{author}{Likharev, K.}, \bibinfo{year}{1986}.
\newblock \bibinfo{title}{Dynamics of Josephson Junctions and Circuits}.
\newblock \bibinfo{publisher}{Gordon and Breach, New York}.
%Type = Inbook
\bibitem[{Luryi and Subashiev(2010)}]{Lur10}
\bibinfo{author}{Luryi, S.}, \bibinfo{author}{Subashiev, A.},
  \bibinfo{year}{2010}.
\newblock \bibinfo{title}{Semiconductor Scintillator for Three-Dimensional
  Array of Radiation Detectors}. \bibinfo{publisher}{John Wiley \& Sons, Inc.}
\newblock pp. \bibinfo{pages}{331--346}.
\newblock \URLprefix \url{http://dx.doi.org/10.1002/9780470649343.ch28},
  \DOIprefix\doi{10.1002/9780470649343.ch28}.
%Type = Article
\bibitem[{Mankin et~al.(2008)Mankin, Soika, Sauga and Ainsaar}]{Man08}
\bibinfo{author}{Mankin, R.}, \bibinfo{author}{Soika, E.},
  \bibinfo{author}{Sauga, A.}, \bibinfo{author}{Ainsaar, A.},
  \bibinfo{year}{2008}.
\newblock \bibinfo{title}{Thermally enhanced stability in fluctuating bistable
  potentials}.
\newblock \bibinfo{journal}{Phys. Rev. E} \bibinfo{volume}{77},
  \bibinfo{pages}{051113}.
\newblock \URLprefix \url{http://link.aps.org/doi/10.1103/PhysRevE.77.051113},
  \DOIprefix\doi{10.1103/PhysRevE.77.051113}.
%Type = Article
\bibitem[{Mantegna and Spagnolo(1996)}]{Man96}
\bibinfo{author}{Mantegna, R.}, \bibinfo{author}{Spagnolo, B.},
  \bibinfo{year}{1996}.
\newblock \bibinfo{title}{Noise enhanced stability in an unstable system}.
\newblock \bibinfo{journal}{Phys. Rev. Lett.} \bibinfo{volume}{76},
  \bibinfo{pages}{563--566}.
\newblock \URLprefix \url{http://link.aps.org/doi/10.1103/PhysRevLett.76.563},
  \DOIprefix\doi{10.1103/PhysRevLett.76.563}.
%Type = Article
\bibitem[{Mantegna and Spagnolo(2000)}]{Man00}
\bibinfo{author}{Mantegna, R.N.}, \bibinfo{author}{Spagnolo, B.},
  \bibinfo{year}{2000}.
\newblock \bibinfo{title}{Experimental investigation of resonant activation}.
\newblock \bibinfo{journal}{Phys. Rev. Lett.} \bibinfo{volume}{84},
  \bibinfo{pages}{3025--3028}.
\newblock \URLprefix \url{http://link.aps.org/doi/10.1103/PhysRevLett.84.3025},
  \DOIprefix\doi{10.1103/PhysRevLett.84.3025}.
%Type = Article
\bibitem[{Marchi et~al.(1996)Marchi, Marchesoni, Gammaitoni, Menichella-Saetta
  and Santucci}]{Mar96}
\bibinfo{author}{Marchi, M.}, \bibinfo{author}{Marchesoni, F.},
  \bibinfo{author}{Gammaitoni, L.}, \bibinfo{author}{Menichella-Saetta, E.},
  \bibinfo{author}{Santucci, S.}, \bibinfo{year}{1996}.
\newblock \bibinfo{title}{Resonant activation in a bistable system}.
\newblock \bibinfo{journal}{Phys. Rev. E} \bibinfo{volume}{54},
  \bibinfo{pages}{3479--3487}.
\newblock \URLprefix \url{http://link.aps.org/doi/10.1103/PhysRevE.54.3479},
  \DOIprefix\doi{10.1103/PhysRevE.54.3479}.
%Type = Article
\bibitem[{Martinis et~al.(1987)Martinis, Devoret and Clarke}]{Mar87}
\bibinfo{author}{Martinis, J.M.}, \bibinfo{author}{Devoret, M.H.},
  \bibinfo{author}{Clarke, J.}, \bibinfo{year}{1987}.
\newblock \bibinfo{title}{Experimental tests for the quantum behavior of a
  macroscopic degree of freedom: The phase difference across a Josephson
  junction}.
\newblock \bibinfo{journal}{Phys. Rev. B} \bibinfo{volume}{35},
  \bibinfo{pages}{4682--4698}.
  \newblock \URLprefix \url{https://journals.aps.org/prb/abstract/10.1103/PhysRevB.35.4682},
  \DOIprefix\doi{10.1103/PhysRevB.35.4682}.
%Type = Article
\bibitem[{McFadden and Smith(1984)}]{McF84}
\bibinfo{author}{McFadden, P.}, \bibinfo{author}{Smith, J.},
  \bibinfo{year}{1984}.
\newblock \bibinfo{title}{Model for the vibration produced by a single point
  defect in a rolling element bearing}.
\newblock \bibinfo{journal}{Journal of Sound and Vibration}
  \bibinfo{volume}{96}, \bibinfo{pages}{69 -- 82}.
\newblock \URLprefix
  \url{http://www.sciencedirect.com/science/article/pii/0022460X84905959},
  \DOIprefix\doi{https://doi.org/10.1016/0022-460X(84)90595-9}.
%Type = Article
\bibitem[{McLaughlin and Scott(1978)}]{McL82}
\bibinfo{author}{McLaughlin, D.W.}, \bibinfo{author}{Scott, A.C.},
  \bibinfo{year}{1978}.
\newblock \bibinfo{title}{Perturbation analysis of fluxon dynamics}.
\newblock \bibinfo{journal}{Phys. Rev. A} \bibinfo{volume}{18},
  \bibinfo{pages}{1652--1680}.
\newblock \URLprefix \url{http://link.aps.org/doi/10.1103/PhysRevA.18.1652},
  \DOIprefix\doi{10.1103/PhysRevA.18.1652}.
%Type = Article
\bibitem[{Mel'Nikov(1991)}]{Mel91}
\bibinfo{author}{Mel'Nikov, V.I.}, \bibinfo{year}{1991}.
\newblock \bibinfo{journal}{Phys. Rep.} \bibinfo{volume}{209},
  \bibinfo{pages}{1}.
  \newblock \URLprefix \url{https://www.sciencedirect.com/science/article/abs/pii/037015739190108X},
  \DOIprefix\doi{10.1016/0370-1573(91)90108-X}.
%Type = Article
\bibitem[{Messin et~al.(2001)Messin, Hermier, Giacobino, Desbiolles and
  Dahan}]{Mes01}
\bibinfo{author}{Messin, G.}, \bibinfo{author}{Hermier, J.P.},
  \bibinfo{author}{Giacobino, E.}, \bibinfo{author}{Desbiolles, P.},
  \bibinfo{author}{Dahan, M.}, \bibinfo{year}{2001}.
\newblock \bibinfo{title}{Bunching and antibunching in the fluorescence of
  semiconductor nanocrystals}.
\newblock \bibinfo{journal}{Opt. Lett.} \bibinfo{volume}{26},
  \bibinfo{pages}{1891--1893}.
\newblock \URLprefix \url{http://ol.osa.org/abstract.cfm?URI=ol-26-23-1891},
  \DOIprefix\doi{10.1364/OL.26.001891}.
%Type = Article
\bibitem[{Mikhaylov et~al.(2020)Mikhaylov, Pimashkin, Pigareva, Gerasimova,
  Gryaznov, Shchanikov, Zuev, Talanov, Lavrov, Demin, Erokhin, Lobov, Mukhina,
  Kazantsev, Wu and Spagnolo}]{Mik20}
\bibinfo{author}{Mikhaylov, A.}, \bibinfo{author}{Pimashkin, A.},
  \bibinfo{author}{Pigareva, Y.}, \bibinfo{author}{Gerasimova, S.},
  \bibinfo{author}{Gryaznov, E.}, \bibinfo{author}{Shchanikov, S.},
  \bibinfo{author}{Zuev, A.}, \bibinfo{author}{Talanov, M.},
  \bibinfo{author}{Lavrov, I.}, \bibinfo{author}{Demin, V.},
  \bibinfo{author}{Erokhin, V.}, \bibinfo{author}{Lobov, S.},
  \bibinfo{author}{Mukhina, I.}, \bibinfo{author}{Kazantsev, V.},
  \bibinfo{author}{Wu, H.}, \bibinfo{author}{Spagnolo, B.},
  \bibinfo{year}{2020}.
\newblock \bibinfo{title}{Neurohybrid memristive CMOS-integrated systems for
  biosensors and neuroprosthetics}.
\newblock \bibinfo{journal}{Frontiers in Neuroscience} \bibinfo{volume}{14},
  \bibinfo{pages}{358}.
\newblock \URLprefix
  \url{https://www.frontiersin.org/article/10.3389/fnins.2020.00358},
  \DOIprefix\doi{10.3389/fnins.2020.00358}.
%Type = Article
\bibitem[{Mikhaylov et~al.(2016)Mikhaylov, Gryaznov, Belov, Korolev, Sharapov,
  Guseinov, Tetelbaum, Tikhov, Malekhonova, Bobrov, Pavlov, Gerasimova,
  Kazantsev, Agudov, Dubkov, Rosário, Sobolev and Spagnolo}]{Mik16}
\bibinfo{author}{Mikhaylov, A.N.}, \bibinfo{author}{Gryaznov, E.G.},
  \bibinfo{author}{Belov, A.I.}, \bibinfo{author}{Korolev, D.S.},
  \bibinfo{author}{Sharapov, A.N.}, \bibinfo{author}{Guseinov, D.V.},
  \bibinfo{author}{Tetelbaum, D.I.}, \bibinfo{author}{Tikhov, S.V.},
  \bibinfo{author}{Malekhonova, N.V.}, \bibinfo{author}{Bobrov, A.I.},
  \bibinfo{author}{Pavlov, D.A.}, \bibinfo{author}{Gerasimova, S.A.},
  \bibinfo{author}{Kazantsev, V.B.}, \bibinfo{author}{Agudov, N.V.},
  \bibinfo{author}{Dubkov, A.A.}, \bibinfo{author}{Rosário, C.M.M.},
  \bibinfo{author}{Sobolev, N.A.}, \bibinfo{author}{Spagnolo, B.},
  \bibinfo{year}{2016}.
\newblock \bibinfo{title}{Field- and irradiation-induced phenomena in
  memristive nanomaterials}.
\newblock \bibinfo{journal}{physica status solidi c} \bibinfo{volume}{13},
  \bibinfo{pages}{870--881}.
\newblock \URLprefix
  \url{https://onlinelibrary.wiley.com/doi/abs/10.1002/pssc.201600083},
  \DOIprefix\doi{https://doi.org/10.1002/pssc.201600083}.
%Type = Article
\bibitem[{Miyamoto et~al.(2010)Miyamoto, Nishiguchi, Ono, Itoh and
  Fujiwara}]{Miy10}
\bibinfo{author}{Miyamoto, S.}, \bibinfo{author}{Nishiguchi, K.},
  \bibinfo{author}{Ono, Y.}, \bibinfo{author}{Itoh, K.M.},
  \bibinfo{author}{Fujiwara, A.}, \bibinfo{year}{2010}.
\newblock \bibinfo{title}{Resonant escape over an oscillating barrier in a
  single-electron ratchet transfer}.
\newblock \bibinfo{journal}{Phys. Rev. B} \bibinfo{volume}{82},
  \bibinfo{pages}{033303}.
\newblock \URLprefix \url{http://link.aps.org/doi/10.1103/PhysRevB.82.033303},
  \DOIprefix\doi{10.1103/PhysRevB.82.033303}.
%Type = Article
\bibitem[{Mohammed et~al.(2015)Mohammed, Koh, Vermeersch, Lu, Burke, Gossard
  and Shakouri}]{Moh15}
\bibinfo{author}{Mohammed, A.M.S.}, \bibinfo{author}{Koh, Y.R.},
  \bibinfo{author}{Vermeersch, B.}, \bibinfo{author}{Lu, H.},
  \bibinfo{author}{Burke, P.G.}, \bibinfo{author}{Gossard, A.C.},
  \bibinfo{author}{Shakouri, A.}, \bibinfo{year}{2015}.
\newblock \bibinfo{title}{Fractal L\'evy heat transport in nanoparticle
  embedded semiconductor alloys}.
\newblock \bibinfo{journal}{Nano Letters} \bibinfo{volume}{15},
  \bibinfo{pages}{4269--4273}.
\newblock \URLprefix \url{https://doi.org/10.1021/nl504466511},
  \DOIprefix\doi{10.1021/nl5044665}.
%Type = Article
\bibitem[{Novikov et~al.(2005)Novikov, Drndic, Levitov, Kastner, Jarosz and
  Bawendi}]{Nov05}
\bibinfo{author}{Novikov, D.S.}, \bibinfo{author}{Drndic, M.},
  \bibinfo{author}{Levitov, L.S.}, \bibinfo{author}{Kastner, M.A.},
  \bibinfo{author}{Jarosz, M.V.}, \bibinfo{author}{Bawendi, M.G.},
  \bibinfo{year}{2005}.
\newblock \bibinfo{title}{L\'evy statistics and anomalous transport in
  quantum-dot arrays}.
\newblock \bibinfo{journal}{Phys. Rev. B} \bibinfo{volume}{72},
  \bibinfo{pages}{075309}.
\newblock \URLprefix \url{http://link.aps.org/doi/10.1103/PhysRevB.72.075309},
  \DOIprefix\doi{10.1103/PhysRevB.72.075309}.
%Type = Article
\bibitem[{Novotn{\'{y}}(2009)}]{Nov09}
\bibinfo{author}{Novotn{\'{y}}, T.}, \bibinfo{year}{2009}.
\newblock \bibinfo{title}{Josephson junctions as threshold detectors of full
  counting statistics: open issues}.
\newblock \bibinfo{journal}{J. Stat. Mech.: Theory Exp.}
  \bibinfo{volume}{2009}, \bibinfo{pages}{P01050}.
\newblock \URLprefix
  \url{https://doi.org/10.1088%2F1742-5468%2F2009%2F01%2Fp01050},
  \DOIprefix\doi{10.1088/1742-5468/2009/01/p01050}.
%Type = Article
\bibitem[{Oelsner et~al.(2017)Oelsner, Andersen, Reh\'ak, Schmelz, Anders,
  Grajcar, H\"ubner, M\o{}lmer and Il'ichev}]{Oel17}
\bibinfo{author}{Oelsner, G.}, \bibinfo{author}{Andersen, C.K.},
  \bibinfo{author}{Reh\'ak, M.}, \bibinfo{author}{Schmelz, M.},
  \bibinfo{author}{Anders, S.}, \bibinfo{author}{Grajcar, M.},
  \bibinfo{author}{H\"ubner, U.}, \bibinfo{author}{M\o{}lmer, K.},
  \bibinfo{author}{Il'ichev, E.}, \bibinfo{year}{2017}.
\newblock \bibinfo{title}{Detection of weak microwave fields with an
  underdamped Josephson junction}.
\newblock \bibinfo{journal}{Phys. Rev. Applied} \bibinfo{volume}{7},
  \bibinfo{pages}{014012}.
\newblock \URLprefix
  \url{https://link.aps.org/doi/10.1103/PhysRevApplied.7.014012},
  \DOIprefix\doi{10.1103/PhysRevApplied.7.014012}.
%Type = Article
\bibitem[{Oelsner et~al.(2013)Oelsner, Revin, Il'ichev, Pankratov, Meyer,
  Gr\"onberg, Hassel and Kuzmin}]{Oel13}
\bibinfo{author}{Oelsner, G.}, \bibinfo{author}{Revin, L.S.},
  \bibinfo{author}{Il'ichev, E.}, \bibinfo{author}{Pankratov, A.L.},
  \bibinfo{author}{Meyer, H.G.}, \bibinfo{author}{Gr\"onberg, L.},
  \bibinfo{author}{Hassel, J.}, \bibinfo{author}{Kuzmin, L.S.},
  \bibinfo{year}{2013}.
\newblock \bibinfo{title}{Underdamped Josephson junction as a switching current
  detector}.
\newblock \bibinfo{journal}{Applied Physics Letters} \bibinfo{volume}{103},
  \bibinfo{pages}{142605}.
\newblock \URLprefix \url{https://doi.org/10.1063/1.4824308},
  \DOIprefix\doi{10.1063/1.4824308}.
%Type = Article
\bibitem[{Ortlepp and Uhlmann(2004)}]{Ort04}
\bibinfo{author}{Ortlepp, T.}, \bibinfo{author}{Uhlmann, H.F.},
  \bibinfo{year}{2004}.
\newblock \bibinfo{title}{Noise analysis for intrinsic and external shunted
  Josephson junctions}.
\newblock \bibinfo{journal}{Superconductor Science and Technology}
  \bibinfo{volume}{17}, \bibinfo{pages}{S112--S116}.
\newblock \URLprefix \url{https://doi.org/10.1088/0953-2048/17/5/004},
  \DOIprefix\doi{10.1088/0953-2048/17/5/004}.
%Type = Article
\bibitem[{Pan et~al.(2009)Pan, Tan, Yu, Sun, Kang, Xu, Chen and Wu}]{Pan09}
\bibinfo{author}{Pan, C.}, \bibinfo{author}{Tan, X.}, \bibinfo{author}{Yu, Y.},
  \bibinfo{author}{Sun, G.}, \bibinfo{author}{Kang, L.}, \bibinfo{author}{Xu,
  W.}, \bibinfo{author}{Chen, J.}, \bibinfo{author}{Wu, P.},
  \bibinfo{year}{2009}.
\newblock \bibinfo{title}{Resonant activation through effective temperature
  oscillation in a josephson tunnel junction}.
\newblock \bibinfo{journal}{Phys. Rev. E} \bibinfo{volume}{79},
  \bibinfo{pages}{030104}.
\newblock \URLprefix \url{https://link.aps.org/doi/10.1103/PhysRevE.79.030104},
  \DOIprefix\doi{10.1103/PhysRevE.79.030104}.
%Type = Article
\bibitem[{Pankratov et~al.(2012)Pankratov, Gordeeva and Kuzmin}]{Pan12}
\bibinfo{author}{Pankratov, A.L.}, \bibinfo{author}{Gordeeva, A.V.},
  \bibinfo{author}{Kuzmin, L.S.}, \bibinfo{year}{2012}.
\newblock \bibinfo{title}{Drastic suppression of noise-induced errors in
  underdamped long Josephson junctions}.
\newblock \bibinfo{journal}{Phys. Rev. Lett.} \bibinfo{volume}{109},
  \bibinfo{pages}{087003}.
\newblock \URLprefix
  \url{https://link.aps.org/doi/10.1103/PhysRevLett.109.087003},
  \DOIprefix\doi{10.1103/PhysRevLett.109.087003}.
%Type = Article
\bibitem[{Pankratov et~al.(2017)Pankratov, Pankratova, Shamporov and
  Shitov}]{Pan17}
\bibinfo{author}{Pankratov, A.L.}, \bibinfo{author}{Pankratova, E.V.},
  \bibinfo{author}{Shamporov, V.A.}, \bibinfo{author}{Shitov, S.V.},
  \bibinfo{year}{2017}.
\newblock \bibinfo{title}{Oscillations in Josephson transmission line
  stimulated by load in the presence of noise}.
\newblock \bibinfo{journal}{Applied Physics Letters} \bibinfo{volume}{110},
  \bibinfo{pages}{112601}.
\newblock \URLprefix \url{https://doi.org/10.1063/1.4978514},
  \DOIprefix\doi{10.1063/1.4978514}.
%Type = Article
\bibitem[{Pankratov and Spagnolo(2005)}]{Pan05}
\bibinfo{author}{Pankratov, E.L.}, \bibinfo{author}{Spagnolo, B.},
  \bibinfo{year}{2005}.
\newblock \bibinfo{title}{Optimization of impurity profile for p-n-junctionin
  heterostructures}.
\newblock \bibinfo{journal}{The European Physical Journal B - Condensed Matter
  and Complex Systems} \bibinfo{volume}{46}, \bibinfo{pages}{15--19}.
\newblock \URLprefix \url{https://doi.org/10.1140/epjb/e2005-00233-1},
  \DOIprefix\doi{10.1140/epjb/e2005-00233-1}.
%Type = Article
\bibitem[{Peacock et~al.(1998)Peacock, Verhoeve, Rando, Erd, Bavdaz, Taylor and
  Perez}]{Pea98}
\bibinfo{author}{Peacock, T.}, \bibinfo{author}{Verhoeve, P.},
  \bibinfo{author}{Rando, N.}, \bibinfo{author}{Erd, C.},
  \bibinfo{author}{Bavdaz, M.}, \bibinfo{author}{Taylor, B.G.},
  \bibinfo{author}{Perez, D.}, \bibinfo{year}{1998}.
\newblock \bibinfo{title}{Recent developments in superconducting tunnel
  junctions for ultraviolet, optical \& near infrared astronomy}.
\newblock \bibinfo{journal}{Astron. Astrophys. Suppl. Ser.}
  \bibinfo{volume}{127}, \bibinfo{pages}{497--504}.
\newblock \URLprefix \url{https://doi.org/10.1051/aas:1998116},
  \DOIprefix\doi{10.1051/aas:1998116}.
%Type = Article
\bibitem[{Pechukas and H\"anggi(1994)}]{Pec94}
\bibinfo{author}{Pechukas, P.}, \bibinfo{author}{H\"anggi, P.},
  \bibinfo{year}{1994}.
\newblock \bibinfo{title}{Rates of activated processes with fluctuating
  barriers}.
\newblock \bibinfo{journal}{Phys. Rev. Lett.} \bibinfo{volume}{73},
  \bibinfo{pages}{2772--2775}.
\newblock \URLprefix \url{http://link.aps.org/doi/10.1103/PhysRevLett.73.2772},
  \DOIprefix\doi{10.1103/PhysRevLett.73.2772}.
%Type = Article
\bibitem[{Pekola(2004)}]{Pek04}
\bibinfo{author}{Pekola, J.P.}, \bibinfo{year}{2004}.
\newblock \bibinfo{title}{Josephson junction as a detector of Poissonian charge
  injection}.
\newblock \bibinfo{journal}{Phys. Rev. Lett.} \bibinfo{volume}{93},
  \bibinfo{pages}{206601}.
\newblock \URLprefix
  \url{http://link.aps.org/doi/10.1103/PhysRevLett.93.206601},
  \DOIprefix\doi{10.1103/PhysRevLett.93.206601}.
%Type = Article
\bibitem[{Pekola et~al.(2005)Pekola, Nieminen, Meschke, Kivioja, Niskanen and
  Vartiainen}]{Pek05}
\bibinfo{author}{Pekola, J.P.}, \bibinfo{author}{Nieminen, T.E.},
  \bibinfo{author}{Meschke, M.}, \bibinfo{author}{Kivioja, J.M.},
  \bibinfo{author}{Niskanen, A.O.}, \bibinfo{author}{Vartiainen, J.J.},
  \bibinfo{year}{2005}.
\newblock \bibinfo{title}{Shot-noise-driven escape in hysteretic Josephson
  junctions}.
\newblock \bibinfo{journal}{Phys. Rev. Lett.} \bibinfo{volume}{95},
  \bibinfo{pages}{197004}.
\newblock \URLprefix
  \url{http://link.aps.org/doi/10.1103/PhysRevLett.95.197004},
  \DOIprefix\doi{10.1103/PhysRevLett.95.197004}.
%Type = Article
\bibitem[{Peltonen et~al.(2007)Peltonen, Timofeev, Meschke, Heikkil\"a and
  Pekola}]{Pel07}
\bibinfo{author}{Peltonen, J.}, \bibinfo{author}{Timofeev, A.},
  \bibinfo{author}{Meschke, M.}, \bibinfo{author}{Heikkil\"a, T.},
  \bibinfo{author}{Pekola, J.}, \bibinfo{year}{2007}.
\newblock \bibinfo{title}{Detecting non-Gaussian current fluctuations using a
  Josephson threshold detector}.
\newblock \bibinfo{journal}{Physica E (Amsterdam)} \bibinfo{volume}{40},
  \bibinfo{pages}{111--122}.
\newblock \URLprefix
  \url{http://www.sciencedirect.com/science/article/pii/S1386947707001245},
  \DOIprefix\doi{http://dx.doi.org/10.1016/j.physe.2007.05.017}.
%Type = Article
\bibitem[{Peotta and Di~Ventra(2014)}]{Peo14}
\bibinfo{author}{Peotta, S.}, \bibinfo{author}{Di~Ventra, M.},
  \bibinfo{year}{2014}.
\newblock \bibinfo{title}{Superconducting memristors}.
\newblock \bibinfo{journal}{Phys. Rev. Applied} \bibinfo{volume}{2},
  \bibinfo{pages}{034011}.
\newblock \URLprefix
  \url{http://link.aps.org/doi/10.1103/PhysRevApplied.2.034011},
  \DOIprefix\doi{10.1103/PhysRevApplied.2.034011}.
%Type = Article
\bibitem[{Pershin and Di~Ventra(2011)}]{Per11}
\bibinfo{author}{Pershin, Y.V.}, \bibinfo{author}{Di~Ventra, M.},
  \bibinfo{year}{2011}.
\newblock \bibinfo{title}{Memory effects in complex materials and nanoscale
  systems}.
\newblock \bibinfo{journal}{Adv. Phys.} \bibinfo{volume}{60},
  \bibinfo{pages}{145--227}.
\newblock \URLprefix \url{http://dx.doi.org/10.1080/00018732.2010.544961},
  \DOIprefix\doi{10.1080/00018732.2010.544961}.
%Type = Article
\bibitem[{Pfeiffer et~al.(2016)Pfeiffer, Egusquiza, Di~Ventra, Sanz and
  Solano}]{Pfe16}
\bibinfo{author}{Pfeiffer, P.}, \bibinfo{author}{Egusquiza, I.L.},
  \bibinfo{author}{Di~Ventra, M.}, \bibinfo{author}{Sanz, M.},
  \bibinfo{author}{Solano, E.}, \bibinfo{year}{2016}.
\newblock \bibinfo{title}{Quantum memristors}.
\newblock \bibinfo{journal}{Scientific Reports} \bibinfo{volume}{6},
  \bibinfo{pages}{29507}.
\newblock \URLprefix \url{https://doi.org/10.1038/srep29507},
  \DOIprefix\doi{10.1038/srep29507}.
%Type = Article
\bibitem[{{Piedjou Komnang} et~al.(2021){Piedjou Komnang}, Guarcello, Barone,
  Gatti, Pagano, Pierro, Rettaroli and Filatrella}]{Pie21}
\bibinfo{author}{{Piedjou Komnang}, A.}, \bibinfo{author}{Guarcello, C.},
  \bibinfo{author}{Barone, C.}, \bibinfo{author}{Gatti, C.},
  \bibinfo{author}{Pagano, S.}, \bibinfo{author}{Pierro, V.},
  \bibinfo{author}{Rettaroli, A.}, \bibinfo{author}{Filatrella, G.},
  \bibinfo{year}{2021}.
\newblock \bibinfo{title}{Analysis of Josephson junctions switching time
  distributions for the detection of single microwave photons}.
\newblock \bibinfo{journal}{Chaos Solitons Fract} \bibinfo{volume}{142},
  \bibinfo{pages}{110496}.
\newblock \URLprefix
  \url{http://www.sciencedirect.com/science/article/pii/S0960077920308882},
  \DOIprefix\doi{https://doi.org/10.1016/j.chaos.2020.110496}.
%Type = Article
\bibitem[{Pizzolato et~al.(2010)Pizzolato, Fiasconaro, Adorno and
  Spagnolo}]{Piz10}
\bibinfo{author}{Pizzolato, N.}, \bibinfo{author}{Fiasconaro, A.},
  \bibinfo{author}{Adorno, D.P.}, \bibinfo{author}{Spagnolo, B.},
  \bibinfo{year}{2010}.
\newblock \bibinfo{title}{Resonant activation in polymer translocation: new
  insights into the escape dynamics of molecules driven by an oscillating
  field}.
\newblock \bibinfo{journal}{Physical Biology} \bibinfo{volume}{7},
  \bibinfo{pages}{034001}.
\newblock \URLprefix \url{https://doi.org/10.1088/1478-3975/7/3/034001},
  \DOIprefix\doi{10.1088/1478-3975/7/3/034001}.
%Type = Article
\bibitem[{Revin et~al.(2020)Revin, Pankratov, Gordeeva, Yablokov, Rakut,
  Zbrozhek and Kuzmin}]{Rev20}
\bibinfo{author}{Revin, L.S.}, \bibinfo{author}{Pankratov, A.L.},
  \bibinfo{author}{Gordeeva, A.V.}, \bibinfo{author}{Yablokov, A.A.},
  \bibinfo{author}{Rakut, I.V.}, \bibinfo{author}{Zbrozhek, V.O.},
  \bibinfo{author}{Kuzmin, L.S.}, \bibinfo{year}{2020}.
\newblock \bibinfo{title}{Microwave photon detection by an Al Josephson
  junction}.
\newblock \bibinfo{journal}{Beilstein Journal of Nanotechnology}
  \bibinfo{volume}{11}, \bibinfo{pages}{960--965}.
\newblock \URLprefix \url{https://www.beilstein-journals.org/bjnano/articles/11/80},
\DOIprefix\doi{10.3762/bjnano.11.80}.
%Type = Book
\bibitem[{Risken(1989)}]{Risken89}
\bibinfo{author}{Risken, H.}, \bibinfo{year}{1989}.
\newblock \bibinfo{title}{The Fokker-Planck Equation: Methods of solution and
  Applications}.
\newblock \bibinfo{publisher}{Springer, Berlin}.
%Type = Inproceedings
\bibitem[{Saadane et~al.(2015)Saadane, Aroussi and Wahbi}]{Saa15}
\bibinfo{author}{Saadane, R.}, \bibinfo{author}{Aroussi, M.E.},
  \bibinfo{author}{Wahbi, M.}, \bibinfo{year}{2015}.
\newblock \bibinfo{title}{Wind turbine fault diagnosis method based on $alpha$
  stable distribution and wiegthed support vector machines}, in:
  \bibinfo{booktitle}{2015 3rd International Renewable and Sustainable Energy
  Conference (IRSEC)}, pp. \bibinfo{pages}{1--5}.
\newblock \DOIprefix\doi{10.1109/IRSEC.2015.7455101}.
%Type = Article
\bibitem[{Saira et~al.(2016)Saira, Zgirski, Viisanen, Golubev and
  Pekola}]{Sai16}
\bibinfo{author}{Saira, O.P.}, \bibinfo{author}{Zgirski, M.},
  \bibinfo{author}{Viisanen, K.L.}, \bibinfo{author}{Golubev, D.S.},
  \bibinfo{author}{Pekola, J.P.}, \bibinfo{year}{2016}.
\newblock \bibinfo{title}{Dispersive thermometry with a Josephson junction
  coupled to a resonator}.
\newblock \bibinfo{journal}{Phys. Rev. Applied} \bibinfo{volume}{6},
  \bibinfo{pages}{024005}.
\newblock \URLprefix
  \url{https://link.aps.org/doi/10.1103/PhysRevApplied.6.024005},
  \DOIprefix\doi{10.1103/PhysRevApplied.6.024005}.
%Type = Article
\bibitem[{Salmilehto et~al.(2017)Salmilehto, Deppe, Di~Ventra, Sanz and
  Solano}]{Sal17}
\bibinfo{author}{Salmilehto, J.}, \bibinfo{author}{Deppe, F.},
  \bibinfo{author}{Di~Ventra, M.}, \bibinfo{author}{Sanz, M.},
  \bibinfo{author}{Solano, E.}, \bibinfo{year}{2017}.
\newblock \bibinfo{title}{Quantum memristors with superconducting circuits}.
\newblock \bibinfo{journal}{Scientific Reports} \bibinfo{volume}{7},
  \bibinfo{pages}{42044}.
\newblock \URLprefix \url{https://doi.org/10.1038/srep42044},
  \DOIprefix\doi{10.1038/srep42044}.
%Type = Book
\bibitem[{Sato(1999)}]{Sat99}
\bibinfo{author}{Sato, K.}, \bibinfo{year}{1999}.
\newblock \bibinfo{title}{L{\'e}vy processes and infinitely divisible
  distributions}.
\newblock \bibinfo{publisher}{Cambridge university press, Cambridge}.
%Type = Article
\bibitem[{Semyonov et~al.(2012)Semyonov, Subashiev, Chen and Luryi}]{Sem12}
\bibinfo{author}{Semyonov, O.}, \bibinfo{author}{Subashiev, A.V.},
  \bibinfo{author}{Chen, Z.}, \bibinfo{author}{Luryi, S.},
  \bibinfo{year}{2012}.
\newblock \bibinfo{title}{Photon assisted L\'evy flights of minority carriers
  in n-InP}.
\newblock \bibinfo{journal}{J. Lumin.} \bibinfo{volume}{132},
  \bibinfo{pages}{1935 -- 1943}.
\newblock \URLprefix
  \url{http://www.sciencedirect.com/science/article/pii/S0022231312001688},
  \DOIprefix\doi{http://dx.doi.org/10.1016/j.jlumin.2012.03.035}.
%Type = Article
\bibitem[{{Sheldon, Forrest} et~al.(2018){Sheldon, Forrest}, {Peotta,
  Sebastiano} and {Di Ventra, Massimiliano}}]{She18}
\bibinfo{author}{{Sheldon, Forrest}}, \bibinfo{author}{{Peotta, Sebastiano}},
  \bibinfo{author}{{Di Ventra, Massimiliano}}, \bibinfo{year}{2018}.
\newblock \bibinfo{title}{Phase-dependent noise in Josephson junctions}.
\newblock \bibinfo{journal}{Eur. Phys. J. Appl. Phys.} \bibinfo{volume}{81},
  \bibinfo{pages}{10601}.
\newblock \URLprefix \url{https://doi.org/10.1051/epjap/2017170297},
  \DOIprefix\doi{10.1051/epjap/2017170297}.
%Type = Article
\bibitem[{Shevchenko et~al.(2016)Shevchenko, Pershin and Nori}]{She16}
\bibinfo{author}{Shevchenko, S.N.}, \bibinfo{author}{Pershin, Y.V.},
  \bibinfo{author}{Nori, F.}, \bibinfo{year}{2016}.
\newblock \bibinfo{title}{Qubit-based memcapacitors and meminductors}.
\newblock \bibinfo{journal}{Phys. Rev. Applied} \bibinfo{volume}{6},
  \bibinfo{pages}{014006}.
\newblock \URLprefix
  \url{https://link.aps.org/doi/10.1103/PhysRevApplied.6.014006},
  \DOIprefix\doi{10.1103/PhysRevApplied.6.014006}.
%Type = Article
\bibitem[{Shimizu et~al.(2001)Shimizu, Neuhauser, Leatherdale, Empedocles, Woo
  and Bawendi}]{Shi01}
\bibinfo{author}{Shimizu, K.T.}, \bibinfo{author}{Neuhauser, R.G.},
  \bibinfo{author}{Leatherdale, C.A.}, \bibinfo{author}{Empedocles, S.A.},
  \bibinfo{author}{Woo, W.K.}, \bibinfo{author}{Bawendi, M.G.},
  \bibinfo{year}{2001}.
\newblock \bibinfo{title}{Blinking statistics in single semiconductor
  nanocrystal quantum dots}.
\newblock \bibinfo{journal}{Phys. Rev. B} \bibinfo{volume}{63},
  \bibinfo{pages}{205316}.
\newblock \URLprefix \url{https://link.aps.org/doi/10.1103/PhysRevB.63.205316},
  \DOIprefix\doi{10.1103/PhysRevB.63.205316}.
%Type = Inbook
\bibitem[{Shongwe et~al.(2015)Shongwe, Ferreira and Han~Vinck}]{Sho15}
\bibinfo{author}{Shongwe, T.}, \bibinfo{author}{Ferreira, H.C.},
  \bibinfo{author}{Han~Vinck, A.J.}, \bibinfo{year}{2015}.
\newblock \bibinfo{title}{Broadband and Narrow-Band Noise Modeling in Powerline
  Communications}. \bibinfo{publisher}{John Wiley \& Sons, Inc.}
\newblock \URLprefix \url{http://dx.doi.org/10.1002/047134608X.W8289},
  \DOIprefix\doi{10.1002/047134608X.W8289}.
%Type = Article
\bibitem[{Smirnov and Pankratov(2010)}]{Smi10}
\bibinfo{author}{Smirnov, A.A.}, \bibinfo{author}{Pankratov, A.L.},
  \bibinfo{year}{2010}.
\newblock \bibinfo{title}{Influence of the size of uniaxial magnetic
  nanoparticle on the reliability of high-speed switching}.
\newblock \bibinfo{journal}{Phys. Rev. B} \bibinfo{volume}{82},
  \bibinfo{pages}{132405}.
\newblock \URLprefix \url{http://link.aps.org/doi/10.1103/PhysRevB.82.132405},
  \DOIprefix\doi{10.1103/PhysRevB.82.132405}.
%Type = Article
\bibitem[{Soloviev et~al.(2013)Soloviev, Klenov, Pankratov, Il'ichev and
  Kuzmin}]{sol13}
\bibinfo{author}{Soloviev, I.I.}, \bibinfo{author}{Klenov, N.V.},
  \bibinfo{author}{Pankratov, A.L.}, \bibinfo{author}{Il'ichev, E.},
  \bibinfo{author}{Kuzmin, L.S.}, \bibinfo{year}{2013}.
\newblock \bibinfo{title}{Effect of Cherenkov radiation on the jitter of
  solitons in the driven underdamped Frenkel-Kontorova model}.
\newblock \bibinfo{journal}{Phys. Rev. E} \bibinfo{volume}{87},
  \bibinfo{pages}{060901}.
\newblock \URLprefix \url{https://link.aps.org/doi/10.1103/PhysRevE.87.060901},
  \DOIprefix\doi{10.1103/PhysRevE.87.060901}.
%Type = Article
\bibitem[{Soloviev et~al.(2015)Soloviev, Klenov, Pankratov, Revin, Il'ichev and
  Kuzmin}]{Sol15}
\bibinfo{author}{Soloviev, I.I.}, \bibinfo{author}{Klenov, N.V.},
  \bibinfo{author}{Pankratov, A.L.}, \bibinfo{author}{Revin, L.S.},
  \bibinfo{author}{Il'ichev, E.}, \bibinfo{author}{Kuzmin, L.S.},
  \bibinfo{year}{2015}.
\newblock \bibinfo{title}{Soliton scattering as a measurement tool for weak
  signals}.
\newblock \bibinfo{journal}{Phys. Rev. B} \bibinfo{volume}{92},
  \bibinfo{pages}{014516}.
\newblock \URLprefix \url{https://link.aps.org/doi/10.1103/PhysRevB.92.014516},
  \DOIprefix\doi{10.1103/PhysRevB.92.014516}.
%Type = Article
\bibitem[{Spagnolo et~al.(2004)Spagnolo, Agudov and Dubkov}]{Spa04}
\bibinfo{author}{Spagnolo, B.}, \bibinfo{author}{Agudov, N.V.},
  \bibinfo{author}{Dubkov, A.A.}, \bibinfo{year}{2004}.
\newblock \bibinfo{title}{Noise enhanced stability}.
\newblock \bibinfo{journal}{Acta Phys. Pol. B} \bibinfo{volume}{35},
  \bibinfo{pages}{1419--1436}.
  \newblock \URLprefix \url{https://www.actaphys.uj.edu.pl/R/35/4/1419}.
%Type = Article
\bibitem[{Spagnolo et~al.(2012)Spagnolo, Caldara, La~Cognata, Augello, Valenti,
  Fiasconaro, Dubkov and Falci}]{Spa12}
\bibinfo{author}{Spagnolo, B.}, \bibinfo{author}{Caldara, P.},
  \bibinfo{author}{La~Cognata, A.}, \bibinfo{author}{Augello, G.},
  \bibinfo{author}{Valenti, D.}, \bibinfo{author}{Fiasconaro, A.},
  \bibinfo{author}{Dubkov, A.A.}, \bibinfo{author}{Falci, G.},
  \bibinfo{year}{2012}.
\newblock \bibinfo{title}{Relaxation phenomena in classical and quantum
  systems}.
\newblock \bibinfo{journal}{Acta Phys. Pol. B} \bibinfo{volume}{43},
  \bibinfo{pages}{1169--1189}.
    \newblock \URLprefix \url{https://www.actaphys.uj.edu.pl/R/43/5/1169}.
%Type = Article
\bibitem[{Spagnolo et~al.(2007)Spagnolo, Dubkov, Pankratov, Pankratova,
  Fiasconaro and Ochab-Marcinek}]{Spa07}
\bibinfo{author}{Spagnolo, B.}, \bibinfo{author}{Dubkov, A.A.},
  \bibinfo{author}{Pankratov, A.L.}, \bibinfo{author}{Pankratova, E.V.},
  \bibinfo{author}{Fiasconaro, A.}, \bibinfo{author}{Ochab-Marcinek, A.},
  \bibinfo{year}{2007}.
\newblock \bibinfo{title}{Lifetime of metastable states and suppression of
  noise in interdisciplinary physical models}.
\newblock \bibinfo{journal}{Acta Phys. Pol. B} \bibinfo{volume}{38},
  \bibinfo{pages}{1925--1950}.
    \newblock \URLprefix \url{https://www.actaphys.uj.edu.pl/R/38/5/1925}.
%Type = Article
\bibitem[{Srokowski(2010)}]{Sro10}
\bibinfo{author}{Srokowski, T.}, \bibinfo{year}{2010}.
\newblock \bibinfo{title}{Nonlinear stochastic equations with multiplicative
  L\'evy noise}.
\newblock \bibinfo{journal}{Phys. Rev. E} \bibinfo{volume}{81},
  \bibinfo{pages}{051110}.
\newblock \URLprefix \url{https://link.aps.org/doi/10.1103/PhysRevE.81.051110},
  \DOIprefix\doi{10.1103/PhysRevE.81.051110}.
%Type = Article
\bibitem[{Subashiev et~al.(2014)Subashiev, Semyonov, Chen and Luryi}]{Sub14}
\bibinfo{author}{Subashiev, A.V.}, \bibinfo{author}{Semyonov, O.},
  \bibinfo{author}{Chen, Z.}, \bibinfo{author}{Luryi, S.},
  \bibinfo{year}{2014}.
\newblock \bibinfo{title}{Temperature controlled L\'evy flights of minority
  carriers in photoexcited bulk n-InP}.
\newblock \bibinfo{journal}{Phys. Lett. A} \bibinfo{volume}{378},
  \bibinfo{pages}{266 -- 269}.
\newblock \URLprefix
  \url{http://www.sciencedirect.com/science/article/pii/S0375960113010566},
  \DOIprefix\doi{http://dx.doi.org/10.1016/j.physleta.2013.11.007}.
%Type = Article
\bibitem[{Subramanian et~al.(2015)Subramanian, Sundaresan and Varshney}]{Sub15}
\bibinfo{author}{Subramanian, A.}, \bibinfo{author}{Sundaresan, A.},
  \bibinfo{author}{Varshney, P.K.}, \bibinfo{year}{2015}.
\newblock \bibinfo{title}{Detection of dependent heavy-tailed signals}.
\newblock \bibinfo{journal}{IEEE Transactions on Signal Processing}
  \bibinfo{volume}{63}, \bibinfo{pages}{2790--2803}.
  \newblock \URLprefix
  \url{https://ieeexplore.ieee.org/document/7054546},
  \DOIprefix\doi{10.1109/TSP.2015.2408570}.
%Type = Article
\bibitem[{Sukhorukov and Jordan(2007)}]{Suk07}
\bibinfo{author}{Sukhorukov, E.V.}, \bibinfo{author}{Jordan, A.N.},
  \bibinfo{year}{2007}.
\newblock \bibinfo{title}{Stochastic dynamics of a Josephson junction threshold
  detector}.
\newblock \bibinfo{journal}{Phys. Rev. Lett.} \bibinfo{volume}{98},
  \bibinfo{pages}{136803}.
\newblock \URLprefix
  \url{http://link.aps.org/doi/10.1103/PhysRevLett.98.136803},
  \DOIprefix\doi{10.1103/PhysRevLett.98.136803}.
%Type = Article
\bibitem[{Sun et~al.(2007)Sun, Dong, Mao, Chen, Xu, Ji, Kang, Wu, Yu and
  Xing}]{Sun07}
\bibinfo{author}{Sun, G.}, \bibinfo{author}{Dong, N.}, \bibinfo{author}{Mao,
  G.}, \bibinfo{author}{Chen, J.}, \bibinfo{author}{Xu, W.},
  \bibinfo{author}{Ji, Z.}, \bibinfo{author}{Kang, L.}, \bibinfo{author}{Wu,
  P.}, \bibinfo{author}{Yu, Y.}, \bibinfo{author}{Xing, D.},
  \bibinfo{year}{2007}.
\newblock \bibinfo{title}{Thermal escape from a metastable state in
  periodically driven Josephson junctions}.
\newblock \bibinfo{journal}{Phys. Rev. E} \bibinfo{volume}{75},
  \bibinfo{pages}{021107}.
\newblock \URLprefix \url{https://link.aps.org/doi/10.1103/PhysRevE.75.021107},
  \DOIprefix\doi{10.1103/PhysRevE.75.021107}.
%Type = Book
\bibitem[{Tafuri(2019)}]{Taf19}
\bibinfo{author}{Tafuri, F.}, \bibinfo{year}{2019}.
\newblock \bibinfo{title}{Fundamentals and Frontiers of the Josephson Effect}.
  volume \bibinfo{volume}{286}.
\newblock \bibinfo{publisher}{Springer International Publishing}.
\newblock \DOIprefix\doi{10.1007/978-3-030-20726-7}.
%Type = Article
\bibitem[{Timofeev et~al.(2007)Timofeev, Meschke, Peltonen, Heikkil\"a and
  Pekola}]{Tim07}
\bibinfo{author}{Timofeev, A.V.}, \bibinfo{author}{Meschke, M.},
  \bibinfo{author}{Peltonen, J.T.}, \bibinfo{author}{Heikkil\"a, T.T.},
  \bibinfo{author}{Pekola, J.P.}, \bibinfo{year}{2007}.
\newblock \bibinfo{title}{Wideband detection of the third moment of shot noise
  by a hysteretic Josephson junction}.
\newblock \bibinfo{journal}{Phys. Rev. Lett.} \bibinfo{volume}{98},
  \bibinfo{pages}{207001}.
\newblock \URLprefix
  \url{http://link.aps.org/doi/10.1103/PhysRevLett.98.207001},
  \DOIprefix\doi{10.1103/PhysRevLett.98.207001}.
%Type = Article
\bibitem[{Tobiska and Nazarov(2004)}]{Tob04}
\bibinfo{author}{Tobiska, J.}, \bibinfo{author}{Nazarov, Y.V.},
  \bibinfo{year}{2004}.
\newblock \bibinfo{title}{Josephson junctions as threshold detectors for full
  counting statistics}.
\newblock \bibinfo{journal}{Phys. Rev. Lett.} \bibinfo{volume}{93},
  \bibinfo{pages}{106801}.
\newblock \URLprefix
  \url{http://link.aps.org/doi/10.1103/PhysRevLett.93.106801},
  \DOIprefix\doi{10.1103/PhysRevLett.93.106801}.
%Type = Article
\bibitem[{Trapanese(2009)}]{Tra09}
\bibinfo{author}{Trapanese, M.}, \bibinfo{year}{2009}.
\newblock \bibinfo{title}{Noise enhanced stability in magnetic systems}.
\newblock \bibinfo{journal}{J. Appl. Phys.} \bibinfo{volume}{105},
  \bibinfo{pages}{--}.
\newblock \URLprefix
  \url{http://scitation.aip.org/content/aip/journal/jap/105/7/10.1063/1.3075864},
  \DOIprefix\doi{http://dx.doi.org/10.1063/1.3075864}.
%Type = Article
\bibitem[{Tsihrintzis and Nikias(1995)}]{Tsi95}
\bibinfo{author}{Tsihrintzis, G.A.}, \bibinfo{author}{Nikias, C.L.},
  \bibinfo{year}{1995}.
\newblock \bibinfo{title}{Performance of optimum and suboptimum receivers in
  the presence of impulsive noise modeled as an alpha-stable process}.
\newblock \bibinfo{journal}{IEEE Transactions on Communications}
  \bibinfo{volume}{43}, \bibinfo{pages}{904--914}.
\newblock \DOIprefix\doi{10.1109/26.380123}.
%Type = Article
\bibitem[{Ullom and Bennett(2015)}]{Ull15}
\bibinfo{author}{Ullom, J.N.}, \bibinfo{author}{Bennett, D.A.},
  \bibinfo{year}{2015}.
\newblock \bibinfo{title}{Review of superconducting transition-edge sensors for
  x-ray and gamma-ray spectroscopy}.
\newblock \bibinfo{journal}{Supercond. Sci. Technol.} \bibinfo{volume}{28},
  \bibinfo{pages}{084003}.
\newblock \URLprefix \url{http://stacks.iop.org/0953-2048/28/i=8/a=084003}.
%Type = Article
\bibitem[{Upadhyaya and Aksamija(2016)}]{Upa16}
\bibinfo{author}{Upadhyaya, M.}, \bibinfo{author}{Aksamija, Z.},
  \bibinfo{year}{2016}.
\newblock \bibinfo{title}{Nondiffusive lattice thermal transport in si-ge alloy
  nanowires}.
\newblock \bibinfo{journal}{Phys. Rev. B} \bibinfo{volume}{94},
  \bibinfo{pages}{174303}.
\newblock \URLprefix \url{https://link.aps.org/doi/10.1103/PhysRevB.94.174303},
  \DOIprefix\doi{10.1103/PhysRevB.94.174303}.
%Type = Article
\bibitem[{Urban and Grabert(2009)}]{Urb09}
\bibinfo{author}{Urban, D.F.}, \bibinfo{author}{Grabert, H.},
  \bibinfo{year}{2009}.
\newblock \bibinfo{title}{Feedback and rate asymmetry of the Josephson junction
  noise detector}.
\newblock \bibinfo{journal}{Phys. Rev. B} \bibinfo{volume}{79},
  \bibinfo{pages}{113102}.
\newblock \URLprefix \url{http://link.aps.org/doi/10.1103/PhysRevB.79.113102},
  \DOIprefix\doi{10.1103/PhysRevB.79.113102}.
%Type = Article
\bibitem[{Ustinov(1998)}]{Ust98}
\bibinfo{author}{Ustinov, A.}, \bibinfo{year}{1998}.
\newblock \bibinfo{title}{Solitons in Josephson junctions}.
\newblock \bibinfo{journal}{Physica D: Nonlinear Phenomena}
  \bibinfo{volume}{123}, \bibinfo{pages}{315--329}.
  \newblock \URLprefix \url{https://www.sciencedirect.com/science/article/pii/S0167278998001316},
  \DOIprefix\doi{10.1016/S0167-2789(98)00131-6}.
%Type = Article
\bibitem[{Valenti et~al.(2014)Valenti, Guarcello and Spagnolo}]{Val14}
\bibinfo{author}{Valenti, D.}, \bibinfo{author}{Guarcello, C.},
  \bibinfo{author}{Spagnolo, B.}, \bibinfo{year}{2014}.
\newblock \bibinfo{title}{Switching times in long-overlap Josephson junctions
  subject to thermal fluctuations and non-Gaussian noise sources}.
\newblock \bibinfo{journal}{Phys. Rev. B} \bibinfo{volume}{89},
  \bibinfo{pages}{214510}.
\newblock \URLprefix \url{https://link.aps.org/doi/10.1103/PhysRevB.89.214510},
  \DOIprefix\doi{10.1103/PhysRevB.89.214510}.
%Type = Article
\bibitem[{Vermeersch et~al.(2015a)Vermeersch, Carrete, Mingo and
  Shakouri}]{Ver15I}
\bibinfo{author}{Vermeersch, B.}, \bibinfo{author}{Carrete, J.},
  \bibinfo{author}{Mingo, N.}, \bibinfo{author}{Shakouri, A.},
  \bibinfo{year}{2015}a.
\newblock \bibinfo{title}{Superdiffusive heat conduction in semiconductor
  alloys. i. theoretical foundations}.
\newblock \bibinfo{journal}{Phys. Rev. B} \bibinfo{volume}{91},
  \bibinfo{pages}{085202}.
\newblock \URLprefix \url{https://link.aps.org/doi/10.1103/PhysRevB.91.085202},
  \DOIprefix\doi{10.1103/PhysRevB.91.085202}.
%Type = Article
\bibitem[{Vermeersch et~al.(2015b)Vermeersch, Mohammed, Pernot, Koh and
  Shakouri}]{Ver15II}
\bibinfo{author}{Vermeersch, B.}, \bibinfo{author}{Mohammed, A.M.S.},
  \bibinfo{author}{Pernot, G.}, \bibinfo{author}{Koh, Y.R.},
  \bibinfo{author}{Shakouri, A.}, \bibinfo{year}{2015}b.
\newblock \bibinfo{title}{Superdiffusive heat conduction in semiconductor
  alloys. ii. truncated L\'evy formalism for experimental analysis}.
\newblock \bibinfo{journal}{Phys. Rev. B} \bibinfo{volume}{91},
  \bibinfo{pages}{085203}.
\newblock \URLprefix \url{https://link.aps.org/doi/10.1103/PhysRevB.91.085203},
  \DOIprefix\doi{10.1103/PhysRevB.91.085203}.
%Type = Article
\bibitem[{Walsh et~al.(2017)Walsh, Efetov, Lee, Heuck, Crossno, Ohki, Kim,
  Englund and Fong}]{Wal17}
\bibinfo{author}{Walsh, E.D.}, \bibinfo{author}{Efetov, D.K.},
  \bibinfo{author}{Lee, G.H.}, \bibinfo{author}{Heuck, M.},
  \bibinfo{author}{Crossno, J.}, \bibinfo{author}{Ohki, T.A.},
  \bibinfo{author}{Kim, P.}, \bibinfo{author}{Englund, D.},
  \bibinfo{author}{Fong, K.C.}, \bibinfo{year}{2017}.
\newblock \bibinfo{title}{Graphene-based Josephson-junction single-photon
  detector}.
\newblock \bibinfo{journal}{Phys. Rev. Applied} \bibinfo{volume}{8},
  \bibinfo{pages}{024022}.
\newblock \URLprefix
  \url{https://link.aps.org/doi/10.1103/PhysRevApplied.8.024022},
  \DOIprefix\doi{10.1103/PhysRevApplied.8.024022}.
%Type = Article
\bibitem[{Weron(1996)}]{Wer96}
\bibinfo{author}{Weron, R.}, \bibinfo{year}{1996}.
\newblock \bibinfo{title}{On the Chambers-Mallows-Stuck method for simulating
  skewed stable random variables}.
\newblock \bibinfo{journal}{Stat. Probab. Lett.} \bibinfo{volume}{28},
  \bibinfo{pages}{165--171}.
  \newblock \URLprefix
  \url{https://www.sciencedirect.com/science/article/pii/0167715295001131},
  \DOIprefix\doi{10.1016/0167-7152(95)00113-1}.
%Type = Article
\bibitem[{White(1984)}]{Whi84}
\bibinfo{author}{White, M.}, \bibinfo{year}{1984}.
\newblock \bibinfo{title}{Simulation and analysis of machinery fault signals}.
\newblock \bibinfo{journal}{Journal of Sound and Vibration}
  \bibinfo{volume}{93}, \bibinfo{pages}{95 -- 116}.
\newblock \URLprefix
  \url{http://www.sciencedirect.com/science/article/pii/0022460X84903535},
  \DOIprefix\doi{https://doi.org/10.1016/0022-460X(84)90353-5}.
%Type = Article
\bibitem[{Yablokov et~al.(2021)Yablokov, Glushkov, Pankratov, Gordeeva, Kuzmin
  and Il’ichev}]{Yab21}
\bibinfo{author}{Yablokov, A.}, \bibinfo{author}{Glushkov, E.},
  \bibinfo{author}{Pankratov, A.}, \bibinfo{author}{Gordeeva, A.},
  \bibinfo{author}{Kuzmin, L.}, \bibinfo{author}{Il’ichev, E.},
  \bibinfo{year}{2021}.
\newblock \bibinfo{title}{Resonant response drives sensitivity of Josephson
  escape detector}.
\newblock \bibinfo{journal}{Chaos Solitons Fract} \bibinfo{volume}{148},
  \bibinfo{pages}{111058}.
\newblock \URLprefix
  \url{https://www.sciencedirect.com/science/article/pii/S0960077921004124},
  \DOIprefix\doi{https://doi.org/10.1016/j.chaos.2021.111058}.
%Type = Article
\bibitem[{Yakimov et~al.(2019)Yakimov, Filatov, Gorshkov, Antonov, Liskin,
  Antonov, Belyakov, Klyuev, Carollo and Spagnolo}]{Yak19}
\bibinfo{author}{Yakimov, A.V.}, \bibinfo{author}{Filatov, D.O.},
  \bibinfo{author}{Gorshkov, O.N.}, \bibinfo{author}{Antonov, D.A.},
  \bibinfo{author}{Liskin, D.A.}, \bibinfo{author}{Antonov, I.N.},
  \bibinfo{author}{Belyakov, A.V.}, \bibinfo{author}{Klyuev, A.V.},
  \bibinfo{author}{Carollo, A.}, \bibinfo{author}{Spagnolo, B.},
  \bibinfo{year}{2019}.
\newblock \bibinfo{title}{Measurement of the activation energies of oxygen ion
  diffusion in yttria stabilized zirconia by flicker noise spectroscopy}.
\newblock \bibinfo{journal}{Applied Physics Letters} \bibinfo{volume}{114},
  \bibinfo{pages}{253506}.
\newblock \URLprefix \url{https://doi.org/10.1063/1.5098066},
  \DOIprefix\doi{10.1063/1.5098066}.
%Type = Article
\bibitem[{Yang and Petropulu(2003)}]{Yan03}
\bibinfo{author}{Yang, X.}, \bibinfo{author}{Petropulu, A.P.},
  \bibinfo{year}{2003}.
\newblock \bibinfo{title}{Co-channel interference modeling and analysis in a
  Poisson field of interferers in wireless communications}.
\newblock \bibinfo{journal}{IEEE Transactions on Signal Processing}
  \bibinfo{volume}{51}, \bibinfo{pages}{64--76}.
\newblock \DOIprefix\doi{10.1109/TSP.2002.806591}.
%Type = Article
\bibitem[{Yoshimoto et~al.(2008)Yoshimoto, Shirahama and Kurosawa}]{Yos08}
\bibinfo{author}{Yoshimoto, M.}, \bibinfo{author}{Shirahama, H.},
  \bibinfo{author}{Kurosawa, S.}, \bibinfo{year}{2008}.
\newblock \bibinfo{title}{Noise-induced order in the chaos of the
  Belousov-Zhabotinsky reaction}.
\newblock \bibinfo{journal}{J. Chem. Phys.} \bibinfo{volume}{129}.
\newblock \URLprefix \url{https://aip.scitation.org/doi/10.1063/1.2946710},
  \DOIprefix\doi{10.1063/1.2946710}.
%Type = Article
\bibitem[{Zaburdaev et~al.(2015)Zaburdaev, Denisov and Klafter}]{Zab15}
\bibinfo{author}{Zaburdaev, V.}, \bibinfo{author}{Denisov, S.},
  \bibinfo{author}{Klafter, J.}, \bibinfo{year}{2015}.
\newblock \bibinfo{title}{L\'evy walks}.
\newblock \bibinfo{journal}{Rev. Mod. Phys.} \bibinfo{volume}{87},
  \bibinfo{pages}{483--530}.
\newblock \URLprefix \url{http://link.aps.org/doi/10.1103/RevModPhys.87.483},
  \DOIprefix\doi{10.1103/RevModPhys.87.483}.

\end{thebibliography}

\end{document}